\documentclass[dvips]{acta}
\usepackage{supertabular,lscape,epsfig}
\usepackage{amssymb}
\usepackage{amsmath}
\usepackage{multirow}
\DeclareSymbolFont{ppa}{OT1}{ppl}{m}{it}
\DeclareMathSymbol{\vv}{\mathalpha}{ppa}{'166}

\newfont{\hb}{rphvb at 10pt}
\newfont{\hbo}{rphvbo at 10pt}
\newfont{\bitt}{rptmbi at 12pt}
\newfont{\bits}{rptmbi at 11pt}

\SetPages{137}{167}

\SetVol{59}{2009}

\begin{document}

\newcommand{\TabCapp}[2]{\begin{center}\parbox[t]{#1}{\centerline{
  \small {\spaceskip 2pt plus 1pt minus 1pt T a b l e}
  \refstepcounter{table}\thetable}
  \vskip2mm
  \centerline{\footnotesize #2}}
  \vskip3mm
\end{center}}

\newcommand{\TTabCap}[3]{\begin{center}\parbox[t]{#1}{\centerline{
  \small {\spaceskip 2pt plus 1pt minus 1pt T a b l e}
  \refstepcounter{table}\thetable}
  \vskip2mm
  \centerline{\footnotesize #2}
  \centerline{\footnotesize #3}}
  \vskip1mm
\end{center}}

\newcommand{\MakeTableSepp}[4]{\begin{table}[p]\TabCapp{#2}{#3}
  \begin{center} \TableFont \begin{tabular}{#1} #4 
  \end{tabular}\end{center}\end{table}}

\newcommand{\MakeTableee}[4]{\begin{table}[htb]\TabCapp{#2}{#3}
  \begin{center} \TableFont \begin{tabular}{#1} #4
  \end{tabular}\end{center}\end{table}}

\newcommand{\MakeTablee}[5]{\begin{table}[htb]\TTabCap{#2}{#3}{#4}
  \begin{center} \TableFont \begin{tabular}{#1} #5 
  \end{tabular}\end{center}\end{table}}

\newfont{\bb}{ptmbi8t at 12pt}
\newfont{\bbb}{cmbxti10}
\newfont{\bbbb}{cmbxti10 at 9pt}
\newcommand{\uprule}{\rule{0pt}{2.5ex}}
\newcommand{\douprule}{\rule[-2ex]{0pt}{4.5ex}}
\newcommand{\dorule}{\rule[-2ex]{0pt}{2ex}}
\begin{Titlepage}
\Title{Galactic Fundamental Mode RR~Lyrae Stars. \\
Period--Amplitude Diagram, Metallicities and Distribution}

\Author{D.\,M.~~S~z~c~z~y~g~i~e~³,~~~G.~~P~o~j~m~a~ñ~s~k~i
~~~and~~B.~~~P~i~l~e~c~k~i}
{Warsaw University Observatory, Al.~Ujazdowskie~4, 00-478~Warsaw, Poland\\ 
e-mail: (dszczyg, gp, pilecki)@astrouw.edu.pl}
\Received{June 8, 2009}
\end{Titlepage}

\Abstract{
We have analyzed 1455 fundamental mode RR~Lyr stars of the Galactic field,
using the All Sky Automated Survey (ASAS) data. The sample covers 75\% of
the sky and contains objects in the close neighborhood of the Sun, within
4~kpc distance. Unlike in the previous analysis of the close field RRab
stars, we see a clear manifestation of the Oosterhoff groups on the
period--amplitude diagram. The relation for Oosterhoff~I type variables
becomes strongly flattened at large {\it V} amplitudes, which was not
observed for globular cluster RR~Lyr. We calculate photometric
metallicities using two available methods: one of Jurcsik and Kov\'acs
(1996) and the other of Sandage (2004). We find significant discrepancies
between results from both methods. Comparison with spectroscopic
metallicities undoubtedly favors the method of Jurcsik and Kov\'acs
(1996). In addition, we notice that RRab stars of Oosterhoff~II type might
follow a different metallicity--period--phase relation than Oosterhoff~I
type variables. The spatial distribution of Galactic field RRab stars does
not show any metallicity gradients with distance from the Galactic center
in either of the Oosterhoff groups. However, both the older, metal poor
Oosterhoff~II variables and the metal rich Oosterhoff~I RRab stars become
more concentrated to the Galactic plane with increasing metal
content.}{Stars: Population II -- Stars: variables: RR~Lyr -- Stars:
abundances -- Stars: fundamental parameters}

\Section{Introduction}
RR~Lyr type stars are radially pulsating giants of the horizontal branch on
the HR diagram. Their periods are in the range of 0.2--1.2 days and the
amplitudes of light variations are typically 0.2--1.6~mag in the {\it
V}-band. Their almost constant absolute mean magnitude (about 0.6 in {\it
V}) makes them good distance indicators and their advanced age (over
10~Gyrs) means they represent a chemical composition of the early stages of
Galaxy formation. Therefore RR~Lyr stars are excellent objects for
investigating the structure and composition of the early nearby Universe.

Since the first detection of RR~Lyr stars in globular clusters, their
number has increased rapidly over the last decade thanks to many sky
surveys. There are approximately 3000 RR~Lyr stars known in the field of
the Milky Way within the radius of $\approx4$~kpc from the Sun, identified
in the ASAS and NSVS catalogs (Pojmañski 2002, Kinemuchi \etal 2006, Wils,
Lloyd and Bernhard 2006), almost 4000 in globular clusters and the halo
(Miceli \etal 2008), 2800 in the Galactic bulge (Mizerski 2003). Recently,
an impressing number of 25\,000 RR~Lyr stars was identified in the OGLE
data of Large Magellanic Cloud by Soszyñski \etal (2009).

An interesting phenomenon concerning RR~Lyr variables is the Oosterhoff
dichotomy (Oosterhoff 1939) observed in globular clusters of the Milky Way
and absent in other nearby galaxies. The dichotomy is believed to
originate from the metallicity and age differences (Lee and Carney 1999,
Clement and Shelton 1999, Cacciari, Corwin and Carney 2005), but the final
explanation still has not been constructed.

In their paper on Galactic field RR~Lyr stars Kinemuchi \etal (2006)
addressed a problem of Oosterhoff dichotomy for field RRab stars, which has
been a matter of argument since the first work of Suntzeff, Kinman and
Kraft (1991) who showed the existence of dichotomy among the Galactic field
RRab stars. Kinemuchi \etal (2006) found that there is no clear separation
in the period--amplitude diagram as observed in the Galactic halo. That is,
the Oosterhoff dichotomy is not a property of a whole Galaxy, but only of
its outer parts. Such conclusion puts important constraints on the early
evolution of the Galaxy and in this paper we attempt to verify this result
by utilizing the Galactic field data from the All Sky Automated Survey.

In Section~2 we describe the data, in Section~3 we investigate the
period--amplitude diagram and color properties of ASAS RRab stars in order
to verify the existence of Oosterhoff dichotomy. Section~4 presents the
metallicity analysis of ASAS RRab stars and a detailed comparison of two
methods of photometric metallicity determination. In Section~5 we
investigate the Galactic distribution and we summarize the results in
Section~6. The Appendix presents the detailed comparison of ASAS and NSVS
catalogs thus explaining discrepancies between some of the results.

\Section{RR~Lyr Data}
\subsection{The ASAS Project}
The All Sky Automated Survey\footnote{http://www.astrouw.edu.pl/asas}
(ASAS) is an ongoing project dedicated to whole sky monitoring and
variability detection. Its southern observing station consists of two small
wide-field telescopes (200/2.8) located at Las Campanas Observatory in
Chile, equipped with standard {\it V} and {\it I} filters, which have been
observing the whole available sky south of $+28\arcd$ since 2000 (and some
parts since 1997). ASAS has identified and classified over 50\,000 variable
stars brighter than 14.5~mag in the {\it V}-band and published the results
in the form of the catalog (Pojmañski 2002, 2003, Pojmañski and
Maciejewski 2004, 2005, Pojmañski, Pilecki and Szczygie³ 2005). The ASAS
Catalog of Variable Stars (hereafter ACVS) contains information on star's
coordinates, {\it V} magnitude, {\it V} amplitude, period of variation,
2MASS colors and variability type. All light curves are available for
download and typically contain 300--500 good quality points. The number of
RR~Lyr stars pulsating in the fundamental mode available in ACVS is
1455. Their distribution in Galactic coordinates is presented in Fig.~1.
\begin{figure}[htb]
\centerline{\includegraphics[width=13.5cm]{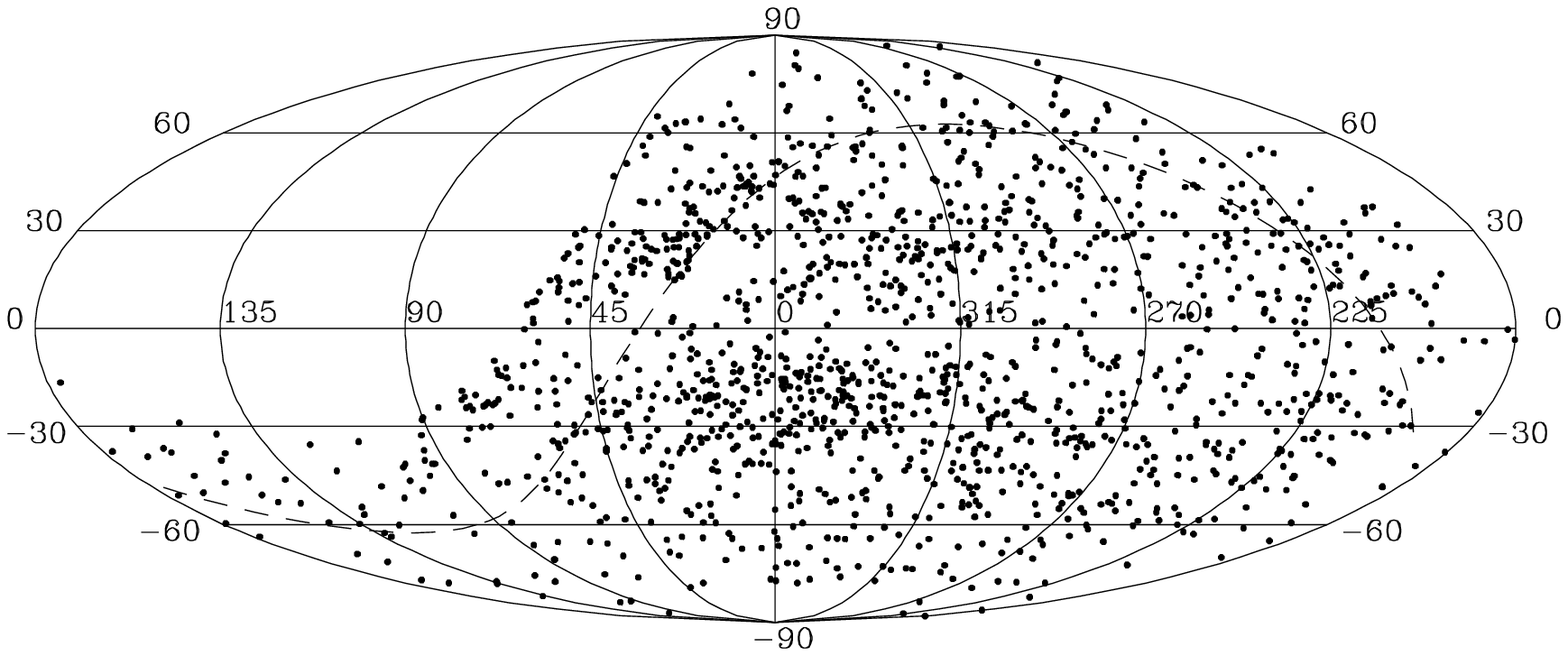}}
\vspace*{-11pt}
\FigCap{Distribution of 1455 fundamental mode RR~Lyr stars from the ASAS 
Catalog of Variable Stars (ACVS) in Galactic coordinates. The dashed line
is the equatorial plane.}
\end{figure}

\subsection{V and I-Band Light Curve Preparation}
The {\it V}-band light curves of all 1455 RRab stars in ACVS were extracted
from the catalog and only points with quality flags 'A' and 'B' were
retained. All light curves were subject to preliminary cleaning and
frequency analysis in order to refine their pulsation period values, which
is especially useful in the case of objects from older parts of the
catalog, when less data points were available (R.A. 0--6h). The light
curves of stars for which new and old periods differed by more than 10\%
were visually examined and the proper period was assigned. A few of them
turned out to be other type variables, usually Cepheids or contact
binaries, a few had poor quality light curves and their variability type
was hard to define. At this stage a chosen period was fixed for the
following analysis. The number of stars was reduced to 1437.

Then an iterative cleaning process was performed. First, all points
deviating more than $3\sigma$ were removed. Then a 6 harmonic sine Fourier
model was fitted to the light curve and again all points deviating by more
than $3\sigma$ were removed. This step was repeated once more. Afterward, a
6 harmonic model was subtracted from the light curve and a linear trend was
subtracted. Then the Fourier model was restored and once again we performed
a 6 harmonic fit but now without a linear trend. Finally, we cleaned the
light curve for the last time and restored the linear model.

The ASAS light curve catalogs contain only {\it V}-band observations, but
ASAS was observing simultaneously in the {\it I}-band also. The {\it
I}-band data has recently been reduced and will soon be publicly
available. For the purpose of this paper we extracted light curves of
RR~Lyr stars in order to obtain {\it I} magnitudes and amplitudes. The
average number of good observations (flags 'A' and 'B') per star is around
150--250, but there were some objects with less than 30 measurements and
those were rejected from further analysis. All 1423 remaining light curves
were cleaned in the same way as in the {\it V}-band.

After the preparation process we once again fitted a sine Fourier series to
all phased {\it V}- and {\it I}-band light curves independently, in the
form:
$$m=a_0+\sum_{i=1}^N a_i\sin(2\pi ix+\varphi_i)\eqno(1)$$
where $x=({\rm HJD-HJD_0})/P$ and $P$ is the pulsation period. ${\rm
HJD_0}$ stands for the epoch of maximum brightness. The number of harmonics
$N$ was set individually for each star, using a slightly modified method of
Kov\'acs (2005) and it varied from 6 to 10, depending on the signal to
noise ratio (smaller order of the fit for more noisy light curves). For
each of 1423 RRab variables all Fourier parameters $a_0\dots a_N$ and
$\varphi_1\dots\varphi_N$ extracted from more numerous {\it V}-band light
curves were stored as well as mean and maximum brightness magnitudes and
amplitudes of variation, in both bands. They are available for download
from the ASAS website. A few exemplary lines of the file containing Fourier
parameters are presented in Table~1.

\renewcommand{\arraystretch}{1.1}
\MakeTable{@{}ccccccccc@{}}{12cm}{Fourier parameters of {\it V}-band 
light curves}
{\hline
\noalign{\vskip2pt}
 ID & N &$a_0$ & $a_{\rm 0,err}$ & $a_1$ &  $a_{\rm 1,err}$ & $\varphi_1$ & 
$\varphi_{\rm 1,err}$ &  ...  \\
\noalign{\vskip2pt}
\hline
\noalign{\vskip2pt}
000036+2639.8 &  6 & 13.218 & 0.019 &  0.331 & 0.026 & 4.078 & 0.086 & ... \\
000248-2456.7 & 10 & 10.297 & 0.002 & -0.403 & 0.002 & 0.825 & 0.005 & ... \\
000301-7041.5 &  6 & 13.707 & 0.011 &  0.333 & 0.015 & 4.228 & 0.048 & ... \\
\noalign{\vskip2pt}
\hline
\noalign{\vskip2pt}
\multicolumn{9}{p{10.5cm}}{First three lines of the file containing Fourier
parameters for 1423 ASAS RRab stars. The columns contain the ASAS ID of a
star, number of harmonics $N$, and a series of Fourier parameters with their
errors, from 0 to $N$: $a_0,a_{\rm 0,err},a_1, a_{\rm 1,err},\varphi_1,
\varphi_{\rm 1,err},\dots,a_N,a_{N,{\rm err}},\varphi_N,\varphi_{N,{\rm
err}}$. The complete file is available for download from the ASAS website.}}
\renewcommand{\arraystretch}{1}

\Section{Period, Amplitude and Color of ASAS RRab Variables}
Fig.~2 shows three histograms of RRab variables: the distribution of {\it
V}-band magnitudes, amplitudes and periods. ASAS limiting magnitude is
roughly $V=14$~mag, putting a distance limit at 4.5~kpc but a limiting
magnitude at which detection efficiency drops significantly is 13.8~mag,
thus we reject all objects fainter than that. This leaves us with 1227
stars. The {\it V}-band amplitudes of ASAS RRab stars are in the range of
0.16--1.5~mag and are defined as a difference between the maximum and the
minimum value of the light curve model as described by Eq.~(1). There is a
visibly lower number of variables with amplitudes smaller than 0.5~mag,
which is an effect of lower detection efficiency. The mean amplitude value
is 0.88~mag.  Pulsation periods are in the range of 0.2 to 1.2 days with
the mean value of 0.56~days, very close to 0.55~days which is an average
period value for Oosterhoff~I type globular cluster RRab stars.
\begin{figure}[htb]
\begin{center}
\begin{tabular}{c c c}
\includegraphics[width=4cm]{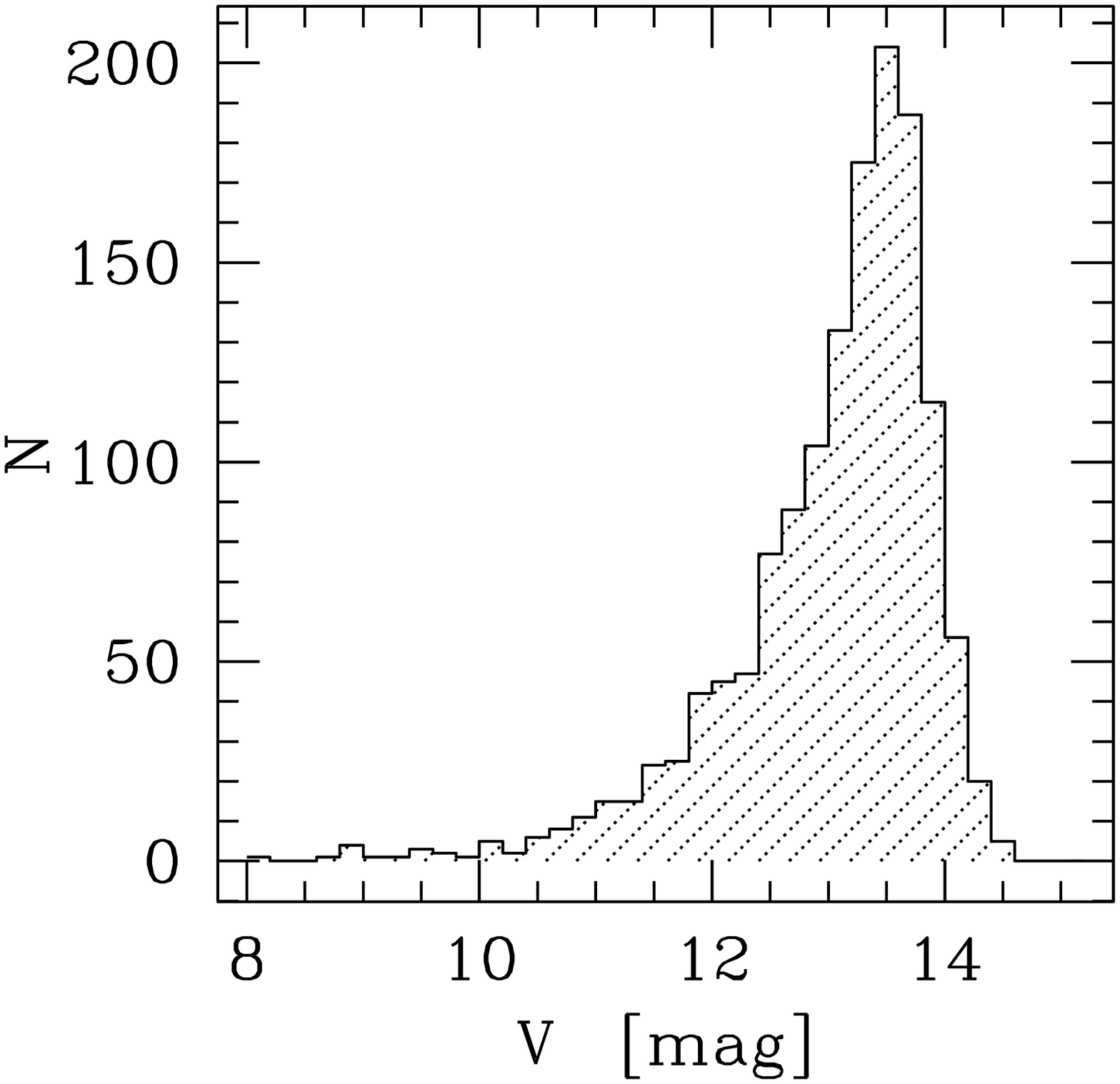} &
\includegraphics[width=4cm]{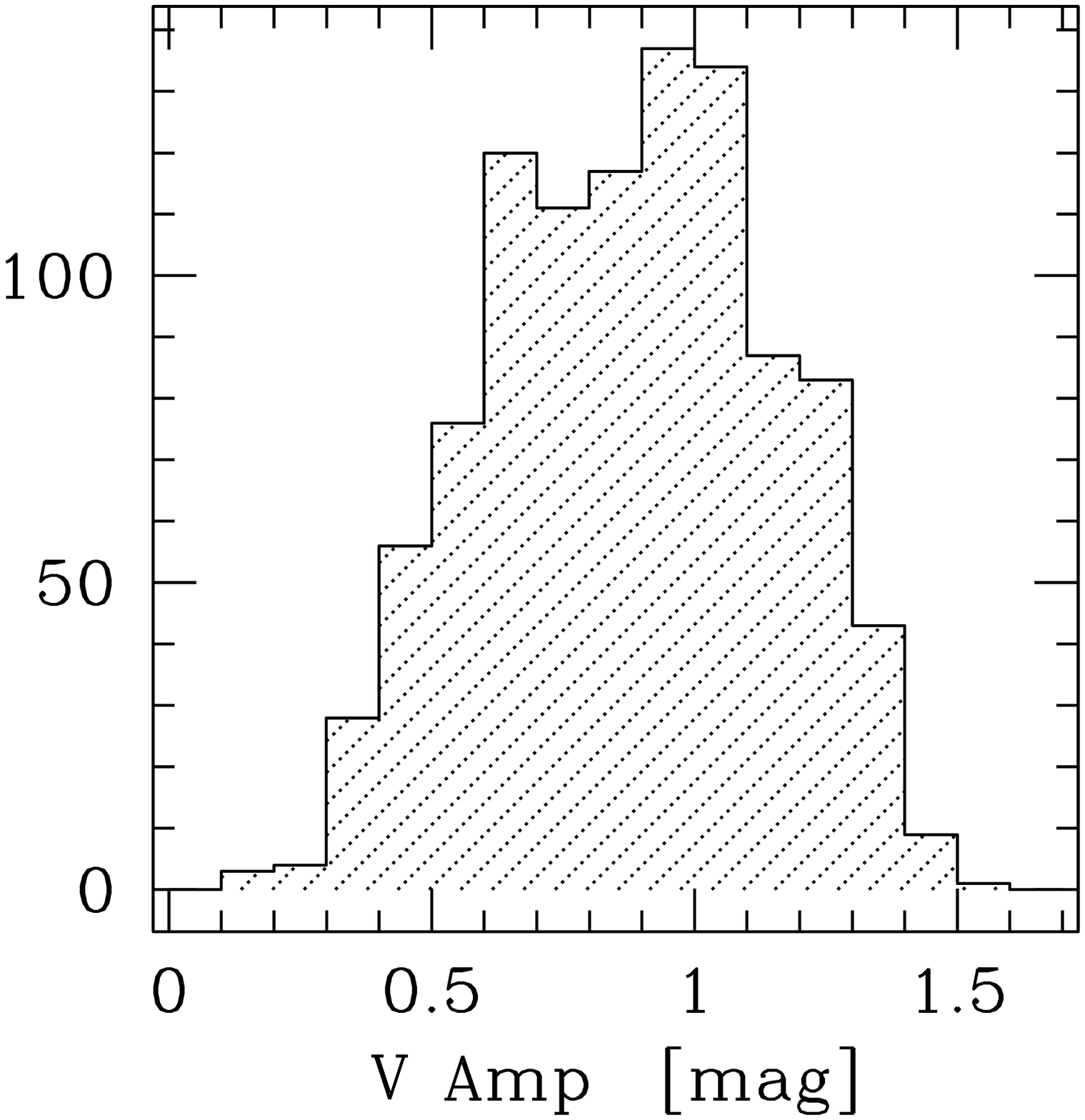} &
\includegraphics[width=4cm]{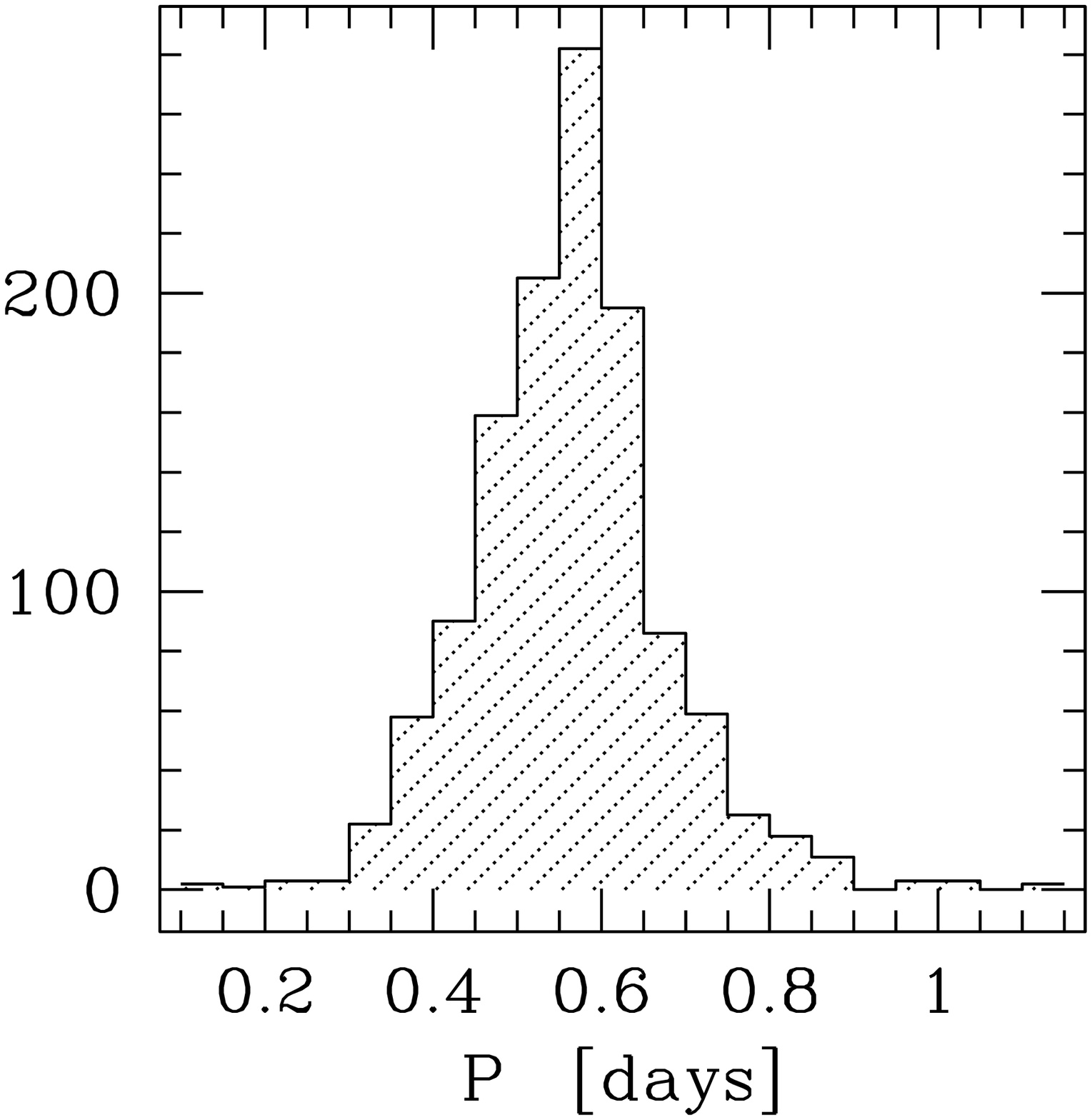} \\
\end{tabular}
\end{center}
\FigCap{Distributions of {\it V}-band average magnitudes ({\it left panel}),
amplitudes ({\it middle panel}) and pulsation periods ({\it right panel})
for 1423 RRab stars from ACVS.}
\end{figure}

\subsection{The Period--Amplitude Diagram}
Soon after the extensive works of Bailey and Pickering (1913) on cluster
variables it has been discovered that globular clusters of our Galaxy fall
into two distinct groups according to their RR~Lyr mean pulsation periods
and a number of RRc variables relative to number of RRab stars. These two
groups had been called Oosterhoff~I and~II (Oosterhoff 1939) and we will
refer to them as Oo~I and Oo~II. Mean periods of RRab pulsators are
$\langle P_{ab}\rangle=0.55$~d (for Oo~I) and $\langle
P_{ab}\rangle=0.64$~d (for Oo~II) and the number of RRc stars relative to
the number of RRab stars is smaller for Oo~I ($N_c/N_{ab}=0.17$) than for
Oo~II ($N_c/N_{ab}=0.44$). It was also noticed that these two groups
possess different metallicities, with Oo~I clusters being more metal rich
than Oo~II. In the later studies it became clear that while the separation
into Oosterhoff groups is significant, there is a continuum of period and
metallicity values within each group (Sandage 1982).

When plotted in the log period--amplitude plane, RRab stars from different
Oosterhoff clusters follow two separate parallel relations such that longer
period stars have smaller amplitudes, and for a given amplitude value Oo~II
RRab stars have longer periods. There have been many attempts to explain
this phenomenon and the research history is well described in the work of
Lee and Carney (1999). For now it is agreed that the main cause of this
period shift is the luminosity difference caused by a different
evolutionary state of RR~Lyr variables -- RRab stars that follow
period--amplitude relation typical of Oo~II globular clusters are believed
to be more evolved than the ones following Oo~I relation (Lee and Carney
1999, Clement and Shelton 1999, Cacciari, Corwin and Carney 2005).

The situation is somewhat different in the case of Galactic halo RR~Lyr
stars. As summarized by Catelan (2009), there has been no agreement whether
the Oosterhoff dichotomy is present in the field of the Galaxy or not. In
the first works regarding this subject, Suntzeff \etal (1991) showed that
it does exist, but later Vivas and Zinn (2003) proved the opposite using
the QUEST data.  Recently Kinemuchi \etal (2006) constructed a
period--amplitude (Bailey) diagram for about 600 close field RRab stars,
within the distance of 4.5~kpc from the Sun, using data of the Northern Sky
Variability Survey\footnote{http://skydot.lanl.gov/nsvs/nsvs.php} (NSVS).
In their Fig.~12 the authors show that there are no clear Oosterhoff groups
visible on the diagram and there is rather a continuum of periods and
amplitudes within their sample, although the majority of RRab stars are
located close to the region occupied by Oo~I globular cluster RR~Lyr
stars. The same diagram was plotted for over 800 RRab stars of the Galactic
halo (excluding globular cluster variables), spanning the distances from
about 4 to 20~kpc from the Sun (Miceli \etal 2008), based on data collected
by the Lowell Observatory Near Earth Objects Survey Phase~I
(LONEOS-I). Here, the majority of RR~Lyr variables are undeniably
concentrated in the regions corresponding to those occupied by Oosterhoff
groups in globular clusters, especially in the region of Oo~I group.

The resulting picture is that RRab variables from the Galactic halo display
the separation into Oosterhoff groups, typical of globular clusters, while
very close field RRab stars do not. According to Lee and Carney (1999) the
age differences between Oo~I and Oo~II groups are about 2--3~Gyrs (Oo~I
being younger), which implies not only different formation times, but
possibly a different origin of Oo~I and Oo~II RR~Lyr stars and sheds some
light on the possible formation history scenarios. In particular, Oo~II
type clusters are believed to have formed early in the proto Galaxy
formation stage, and Oo~I type 2--3~Gyrs later. Also, some Oo~I clusters
could have joined the Galaxy by accretion processes (van den Bergh 1993),
but this hypothesis has been recently questioned, since RR~Lyr stars in
neighboring dwarf galaxies do not display the Oosterhoff dichotomy, but
quite the contrary -- they fall into the Oosterhoff gap on the Bailey
diagram. Absence of the Oosterhoff effect in Sun's proximity would imply
that there was rather a continuous creation rate of Galactic RRab stars,
without particular ``events'' such as accretion processes, at least in
regions close to the Galactic plane. Another scenario could allow for the
accretion processes to add up to the Galactic halo content.

We construct a Bailey diagram for the ASAS fundamental mode RR~Lyr stars in
Fig.~3. There is a continuum of periods and amplitudes, but unlike in the
NSVS sample, we see a clear manifestation of the Oosterhoff groups with the
majority of objects lying in the Oo~I group region. Given the similarity of
ASAS and ROTSE-I telescopes, the ACVS and NSVS catalogs have practically
the same angular resolution, magnitude range and thus distance limit to
RR~Lyr stars. In this case we would expect them to have uniform RR~Lyr
samples exhibiting same properties, which is not observed on the Bailey
diagram. In order to check the conformity of both catalogs we investigated
RRab variables in the overlapping region of ASAS and NSVS, between
declinations $-38\arcd\div+28\arcd$. We found that corresponding objects
from ASAS and NSVS catalogs often have significantly different amplitudes,
even up to 0.5~mag.  Thus, different distributions in period--amplitude
diagram most probably originate from wrongly determined amplitudes of NSVS
RR~Lyr stars, possibly a result of unfiltered ROTSE observations. The
details of ASAS and NSVS catalogs comparison are presented in the Appendix.
\begin{figure}[htb]
\begin{center}
\begin{tabular}{@{}c@{}c@{}}
\includegraphics[width=6cm]{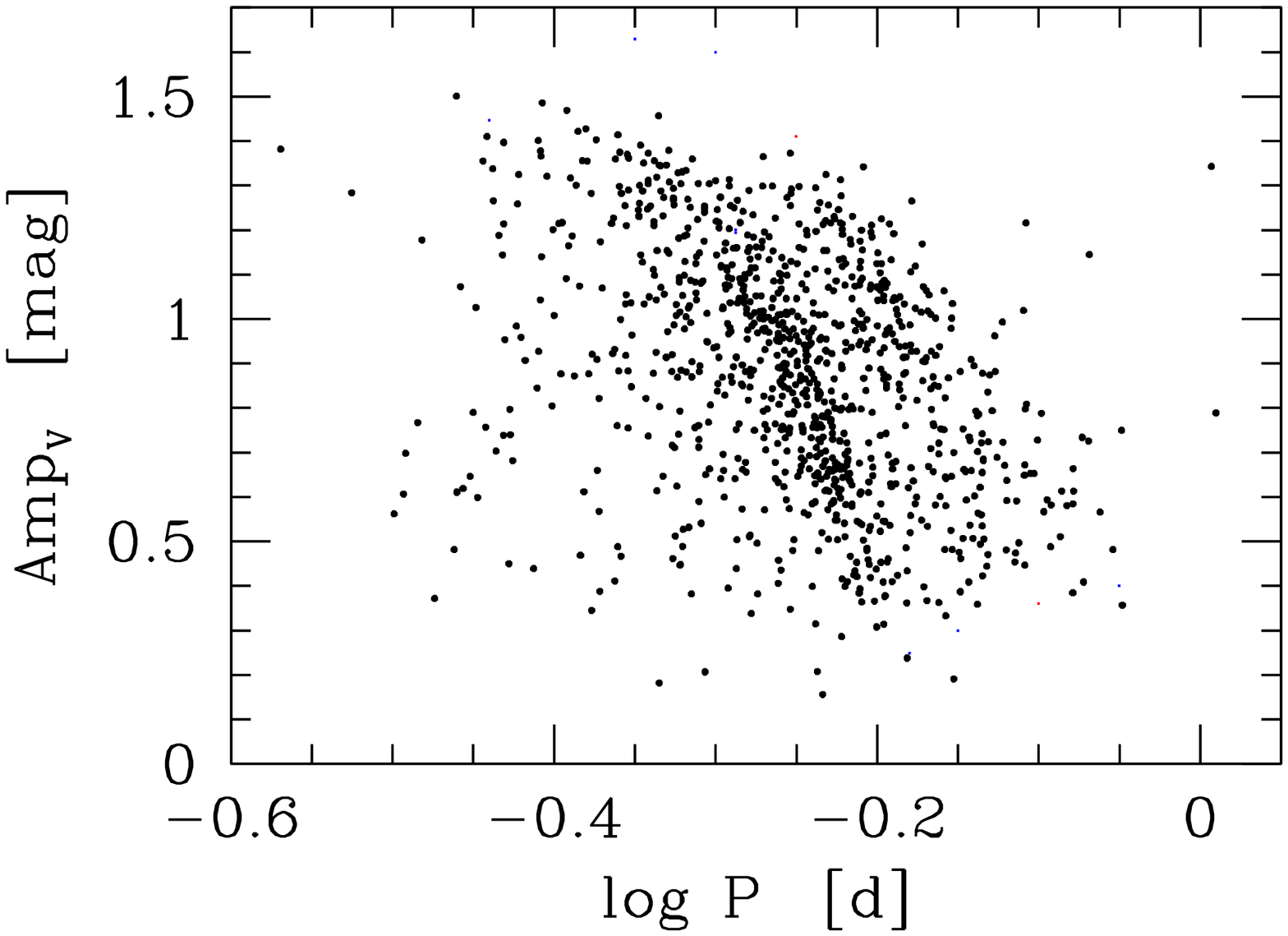} &
\includegraphics[width=6cm]{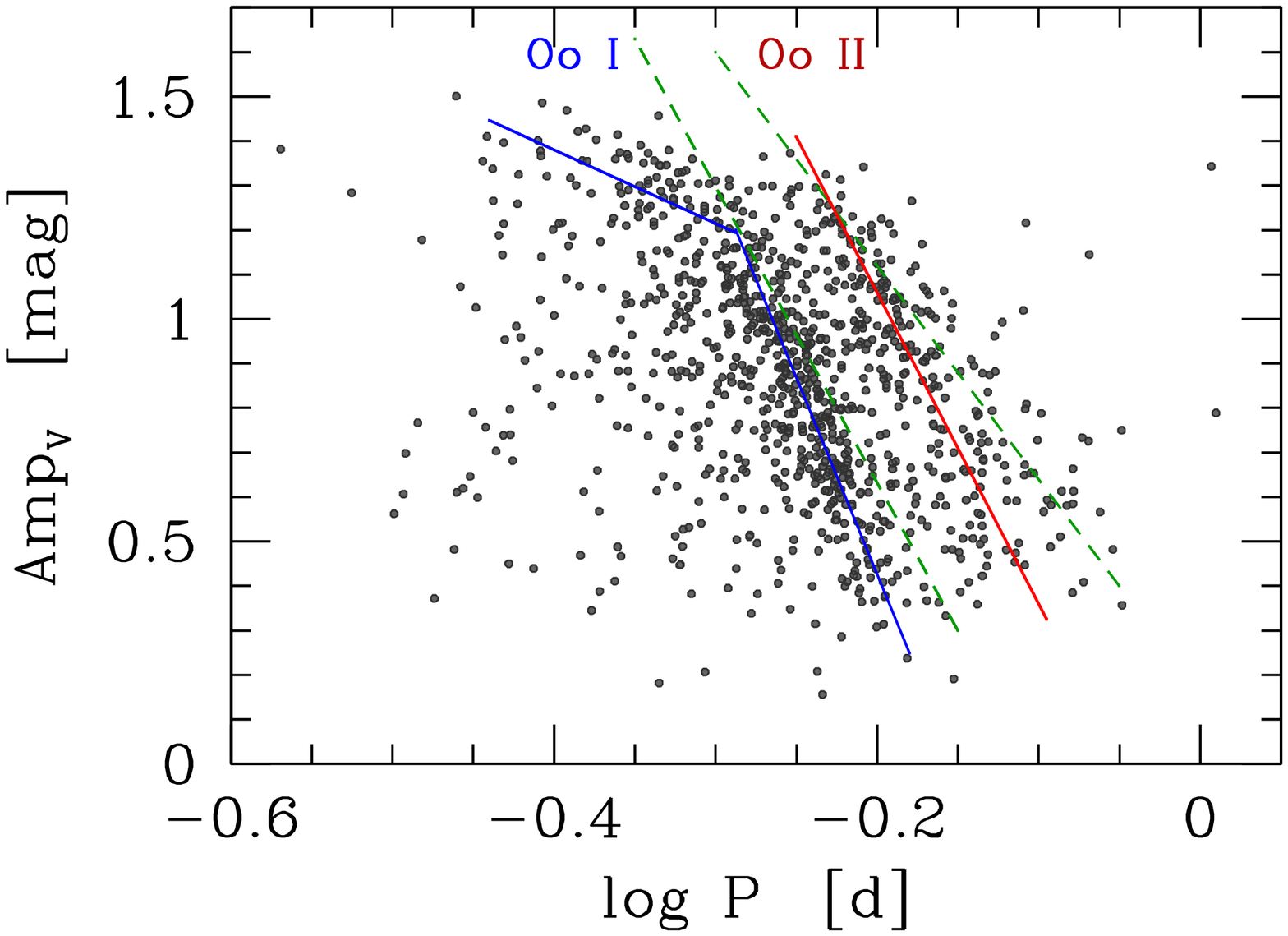} \\
\end{tabular}
\end{center}
\vspace*{-11pt}
\FigCap{{\it Both panels} present a period--amplitude (Bailey) diagram for ASAS
RRab stars. The green dashed lines overplotted on the {\it right panel}
image represent relations for M3 (Oo~I globular cluster) and $\omega$~Cen
(Oo~II globular clusters) derived by Clement and Rowe (2000). The solid
lines are the fits to Oo~I (blue) and Oo~II (red) groups derived in the
course of this work.}
\end{figure}

For reference, in the right panel of Fig.~3 we overplot the Oosterhoff
relations for Oo~I and Oo~II globular clusters used by Kinemuchi \etal
(2006) in the NSVS plots (their Figs.~12, 13 and 26), and derived by
Clement and Rowe (2000) for clusters M3 (Oo~I) and $\omega$~Cen (Oo~II). It
is well visible that these relations are somewhat shifted with reference to
the main concentration of ASAS data points. That is, at constant amplitude
value ASAS Oosterhoff groups have shorter average periods than globular
cluster Oosterhoff groups. Since this $\log P$ shift is explained as the
difference in age and/or evolutionary advancement this would imply that the
evolutionary advancement is different among field RRab stars of the two
Oosterhoff types and typical globular cluster RRab stars. If the Oosterhoff
concentrations can be credited to some episodes in the Galaxy formation
(accretion from neighboring galaxies has been suggested) then their
different characteristics among ASAS RRab stars means that some other
processes took place in the proximity of the Galactic plane. Of course
there is a continuum of points on the Bailey diagram indicating that there
was an RR~Lyr stars production at all times of Galaxy formation. The same
plot constructed for a subsample of bright RR~Lyr stars with well defined
amplitudes does not change the location and the shape of the two Oosterhoff
relations.

The period--amplitude relation for Oosterhoff~I type variables becomes
strongly flattened at large {\it V} amplitude values, which was not
observed for globular cluster RR~Lyr before. While it is true that Cacciari
\etal (2005) noticed that a quadratic relation provides a better fit to
their data for Oo~I type cluster M3, the curvature of their relation is
small. In the case of ASAS RRab stars the best fit to Oo I group is a
fourth order polynomial, but we choose a much simpler two line fit, as the
relation becomes strongly rather than smoothly flattened at high
amplitudes:
$${\rm Amp_V}=-8.844\times\log P-1.343\qquad {\rm Amp_V}<1.2,\eqno(2)$$
$${\rm Amp_V}=-1.654\times\log P+0.719 \qquad{\rm Amp_V}\ge 1.2.\eqno(3)$$

The above relations are overplotted in Fig.~3 with blue solid line. The
reason that this ``flattening'' effect was not observed before might
indicate that there is a lack of high amplitude RRab satrs in globular
clusters.  Another possible reason is small number statistics for a given
cluster. We also checked that this effect is not due to wrongly determined
RR~Lyr stars amplitudes.

The relation for Oosterhoff II type RRab stars remains simple, being a one
linear equation of the form (red solid line in Fig.~3):
$${\rm Amp_V}=-7.007\times\log P-0.343.\eqno(4)$$
\vspace*{-27pt}
\begin{figure}[h]
\begin{center}
\includegraphics[width=6.5cm]{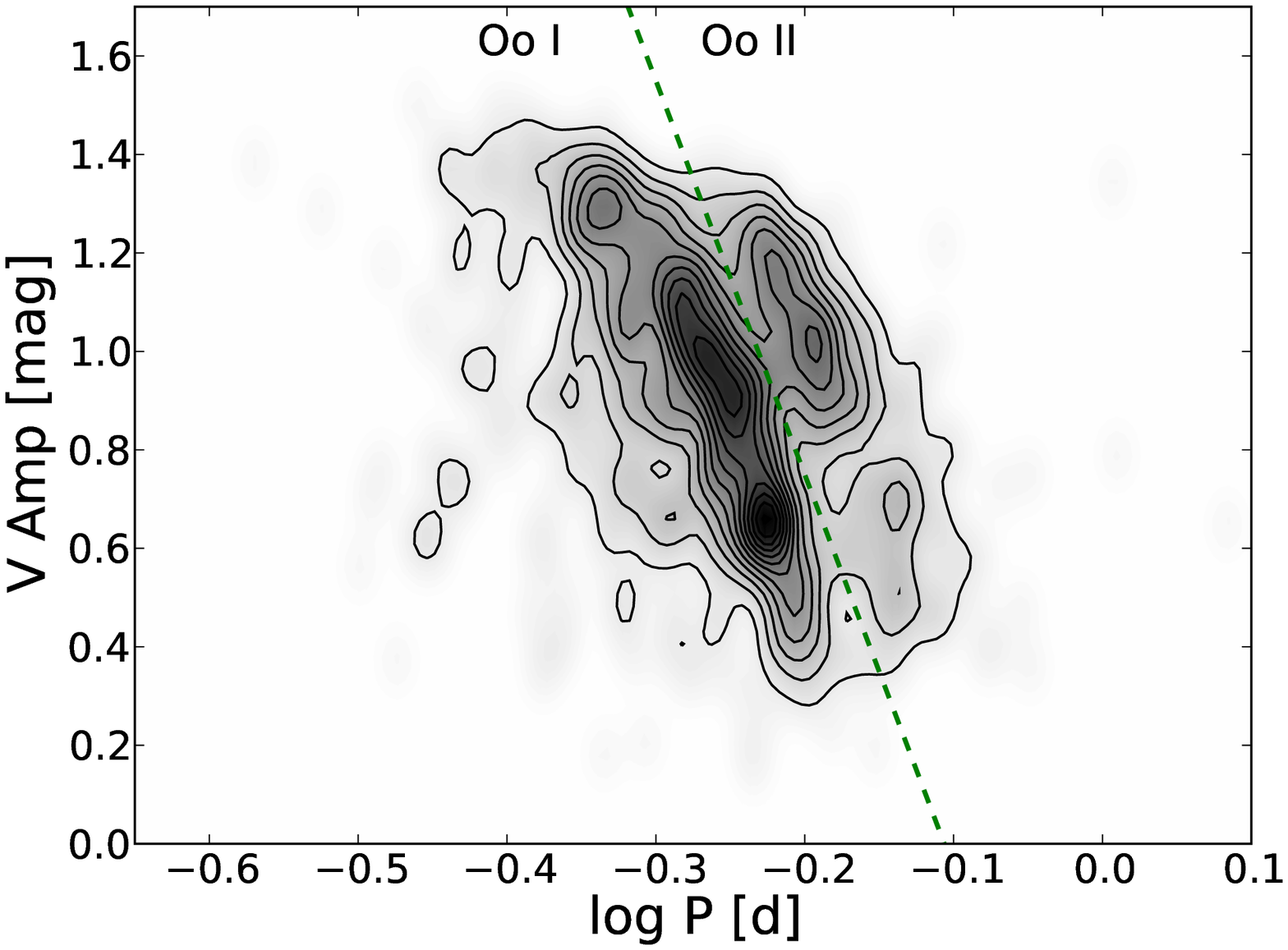} 
\end{center}
\vspace*{-11pt}
\FigCap{Density maps of the period--amplitude (Bailey) diagram for
ASAS RRab stars. The dashed line separates Oo~I and Oo~II type
variables and it follows Eq.~(5).}
\end{figure}

We draw density maps of the Bailey diagram in Fig.~4 to show that the
highest concentration of RRab stars in Oosterhoff~I group indeed follows a
relation of different shape than in the case of globular cluster or distant
halo RRab stars. As for Oosterhoff~II group it seems that the relation is
linear although there are no high amplitude Oo~II type RRab stars. Another
interesting feature in Fig.~4 are higher concentrations of ASAS RRab stars
at certain values of period and amplitude. We do not know whether this is
simply an observational effect or RRab stars indeed prefer certain values
of period and amplitude and are likely to adopt them. However, these
concentrations are present in density maps plotted for limited subsamples
of RRab stars which suggests this effect is real.

For further analysis we separate Oosterhoff I and II groups with a
dashed line in Fig.~4, described by an equation:
$${\rm Amp_V}=-8.0\times\log P-0.85.\eqno(5)$$

\subsection{The Period--Color Diagram}
We plot ASAS fundamental mode RR~Lyr stars on a period--color diagram in
Fig.~5. We use two $V-I$ color values: at maximum light (left panel) and at
minimum light (right panel). There is no evident dependence of $V-I$ color
on period at minimum light but it appears at maximum light as was
already observed for MACHO RRab stars by Kanbur and Fernando (2005) and is
explained by interaction between the photosphere and ionization front. In
addition, our plots show a separation into Oosterhoff groups visible only
at maximum light, at minimum light both groups are mixed. Oo~I RRab stars are
marked with blue open squares and Oo~II with red filled circles. The groups
were separated with the line described by Eq.~(5).
\begin{figure}[htb]
\begin{center}
\begin{tabular}{@{}c@{}c@{}}
\includegraphics[width=6.4cm]{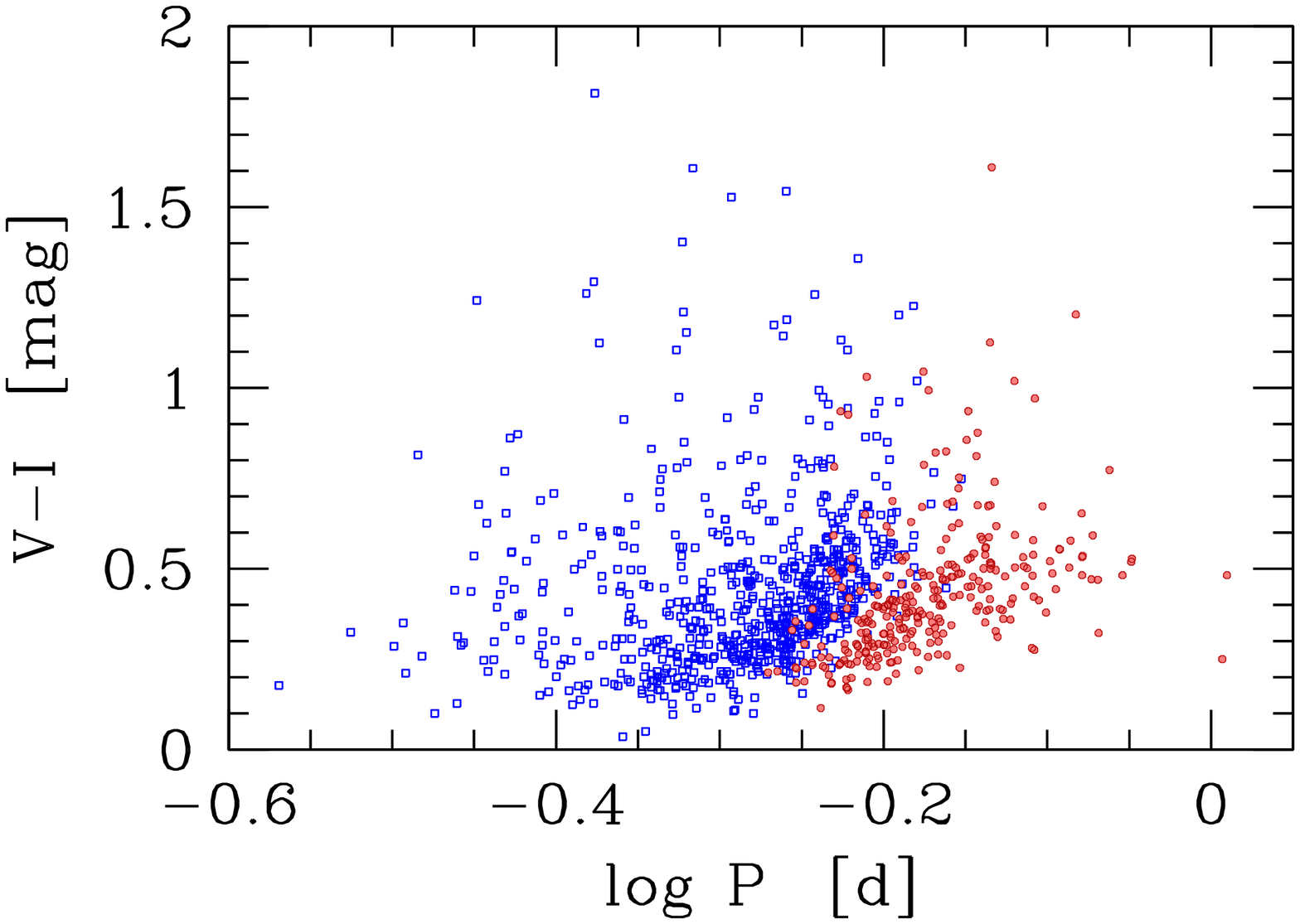} &
\includegraphics[width=6.4cm]{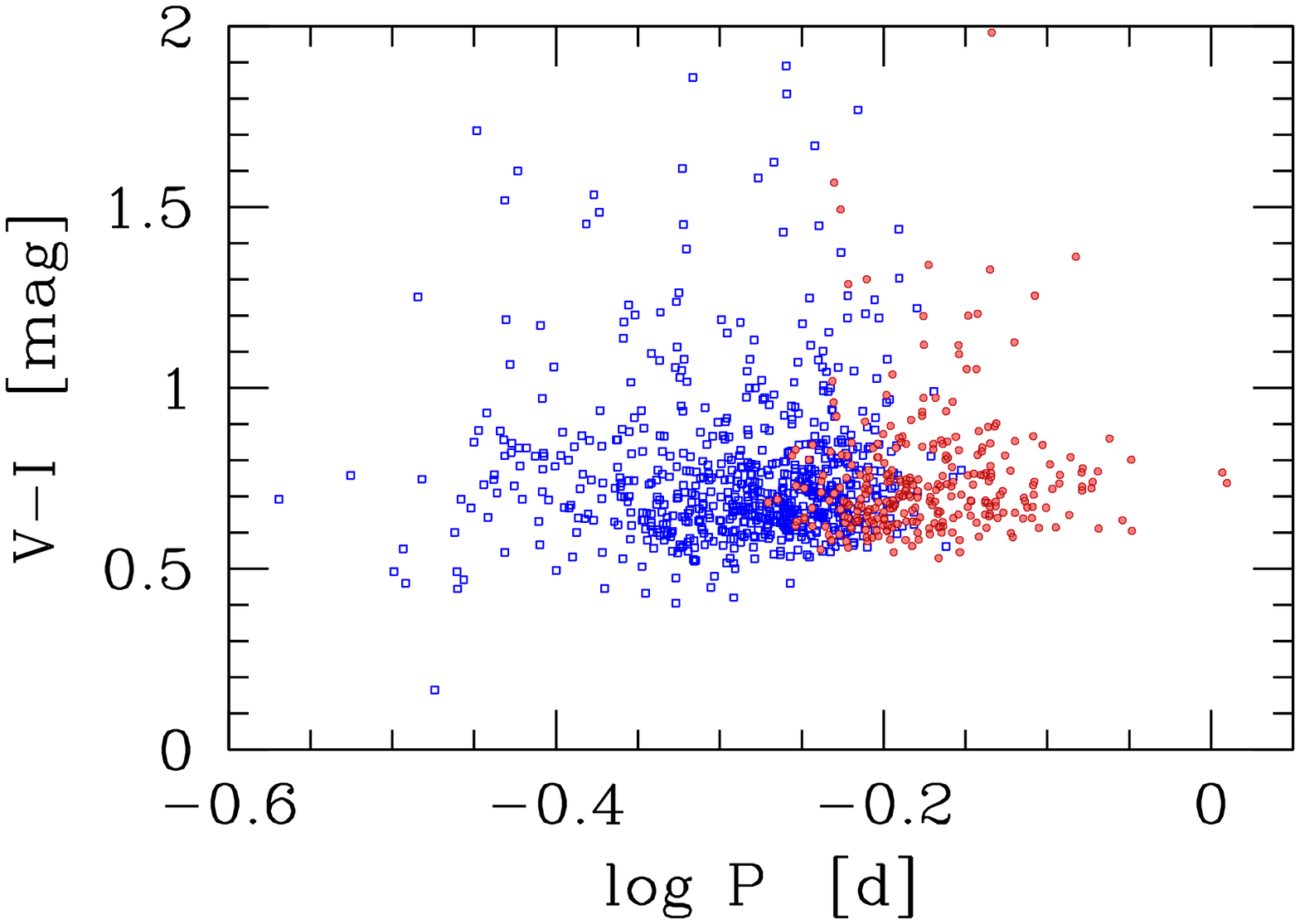} \\
\end{tabular}
\end{center}
\vspace*{-11pt}
\FigCap{Period--color diagram for ASAS RRab stars. {\it Left panel} presents 
$V-I$ color at maximum light, and the {\it right panel} at minimum light.}
\end{figure}

The data used in this investigation is available for download from the ASAS
website. The form of the file is presented in Table~2.

\MakeTable{@{}c@{\ \ \ }c@{\ \ \ }c@{\ \ \ }c@{\ \ \ }c@{\ \ \
}c@{}c@{}c@{}}{12cm}{ASAS RRab stars data}
{\hline
\noalign{\vskip2pt}
 ID & RA  & Dec   & l     & b     & P      & {$\rm HJD_0 - 2450000$} &  \\
    & [h] & [deg] & [deg] & [deg] & [days] & [days]        &      \\
\noalign{\vskip2pt}
\hline
\noalign{\vskip2pt}
000248-2456.7 & 0.046667 & $-24.945$  & 41.513967 & $-78.863793$ & 0.4933545 & 1870.54 & \\
000301-7041.5 & 0.050278 & $-70.69167$ & 308.649749 & $-45.895328$ & 0.5538941 & 1872.08 & \\
000321+0323.9 & 0.055833 & 3.39833 & 100.269549 & $-57.342245$ & 0.5790441 & 1870.42 & \\
&&&&&&&\\
\noalign{\vskip2pt}
\hline
\noalign{\vskip2pt}
 $V_{\rm min}$ & $V_{\rm max}$ & $V_{\rm avg}$ & ${\rm Amp}_V$  &
 $I_{\rm min}$ & $I_{\rm max}$ & $I_{\rm avg}$ & ${\rm Amp}_I$ \\ 
 {\rm [mag]} & [mag] & [mag] & [mag] & [mag] & [mag] & [mag] & [mag] \\
\noalign{\vskip2pt}
\hline
\noalign{\vskip2pt}
10.749 &  9.543 & 10.307 & 1.248 & 10.161 &  9.358 &  9.820 & 0.833 \\
14.034 & 13.139 & 13.683 & 0.877 & 13.392 & 12.846 & 13.190 & 0.572 \\
13.924 & 13.174 & 13.612 & 0.842 & 13.375 & 12.828 & 13.069 & 0.640 \\
\noalign{\vskip2pt}
\hline
\noalign{\vskip2pt}
\multicolumn{8}{p{12cm}}{The file contains 15 columns described in the
header of the table and its full version is available for download from the ASAS
website.}}

\Section{Metallicity Determination}
In the era of photometric sky surveys, the number of spectroscopic
observations from which metallicities are usually extracted is a few orders
of magnitude lower than the amount of photometric observations. Fortunately
for fundamental mode RR~Lyr variables photometry itself can be used to
calculate their metallicities with satisfactory accuracy. There are two
widely used methods of metallicity determination for RR~Lyr stars using
their light curves: one by Jurcsik and Kov\'acs (1996) and one by Sandage
(2004). We will refer to them as JK96 and S04.

\subsection{Methods of Metallicity Determination}
In the method of JK96 metallicity is calculated from the period of the star
$P$ and the phase combination $\varphi_{31}=\varphi_3-3\varphi_1$, where
$\varphi_3$ and $\varphi_1$ phases come from the Fourier sine decomposition
of the {\it V}-band light curve (see Eq.~1):
$${\rm [Fe/H]_{\rm JK}}=-5.038-5.394P+1.345\varphi_{31}.\eqno(6)$$
S04 transformed ${\rm [Fe/H]_{\rm JK}}$ to the Zinn and West (1984) system that
will be used in this paper:
$${\rm [Fe/H]_{\rm JKZW}}=1.05\times{\rm [Fe/H]_{\rm JK}}-0.20.\eqno(7)$$
This method was already successfully applied to a small sample of ASAS RRab
stars by Kov\'acs (2005). In Fig.~6 we compare metallicities calculated by
us (details will follow in the next section) with the ones computed by
Kov\'acs (2005) for a set of ASAS light curves. The whole sample consists
of 101 variables, and the filled circles (28 stars) represent a subsample
that was also present in the paper of JK96.  The average difference between
metallicity values is $\langle{\rm [Fe/H]_{\rm diff}}\rangle=\langle{\rm
[Fe/H]_{\rm JK}-[Fe/H]_{\rm Kov}}\rangle=0.02$~dex and the scatter is
$\sigma_{\rm [Fe/H]_{diff}}=0.29$~dex. The number of objects falling in the
$\pm0.3$~dex region (marked with dotted lines) is 73 which constitutes
72\%. Given that metallicity values were obtained by means of the same
method and applied to almost the same light curves, we would expect to
obtain much more consistent results. However, light curves in the sample of
Kov\'acs contain about 30\% less points than our sample and this may be the
source of the discrepancy. Nevertheless, the inconsistency is still
disturbing. For now we can assume that the 0.29 dex scatter is what we can
expect from the Fourier decomposition procedure uncertainties. Later in the
paper we will assume that 0.3~dex difference in metallicities is an
intrinsic JK96 method error.
\begin{figure}[htb]
\begin{center}
\includegraphics[width=7cm]{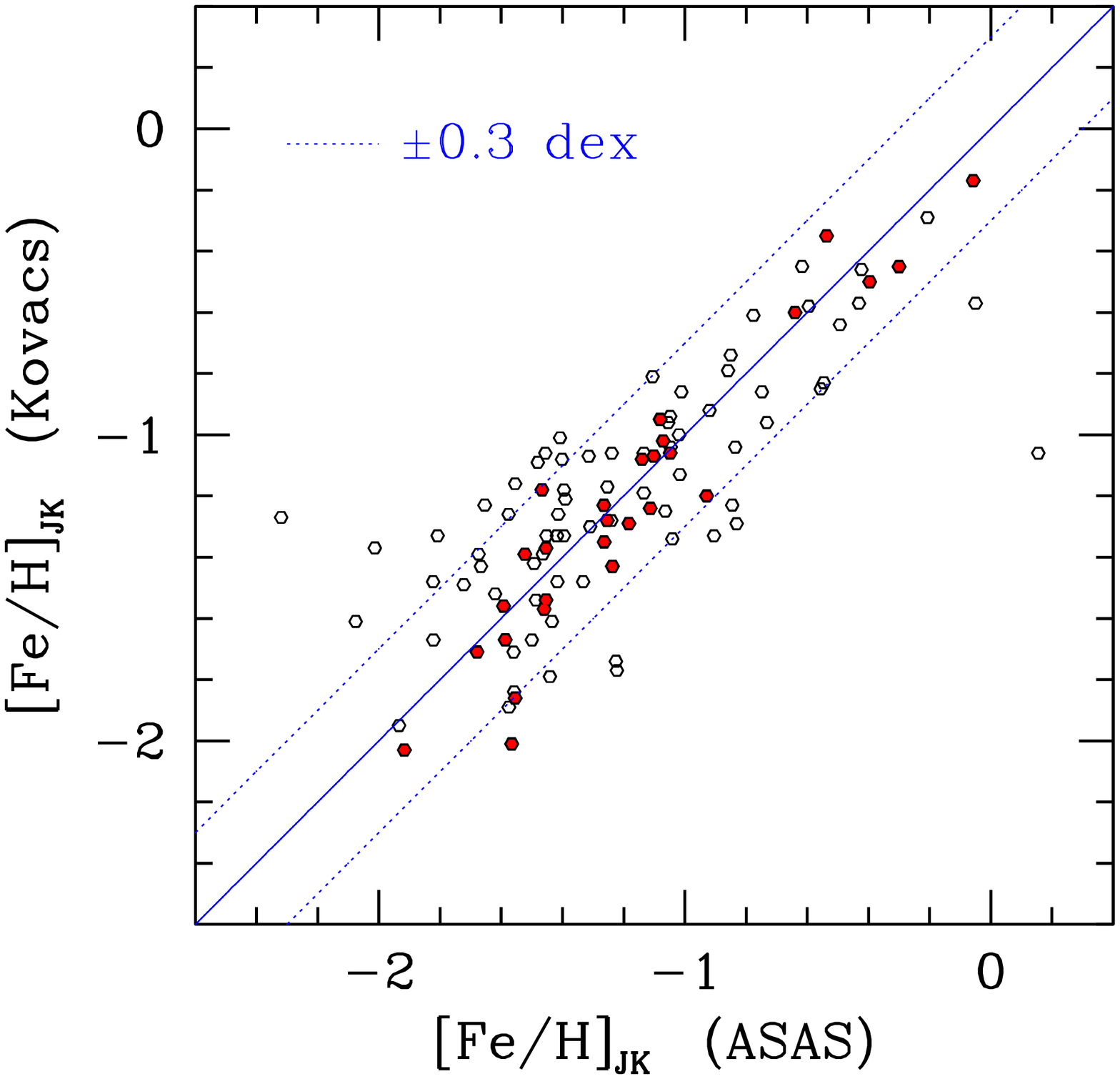} 
\end{center}
\vspace*{-7pt}
\FigCap{Comparison of [Fe/H] values from Kov\'acs (2005) and this paper.
Both were computed with the method of JK96 (Eq.~6) for a set of 101 ASAS
light curves. Filled circles (28 stars) represent a subsample that was also
present in the paper of JK96. The average difference between the two
determinations is 0.02~dex and the scatter is 0.29. The dotted lines
delimit the $\pm0.3$~dex region.}
\end{figure}

The method of S04 employs star's period and {\it V}-band amplitude:
$${\rm [Fe/H]_S}=-1.453A_V-7.990\log P-2.145\eqno(8)$$ and is less
sensitive to light curve quality -- it is always easier to determine
accurately a total amplitude than phases of Fourier harmonics. However, an
error of 0.1~mag in amplitude gives an error of almost 0.15~dex in
metallicity. Such uncertainties usually occur for Blazhko RR~Lyr stars, as
the amplitudes averaged over several cycles are usually lower than for non
Blazhko variables producing larger metallicity values.

We will compare these two methods in the next Sections.

\subsection{Metallicity Calculation for the Sample of ASAS RRab Stars}
We compute photometric metallicities for the 1227 ASAS RRab stars using
both the method of JK96 and S04 (Eqs.~7 and~8). We have already refined
(and in some cases recalculated) pulsation periods supplied by the ACVS and
we can assume that they are correct to very high accuracy. In order to
estimate light curve parameters $A_V$ and $\varphi_{31}$ we decomposed the
light curves into Fourier sine series, as described in the last paragraph
of Section~2.2. The amplitude $A_V$ is defined as the difference between
the maximum and minimum values of the Fourier model. In addition we
calculated phase errors $\varphi_{31,{\rm err}}$ and we will consider them
as an uncertainty measure in metallicity determination.

We tried to apply a compatibility test used by JK96 to verify whether a
light curve had a good shape, \ie was in agreement with their ``basic
sample'' of the light curves. However, most of our light curves did not
pass this test, having a maximum deviation parameter $D_{\rm max}$ much
higher than the acceptable value 3.0. There were cases when calculated
metallicity was in good agreement with the spectroscopic value and the
light curve was of high quality but the deviation parameter was very
high. For this reason and to avoid detailed investigation of each of 1227
light curves we used $\varphi_{31,{\rm err}}$ alone as a measure of
metallicity error, as we can ignore the error in period estimation.

\begin{figure}[htb]
\begin{center}
\begin{tabular}{@{}c@{}c@{}}
\includegraphics[width=4cm]{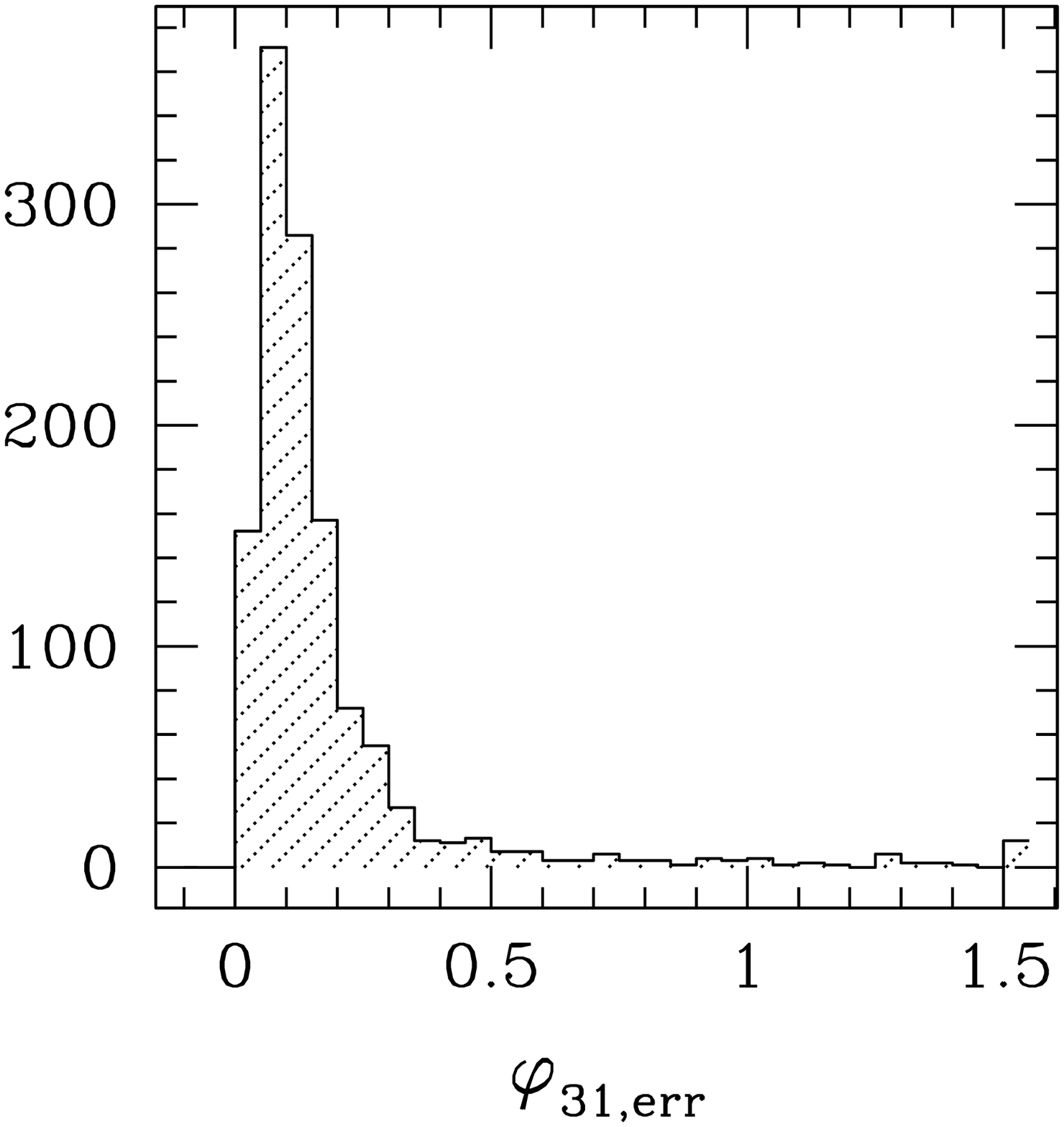} &
\includegraphics[width=4cm]{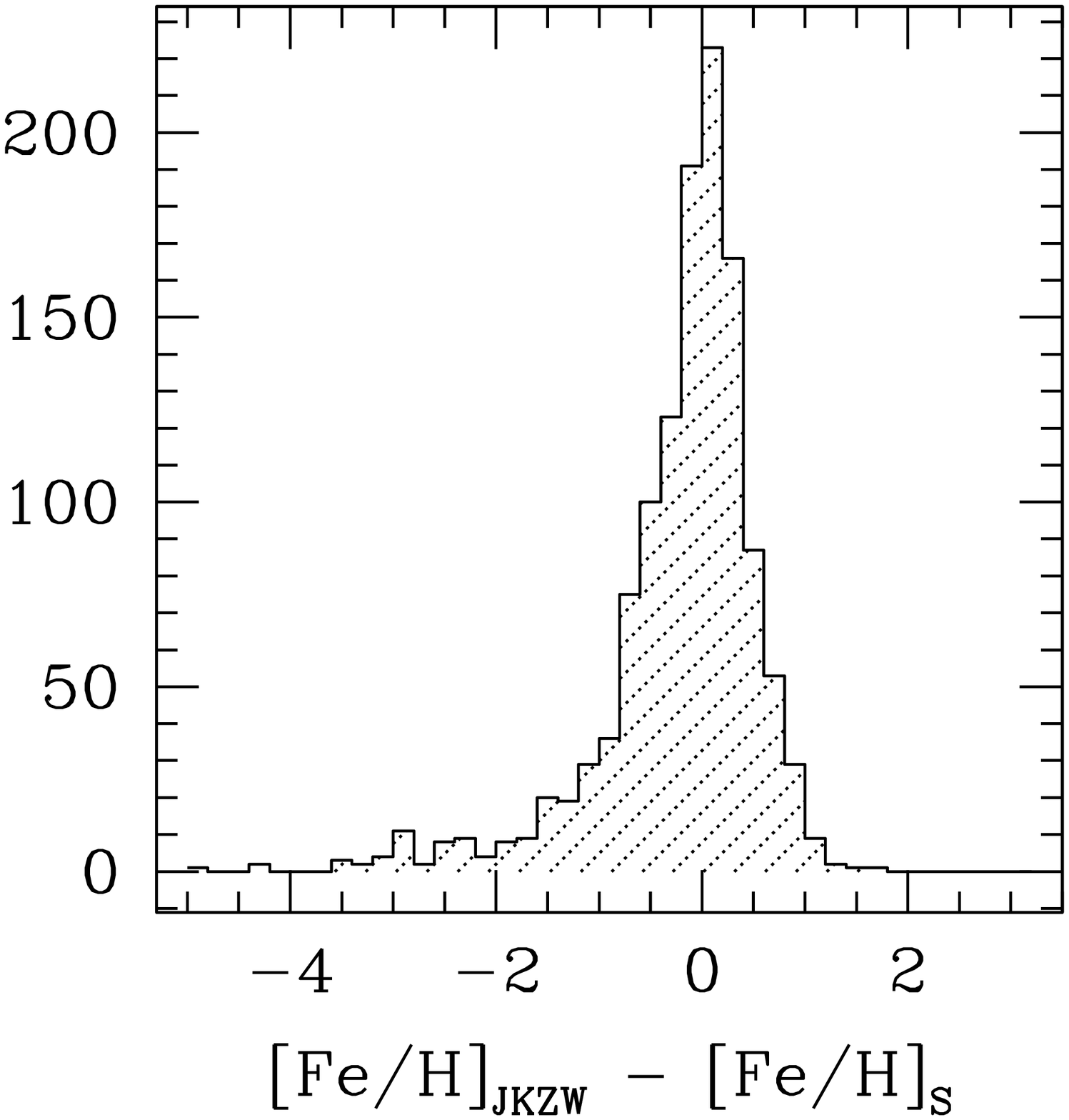} \\
\end{tabular}
\end{center}
\FigCap{{\it Left panel} shows distribution of phase errors
$\varphi_{31,{\rm err}}$ and the {\it right panel} the distribution of
differences between metallicities calculated from method of JK96 and
S04. The majority of objects have well defined $\varphi_{31}$ values (below
0.3) but nevertheless the metallicity differences are significant.}
\end{figure}

Fig.~7 presents a histogram of $\varphi_{31,{\rm err}}$ (left panel) and a
histogram of differences of metallicities calculated with two methods ${\rm
[Fe/H]_{\rm diff}{=}[Fe/H]_{\rm JKZW}{-}[Fe/H]_S}$ (right panel). The
average difference is $\langle{\rm [Fe/H]_{\rm diff}}\rangle=-0.17$~dex
with the majority of the outliers located at ${\rm [Fe/H]_{\rm
diff}}<-2.0$~dex. When we discard objects with $|{\rm [Fe/H]_{\rm
diff}}|>1.5$~dex the average is $\langle{\rm [Fe/H]_{\rm
diff}}\rangle=-0.04$. The discrepancy between methods is not unexpected, as
most stars in the sample are faint and have noisy light curves and both
$\varphi_{31,{\rm err}}$ and the scatter of ${\rm [Fe/H]_{\rm diff}}$
increase with magnitude, as shown in Fig.~8.
\begin{figure}[htb]
\begin{center}
\begin{tabular}{@{}c@{}c@{}}
\includegraphics[width=4.6cm]{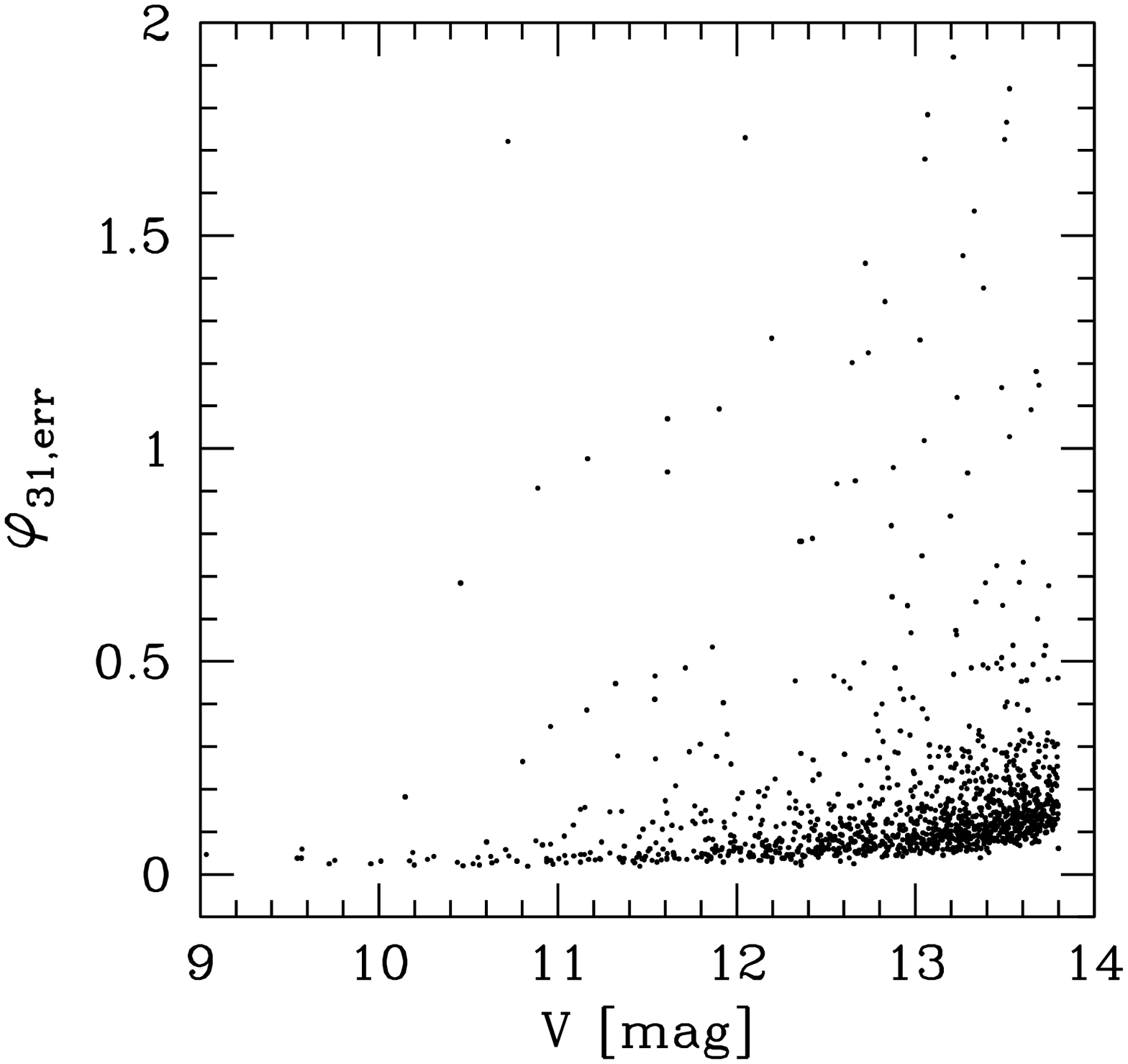} &
\includegraphics[width=4.6cm]{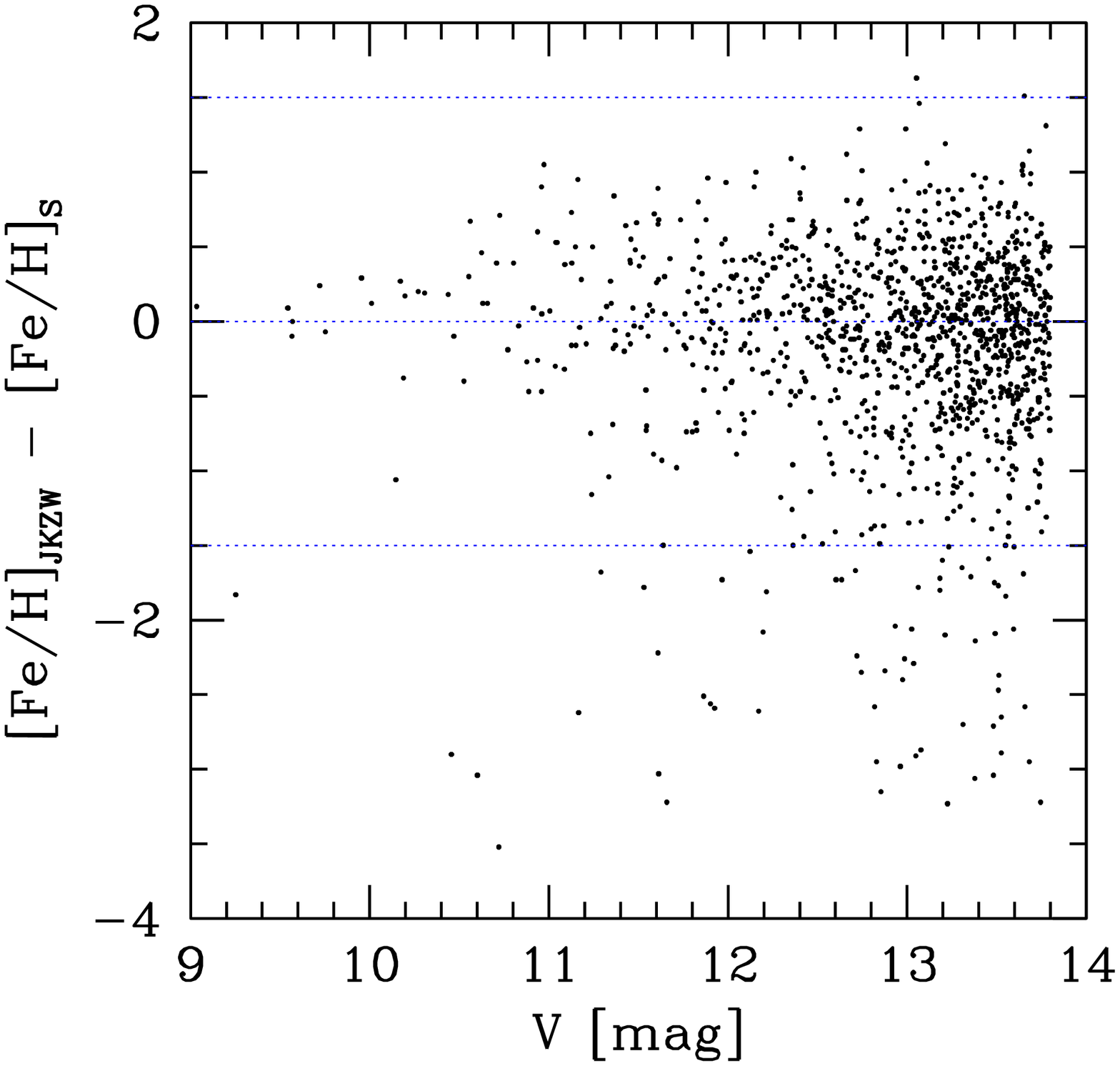} \\
\end{tabular}
\end{center}
\vspace*{-11pt}
\FigCap{Photometric metallicity estimation
dependence on star magnitudes -- the fainter the star the larger is the phase
error ({\it left panel}), and similarly -- the fainter the star, the larger
is the discrepancy between metallicities calculated with different methods
({\it right panel}).}
\end{figure}

After visual inspection of the light curves with large errors and
metallicity differences, we decided to reject objects with
$\varphi_{31,{\rm err}}>0.3$ (131 stars) and with $|{\rm [Fe/H]_{\rm
diff}}|> 1.5$ (73 stars) which altogether gives 155 objects. There still
remain objects that have large metallicity differences, namely 52 RRab
stars with $|{\rm [Fe/H]_{\rm diff}}|>1$. Some of them have quite noisy
light curves and we could accept this as a satisfactory explanation but
some objects have well defined light curves and both amplitude and
$\varphi_{31}$ phase combination possess small errors, for example ASAS
135141+0625.9 and 221635-0349.0 pictured in Fig.~9. We decided to leave all
these stars in the sample for further analysis. In 43 (out of 52) such
cases ${\rm [Fe/H]_{\rm JKZW}}$ is smaller than ${\rm [Fe/H]_S}$.

\begin{figure}[htb]
\begin{center}
\begin{tabular}{@{}c@{}c@{}}
\includegraphics[width=6cm]{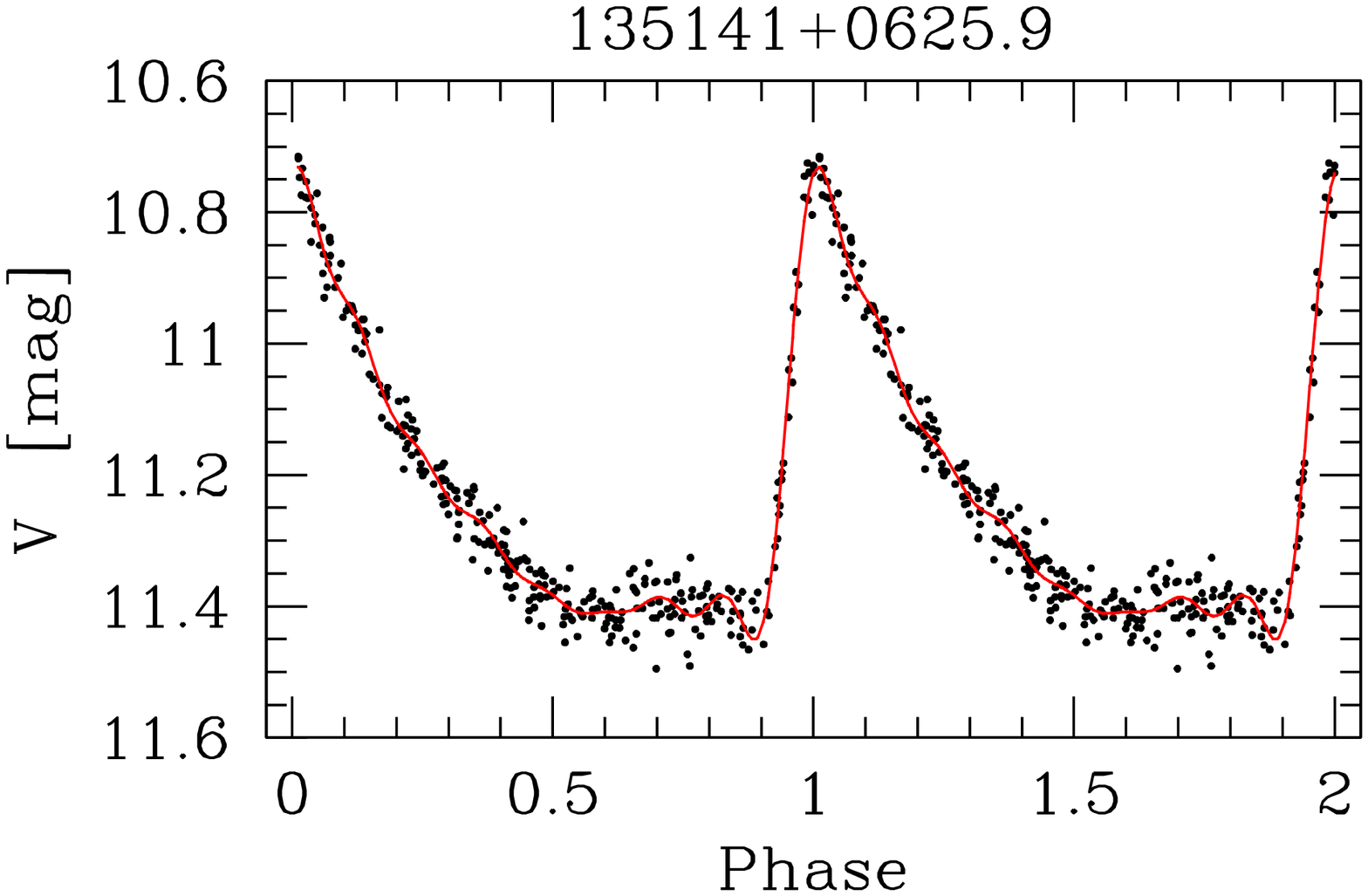} &
\includegraphics[width=6cm]{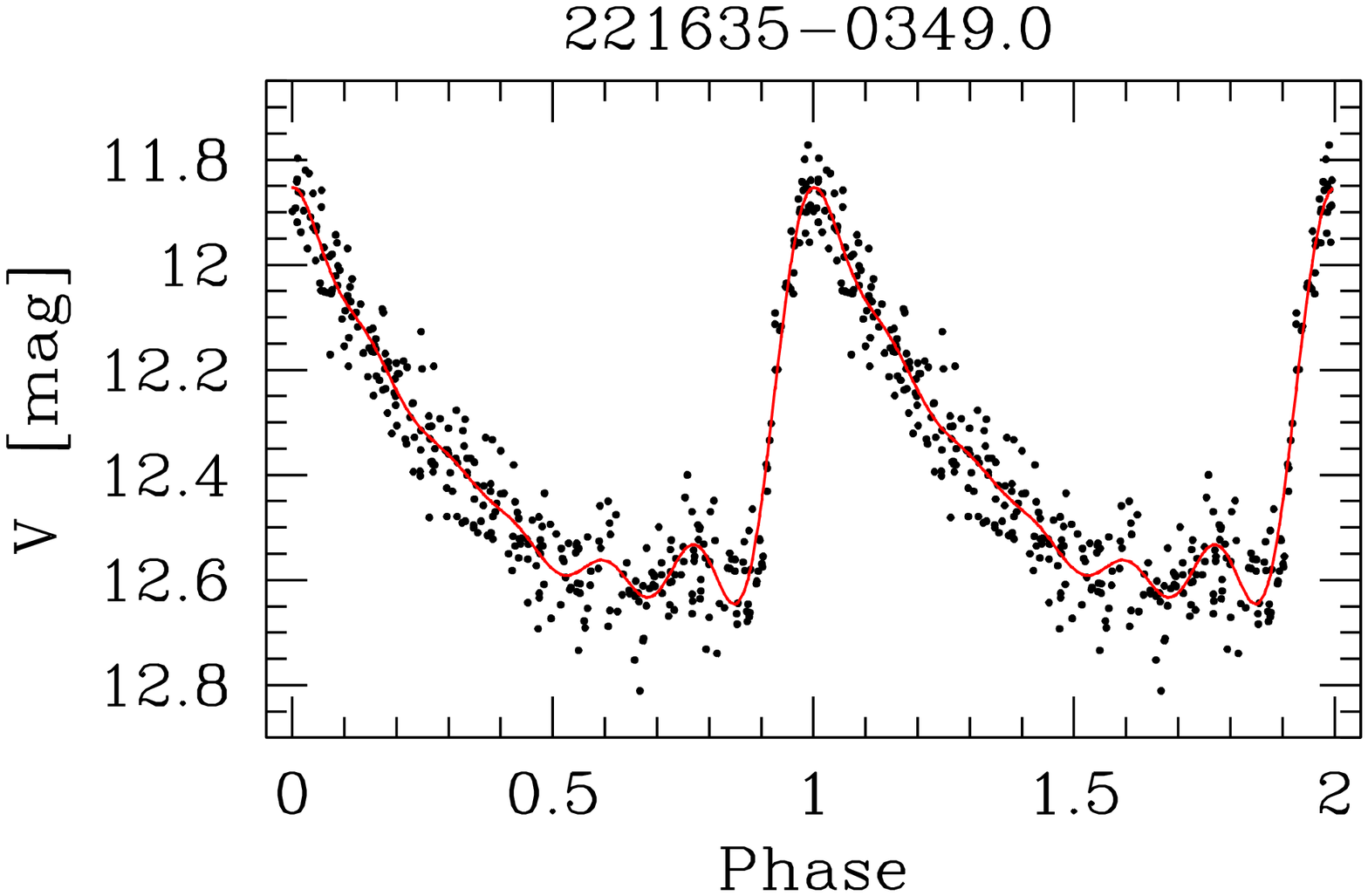} \\
\end{tabular}
\end{center}
\vspace*{-11pt}
\FigCap{Two examples of well defined RRab stars light curves showing
large differences in photometric metallicity determination. For ASAS
135141+0625.9 ({\it left panel}) ${\rm [Fe/H]_{\rm JKZW}}=-1.74$ and ${\rm
[Fe/H]_S}=-0.55$ and for ASAS 221635-0349.0 ({\it right panel}) ${\rm
[Fe/H]_{\rm JKZW}}=-4.63$ and ${\rm [Fe/H]_S}=-3.37$.  The line is a 6th order
Fourier fit to the light curve.}
\end{figure}

Szczygie³ and Fabrycky (2007) identified 180 Blazhko or double mode RR~Lyr
stars in the ASAS data. These light curves display a large scatter which
strongly affects amplitude and Fourier parameters determination and would
change their true photometric metallicities. However, Smolec (2005) showed
that after prewhitening the data with additional close frequencies reliable
photometric metallicity values can be obtained. Given that the check was
performed only on five well observed stars from a narrow metallicity range,
we decide to exclude all multiperiodic RRab stars from the final sample.

\subsection{Comparison of Photometric Metallicities from Two Methods}
Now we compare ${\rm [Fe/H]_{\rm JKZW}}$ and ${\rm [Fe/H]_S}$ metallicities
of the remaining 1008 RRab stars by plotting the two values against each
other in the left part of Fig.~10 and taking into account Oosterhoff type
as assigned in Section~3.1. Oosterhoff~I type RRab stars are marked with
blue squares and Oosterhoff~II with red circles. While there is a rough
agreement between methods, we see two distinct concentrations of points
which are apparently a result of Oosterhoff dichotomy. As previously shown
in Fig.~8, the largest errors in metallicity estimation and the largest
differences between methods are at fainter magnitudes. However, plotting a
subsample of variables brighter than $V=12.5$~mag (255 objects) does not
change the overall structure of the image.

The distributions of ${\rm [Fe/H]_{\rm JKZW}}$ and ${\rm [Fe/H]_S}$ are
presented in the right part of Fig.~10. There is a clear bimodal
distribution of S04 metallicities, while the JK96 distribution is uniform.
In the case of S04 method a separation into Oosterhoff groups is
responsible for the bimodality: almost all Oo I RRab stars have
metallicities larger than $-1.75$~dex while Oo~II smaller than
$-1.6$~dex. This is not observed in the case of JK96 method, where Oo~I
type RRab stars adopt metallicities from the wide range starting at
$-2.5$~dex, while Oo~II RRab stars stay in the low metallicity regime below
$-1.0$~dex. Average metallicity values differ significantly between methods
and in the case of S04 method are $-2.06$~dex for Oo~II and $-1.12$~dex for
Oo~I and in the case of JK96 $-1.85$~dex for Oo~II and $-1.27$~dex for
Oo~I. This sharp division in the case of S04 method is not physical and is
not in agreement with metallicity distribution of spectroscopically
observed ASAS RRab stars, as described in the following section.

Another observation from Fig.~10 is that the most extreme differences in
[Fe/H] at very low and very high metallicities have larger ${\rm [Fe/H]_S}$
values, which again suggests that one of the methods does not work well at
untypical metallicities.

\begin{figure}[p]
\begin{center}
\vglue-11mm
\begin{tabular}{@{}c@{}c@{}}
\multirow{3}{*}{\includegraphics[width=9.1cm]{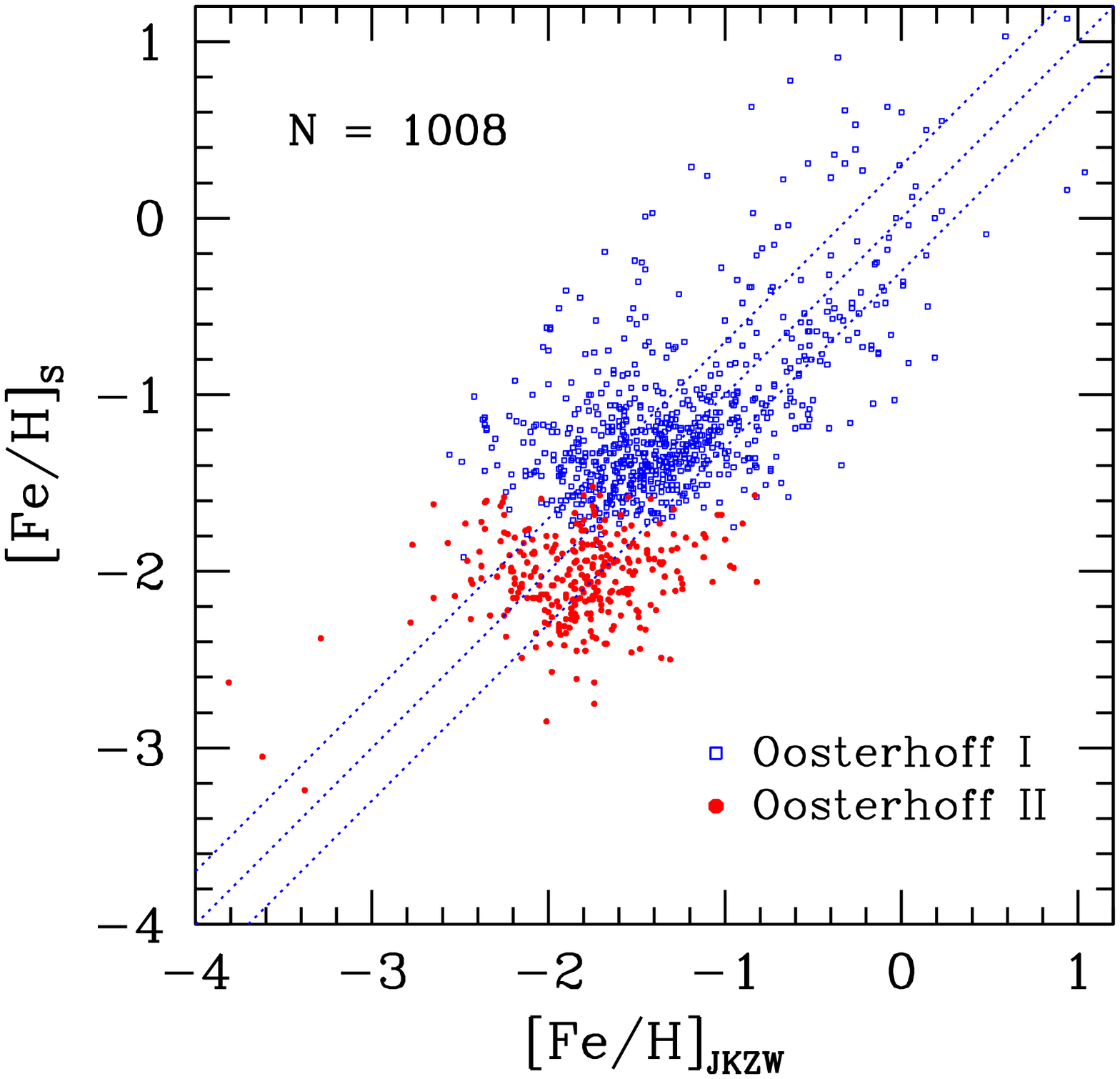}} & \\
 & \includegraphics[width=3.9cm]{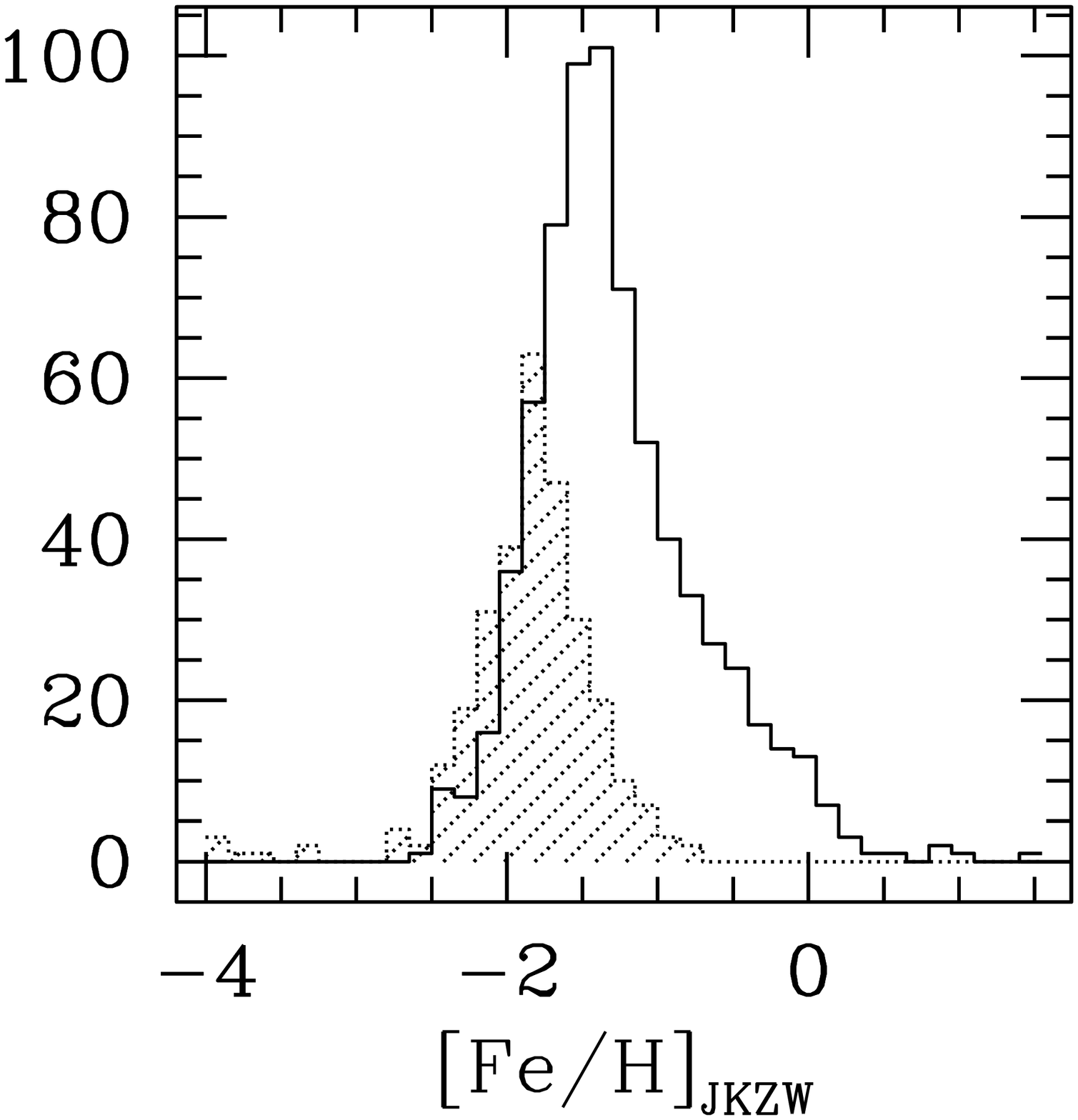} \\
 & \includegraphics[width=3.9cm]{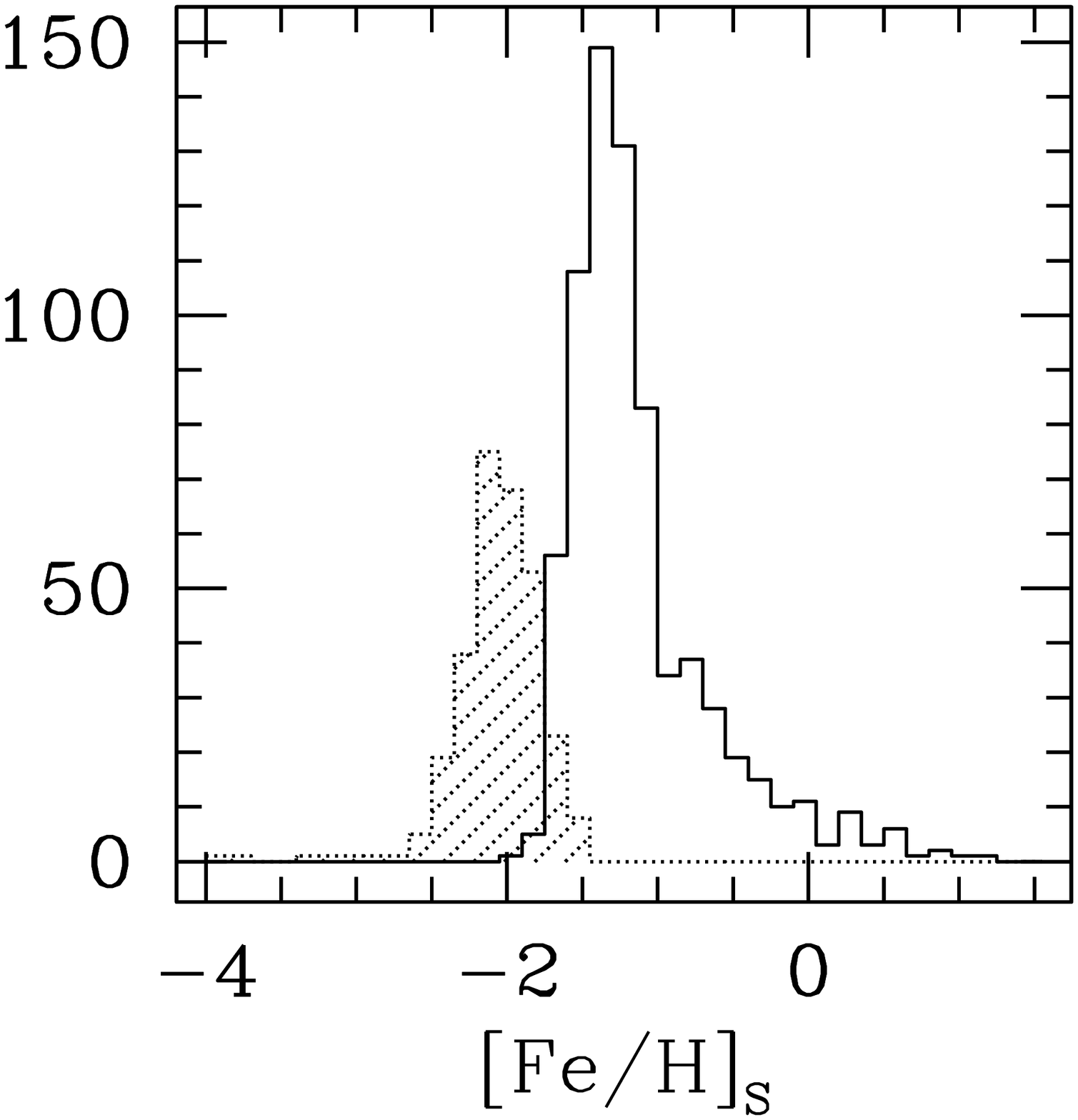} \\
\end{tabular}
\end{center}
\vspace*{5pt}
\FigCap{
Comparison of two methods of photometric metallicity estimation is
presented on the {\it left plot}. ${\rm [Fe/H]_{\rm JKZW}}$ is computed by
the method of Jurcsik and Kov\'acs (1996) and ${\rm [Fe/H]_S}$ by that of
Sandage (2004). Oosterhoff groups are marked with different symbols: Oo~I
with blue squares and Oo~II with red circles. The three dotted lines are
separated by 0.3~dex. Two distinct concentrations of points are visible and
are apparently a result of Oosterhoff dichotomy. Two histograms on the {\it
right} present distribution of photometric metallicities from JK96 ({\it
top panel}) and S04 ({\it bottom panel}) methods. Shaded areas on the
histograms correspond to Oo~II group (296 stars) and the clear ones to Oo~I
group (712 stars). While the distribution of ${\rm [Fe/H]_{\rm JKZW}}$ is
uniform, the ${\rm [Fe/H]_S}$ metallicity distribution is bimodal and
corresponds to the Oosterhoff separation. The mean metallicities are ${\rm
\langle[Fe/H]_{\rm JKZW}\rangle}=-1.85$~dex for Oo~II and 
${\rm \langle [Fe/H]_{\rm JKZW}\rangle}=-1.27$~dex for Oo~I ({\it top panel}),
and ${\rm\langle [Fe/H]_S\rangle}=-2.06$~dex for Oo~II and ${\rm
\langle[Fe/H]_S\rangle}=-1.12$~dex for Oo~I.}

\vspace*{1mm}
\begin{center}
\begin{tabular}{@{}c@{}c@{}c@{}}
\includegraphics[width=4.3cm]{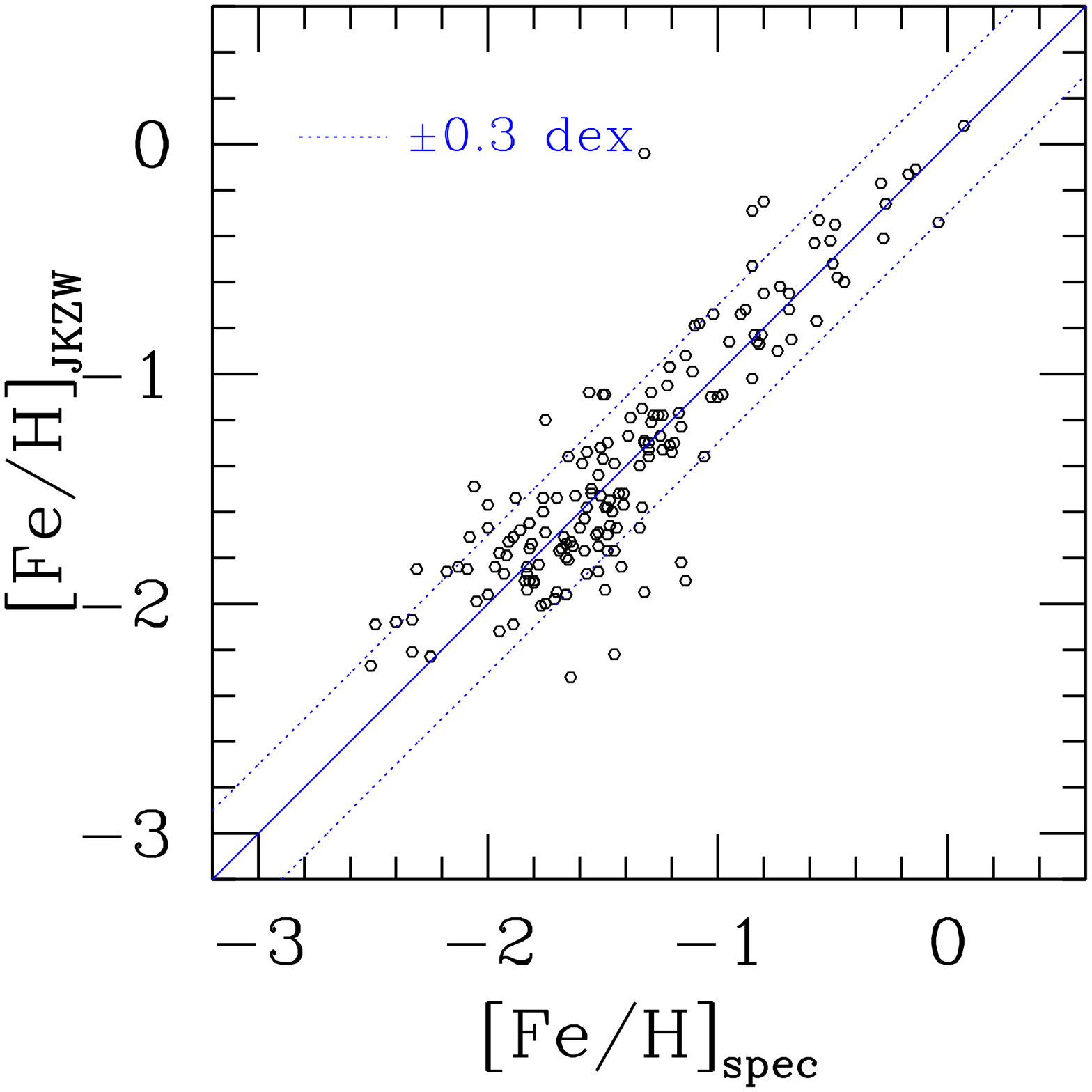} &
\includegraphics[width=4.3cm]{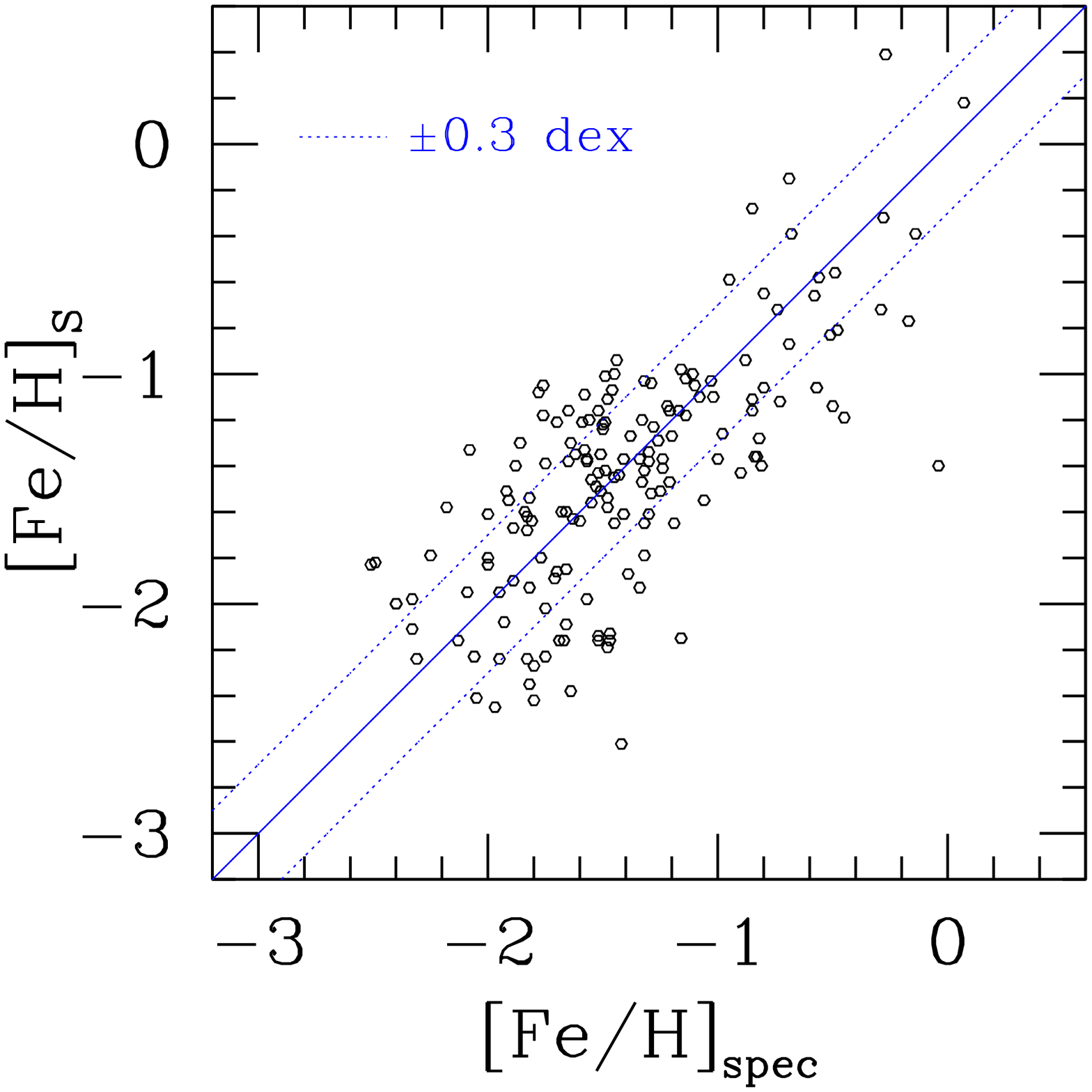} &
\includegraphics[width=4.3cm]{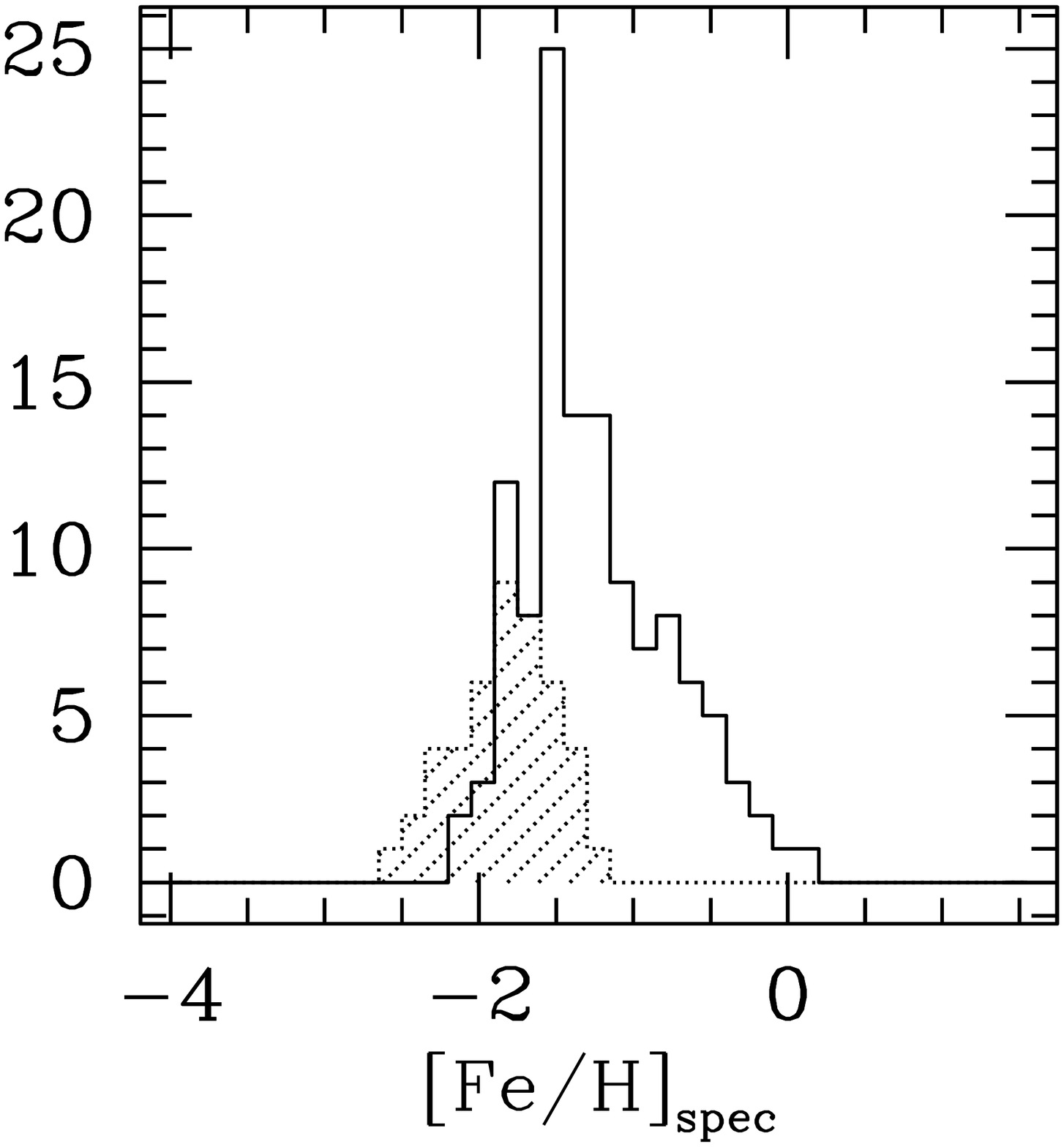} \\
\end{tabular}
\end{center}
\vspace*{-3pt}
\FigCap{Comparison of spectroscopically determined metallicity values from
Layden (1994) with photometrically calculated ones using the method of JK96
({\it left panel}) and S04 ({\it middle panel}), for the sample of 186 RRab
stars. The dotted lines on both images delimit the $\pm0.3$~dex region. The
{\it right panel} presents a distribution of spectroscopic metallicities
for ASAS RRab stars -- the shaded area corresponds to Oo~II group and the
clear area to Oo I~group.}
\end{figure}

\subsection{Comparison with Spectroscopic Metallicities}
\hglue-6pt We combine spectroscopically obtained metallicities published by Layden
(1994) with photometric metallicities calculated for the ASAS RRab stars
sample. The combination results in 186 objects and is presented in
Fig.~11. Left panel shows the comparison with method of JK96 and the middle
panel with S04. The dotted lines on both images delimit the $\pm0.3$~dex
region. It is clear that the method of S04 produces larger scatter than the
method of JK96 and two concentrations described in the previous section are
present in the middle panel, but poorly visible.

We perform a star to star comparison for both sets in order to see whether
the method of JK96 is significantly better than S04. The average difference
between photometric and spectroscopic metallicity values in case of JK96
method is 0.03 dex and the dispersion is 0.30~dex. 75\% stars lie within
the $\pm0.3$~dex lines. In case of S04 method the difference is 0.02~dex
and the scatter is 0.43~dex. Only 50\% stars fall within $\pm0.3$~dex
region. Summarizing, JK96 method is better calibrated against spectroscopic
metallicities and the conformity holds reasonably well within the whole
metallicity range. Photometric metallicities from S04 method seem to be
underestimated in low metallicity range and overestimated at high
metallicity range, suggesting that there should be separate formulae for
the two Oosterhoff groups.

Finally, right panel of Fig.~11 presents spectroscopically obtained
metallicity distribution of ASAS RRab stars, with distinguished Oosterhoff
groups -- the shaded area corresponds to Oo~II group and the solid line
defines the area of Oo~I group. Similarly to the histogram of photometric
JK96 metallicities (Fig.~10), Oo~I RRab stars cover metallicity values from
a wide range between $-2.2$ and 0.2~dex, overlapping with metallicities
typical for Oo~II group, between $-2.6$ and $-1$~dex. This disagrees with a
sharp separation of photometric metallicity values of S04 among Oosterhoff
groups, visible on the bottom histogram of Fig.~10.

\begin{figure}[htb]
\begin{center}
\begin{tabular}{@{}c@{}c@{}}
\includegraphics[width=5.7cm]{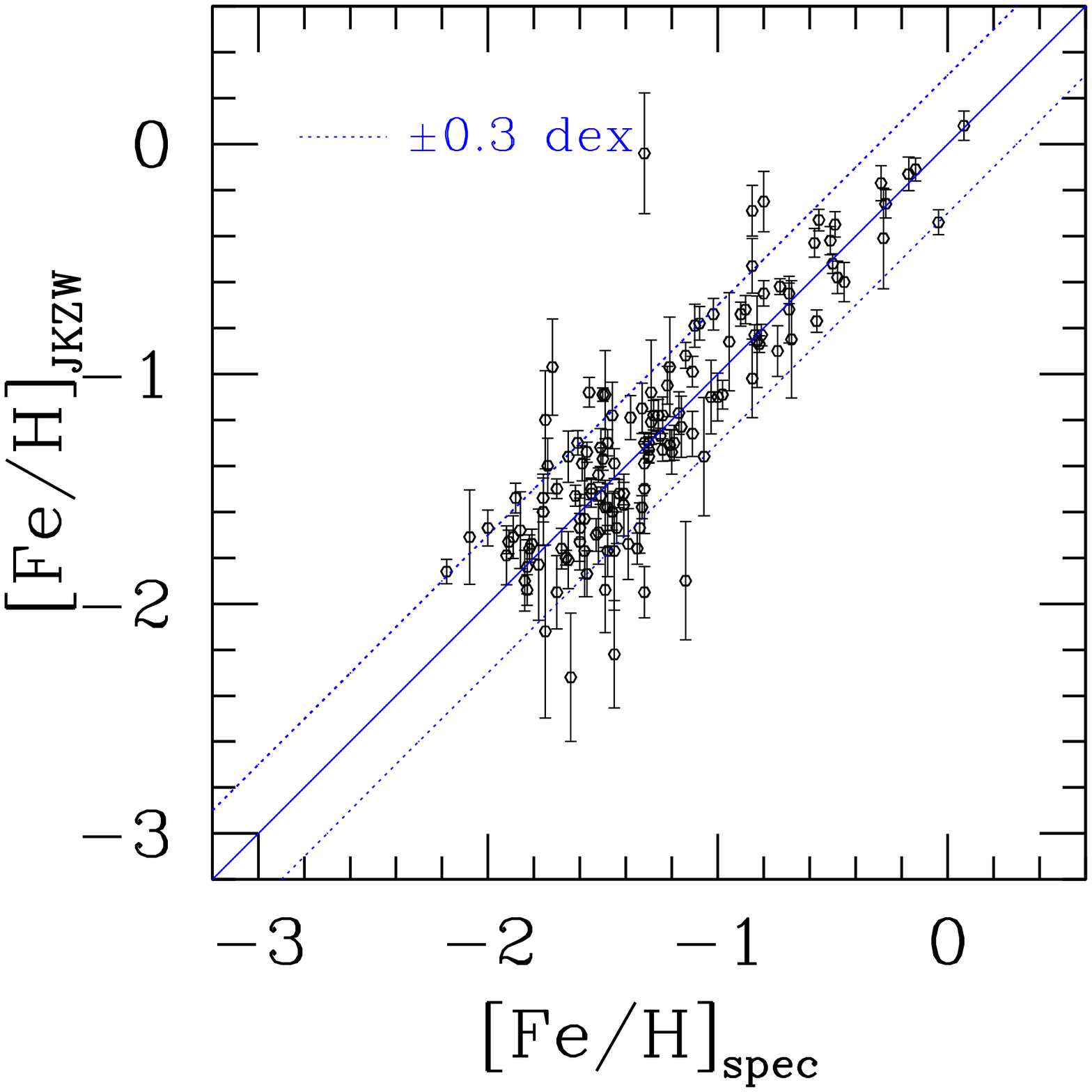} &
\includegraphics[width=5.7cm]{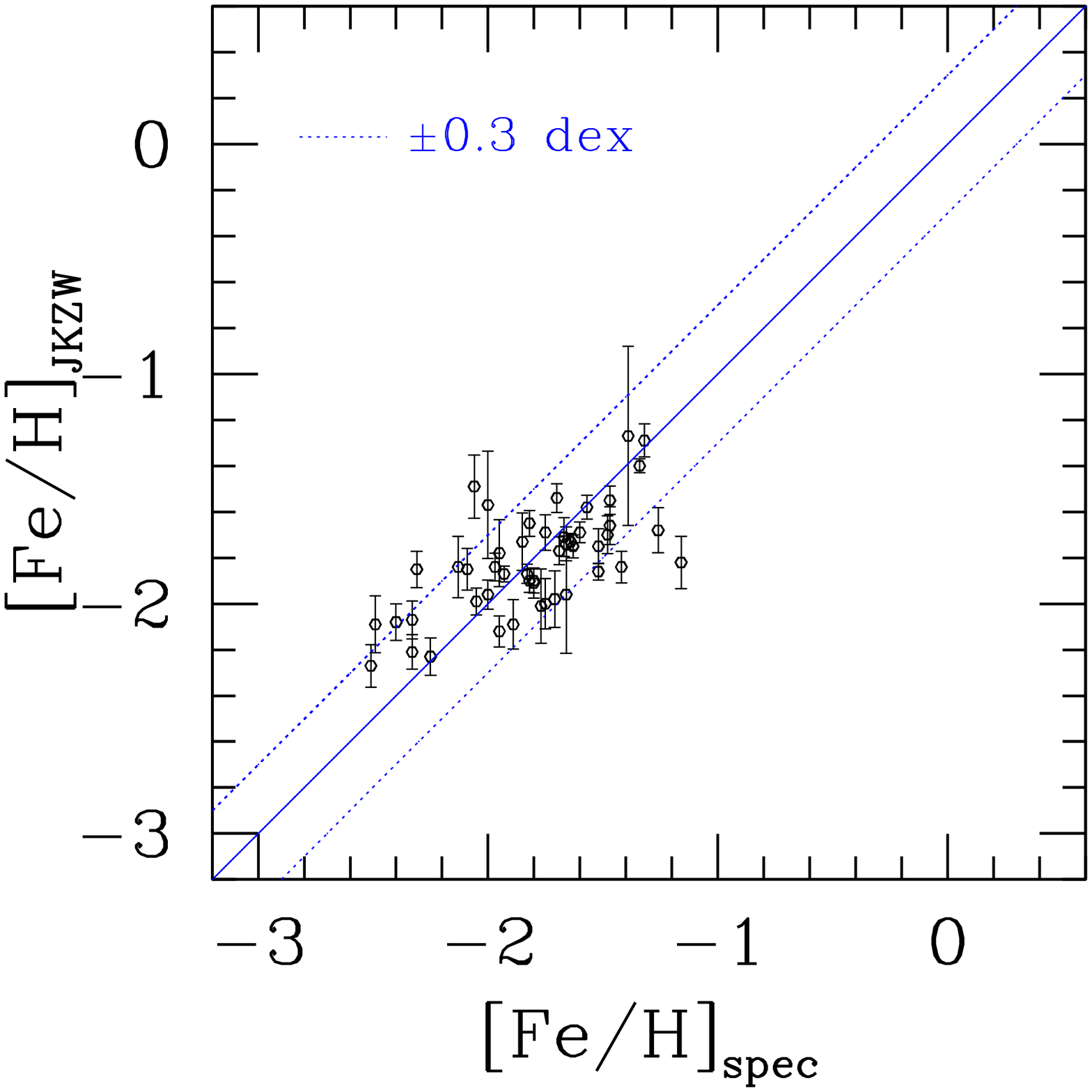} \\
\end{tabular}
\end{center}
\vspace*{-11pt}
\FigCap{Comparison of spectroscopically determined metallicity values
from Layden (1994) with photometrically calculated ones using the method of
JK96. {\it Left panel} presents Oosterhoff I RRab stars and the {\it right
panel} Oosterhoff~II RRab stars. The dotted lines on both images delimit
the $\pm0.3$~dex region.}
\end{figure}
Although the method of JK96 produces metallicities in good agreement with
spectroscopic ones, there is an increased scatter in the lower part of the
left plot in Fig.~11, for ${\rm [Fe/H]_{\rm JKZW}}<1.5$~dex. We plot a
comparison of JK96 metallicities with spectroscopic ones again (Fig.~12),
but this time separately for each Oosterhoff group (Oo~I in the left panel
and Oo~II in the right).  It is visible that most of the scatter is
introduced by the Oo~II group, for which a correlation is not as tight as
for Oo~I variables. There are also several outliers in the low metallicity
range in case of Oo~I type RRab stars, but these measurements have high
${\rm [Fe/H]_{\rm JKZW}}$ errors, while the errors of the outliers in Oo II
group are small. This would imply that separate metallicity estimation
formulae should be constructed for different Oosterhoff groups, not only
for S04 method, but also for JK96. We think this is conceivable, given that
Oosterhoff~I and II type RRab stars are of different age, Oo~II being more
advanced in their evolution. It is possible that their pulsation has
changed producing slightly different light curve shape, which would affect
metallicity dependence on Fourier parameters of the light curve
decomposition. We tried to modify the existing formula (Eq.~6), with no
satisfactory results -- the obtained relations were very similar to the
present one and did not reduce the scatter when compared with spectroscopic
metallicities. It looks as a linear dependence of the metallicity on period
and phase parameter is not sufficient to describe this relation and other
components might have to be included. In any event, more spectroscopic
observations of RRab stars with small metal content would be desirable.

\subsection{Final Metallicity Values}
In previous subsections we presented two methods of photometric metallicity
determination and compared them with each other showing, that there is a
significant discrepancy especially in two regions of metallicities.
Comparison of both methods with spectroscopic results undoubtedly favors
the JK96 method, which shows better conformity and smaller scatter and is
less sensitive to Oosterhoff type. For this reason we will use JK96
metallicities only.

\MakeTableee{@{}c@{\ \ }c@{\ \ }c@{\ \ }c@{\ \ }c@{\ \ }c@{\ \ }c@{\ \ }@{}}
{12cm}{Metallicity values of selected ASAS RRab stars}
{\hline
\noalign{\vskip2pt}
ID & ${\rm[Fe/H]_{\rm JKZW}}$ & ${\rm[Fe/H]_{\rm JKZW,err}}$ & ${\rm[Fe/H]_S}$ &
${\rm[Fe/H]_{\rm NSVS}}$  & ${\rm[Fe/H]_{\rm spec}}$ & ${\rm[Fe/H]_{\rm spec,err}}$ \\
\noalign{\vskip2pt}
\hline
\noalign{\vskip2pt}
000248-2456.7 & $-1.25$ & 0.028 & $-1.44$ & $-    $ & $-1.25$ & $0.130$\\
000301-7041.5 & $-1.35$ & 0.230 & $-1.41$ & $-    $ & $-    $ & $-    $\\
000321+0323.9 & $-1.86$ & 0.281 & $-1.36$ & $-1.69$ & $-    $ & $-    $\\
\noalign{\vskip2pt}
\hline
\noalign{\vskip2pt}
\multicolumn{7}{p{12cm}}{The first column contains ASAS ID.
The following columns contain metallicity values: from the JK96 method and
its error, from the S04 method, for the NSVS counterpart calculated by 
Kinemuchi \etal (2006), and finally spectroscopic value from Layden (1994)
and its error. The full version of this Table is available for download from
the ASAS website.}}
Kinemuchi \etal (2006) also determined photometric metallicities for their
589 fundamental mode RR~Lyr stars, using both methods. For 363 stars that
had sufficient number of observations and a good phase coverage, they took
an average from both methods as their metallicity value. For the remaining
226 objects they applied the metallicity of S04 alone, assuming it is more
reliable in case of lower quality light curves. However, as shown in the
Appendix, NSVS amplitudes are not always well determined and the S04 method
itself is not reliable enough, thus we expect a majority of NSVS
metallicity values to be highly uncertain. We compared ASAS and NSVS
metallicity values for stars from the overlapping region and concluded,
that the mean absolute difference in metallicities is fairly high, around
0.3~dex. The details of this comparison are presented in the Appendix. This
is a serious argument for not combining both samples to investigate
metallicity based properties of all Galactic RRab stars from both
hemispheres.

Calculated metallicities are available for download from the ASAS website.
The exemplary lines of the file are presented in Table~3.

\Section{Galactic Distribution}
It would be interesting to combine ASAS and NSVS RR~Lyr stars catalogs in
order to investigate Galactic distribution of the complete RRab stars
sample in the Solar neighborhood and we initially intended to do so. The
data from both projects were obtained with similar equipment and cover very
similar magnitude range, but telescopes observed from different hemispheres
thus providing a complete sky coverage of the whole sphere. Unfortunately,
when comparing the ASAS and NSVS data in the overlapping region we
encountered several inconsistencies concerning RR~Lyr stars detection
efficiency in both catalogs -- among 967 ASAS objects and 806 NSVS objects
in this area only 446 are common in the whole magnitude range. When we
narrow the magnitude range to 10--12~mag (in which both projects should
have complete samples) these numbers are 114 for ASAS and 72 for NSVS, with
63 common stars, showing that the catalogs are neither complete nor
compatible. The more detailed comparison of the catalogs is presented in
the Appendix. Such inconsistency excludes any statistical conclusions based
on RR~Lyr stars absolute numbers.

\begin{figure}[htb]
\begin{center}
\includegraphics[width=8.7cm]{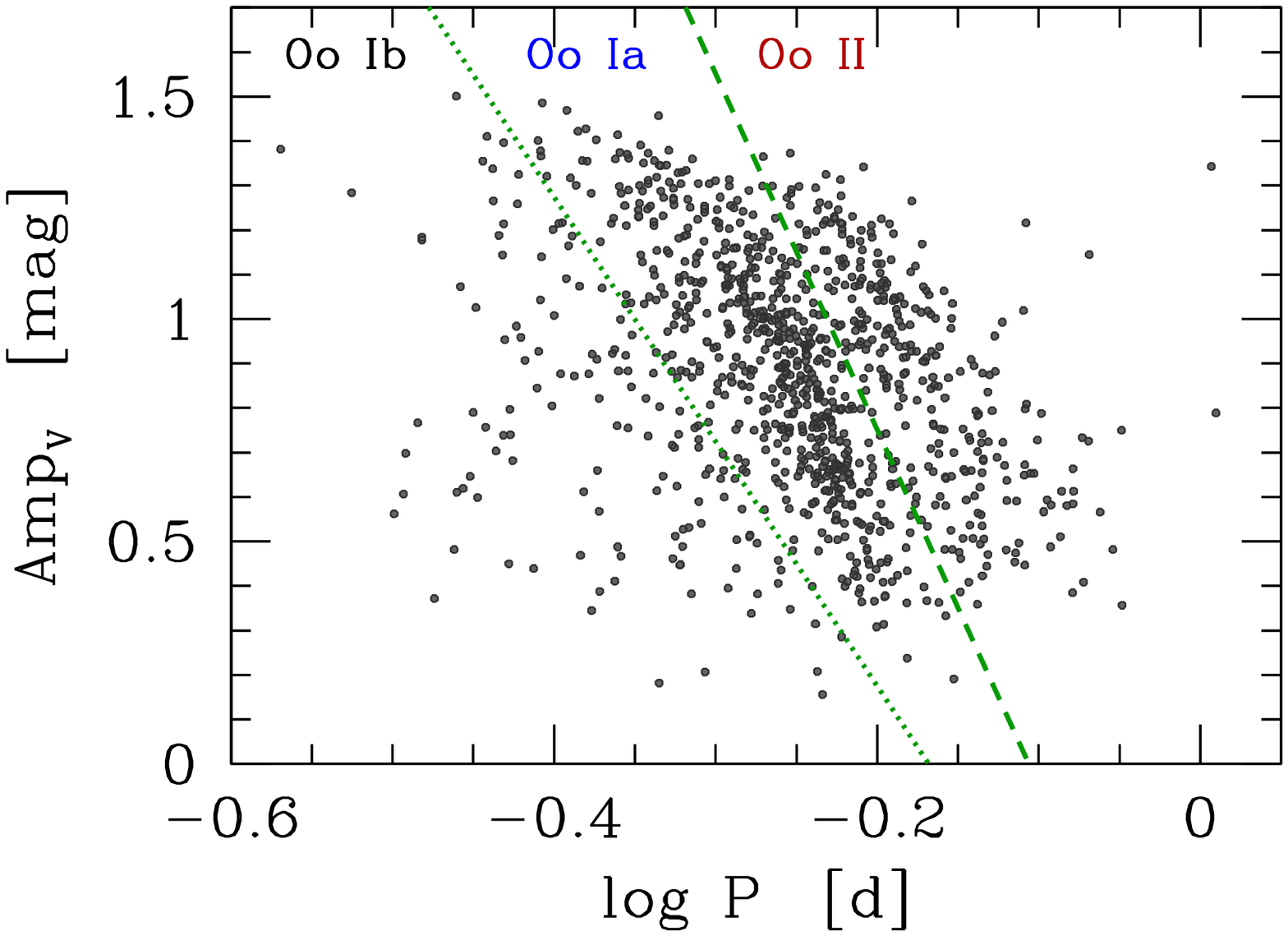}
\end{center}
\vspace*{-11pt}
\FigCap{Period--amplitude (Bailey) diagram for ASAS RRab stars. Lines
divide points into three groups: right side of the dashed line contains
points from Oosterhoff~II group, the points between dashed and doted lines
(being the main part of Oosterhoff~I group) are assigned to Oosterhoff~Ia
group while the remaining points to the left side of the dotted line are
the Oosterhoff~Ib group.}
\end{figure}
Additional difficulties were met when comparing photometric metallicities,
as described in the previous section. A mean absolute difference in
metallicities of $\approx0.3$~dex induces a difference in calculated
distance $\approx130$~pc for $V=12$ magnitude star and $\approx340$~pc for
$V=14$ magnitude star. This is another argument against combining both
databases and investigating total distance distributions or density
functions.

These differences, preventing from combining both catalogs, at the same
time encouraged us to inspect distribution of ASAS RRab variables and to
compare it to results of Kinemuchi \etal (2006) for NSVS data. For the
following analysis we divided ASAS RRab stars into three groups, as
pictured in Fig.~13. The Oosterhoff~II group remains unchanged and consists
of all points to the right of the dashed line given by Eq.~(5). Similarly
to Kinemuchi \etal (2006), we have plotted in Fig.~13 the dotted line
corresponding to S04 constant metallicity line of ${\rm [Fe/H]=-0.8}$.

This line splits Oosterhoff~I group into two subgroups that we will call
Oo~Ia (numerous, longer period, lower metallicity stars located between
dotted and dashed lines), and Oo~Ib (sparse, shorter period, higher
metallicity stars located on the left side of the dotted line).

\subsection{Distance Determination}
Distances to ASAS RRab stars were determined using the distance modulus,
without correction for extinction. We assume that at short distances such
as exhibited by ASAS RR~Lyr stars extinction does not play a significant
role. The distance $d$ from the Sun is calculated as follows:
$$d=10^{0.2(m_V-M_V+5)}\eqno(9)$$
where $m_V$ and $M_V$ are apparent and absolute average magnitudes in {\it
V}-band, respectively. Absolute magnitudes $M_V$ were calculated using
formula from Bono, Caputo and Di Criscienzo (2007):
$$M_V=1.19+0.5{\rm [Fe/H]_{\rm JKZW}+0.09 [Fe/H]_{\rm JKZW}^2}.\eqno(10)$$

\begin{figure}[htb]
\begin{center}
\begin{tabular}{@{}c@{\ }c@{\ \ }c@{}}
\includegraphics[width=4cm]{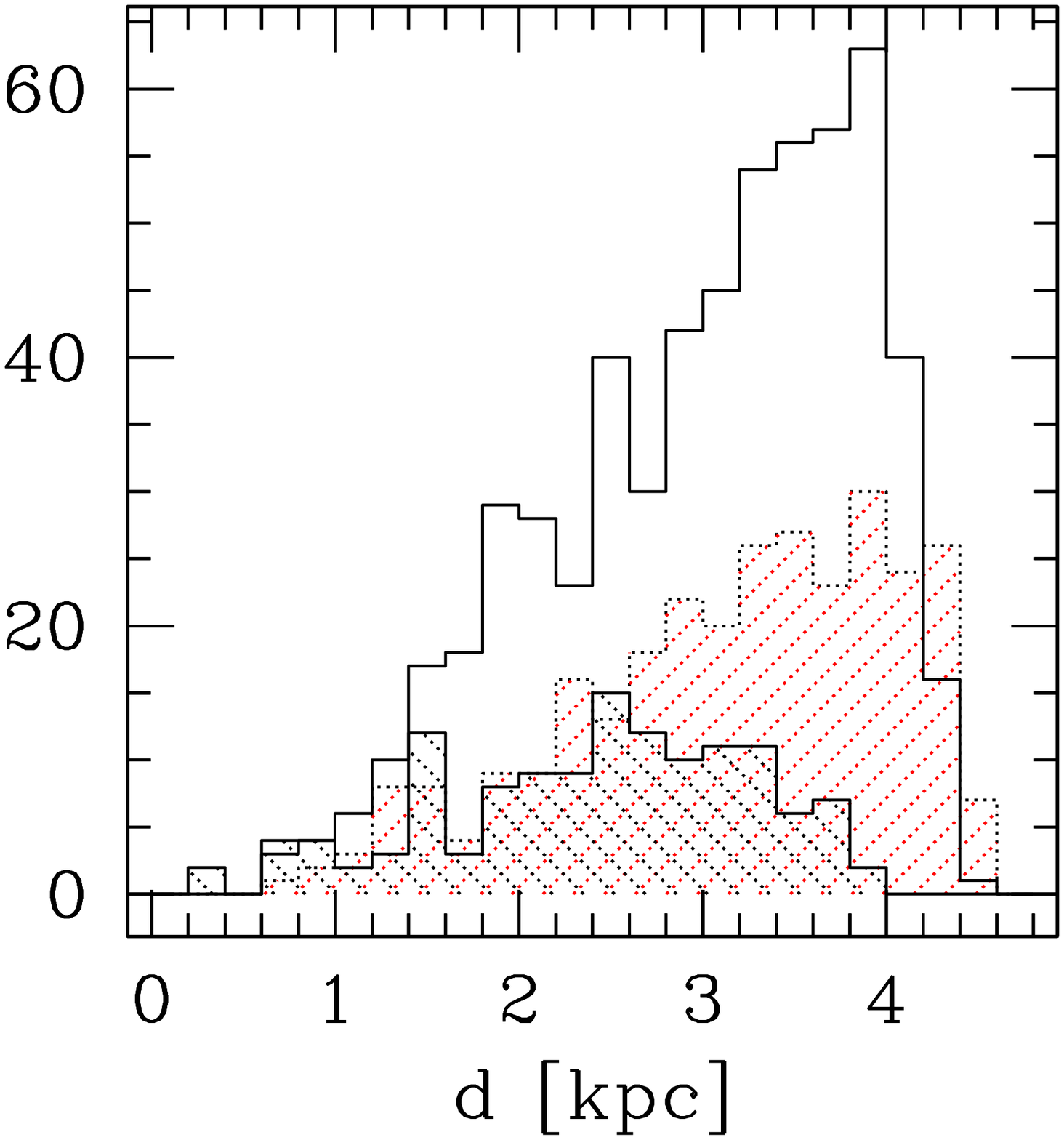} &
\includegraphics[width=4cm]{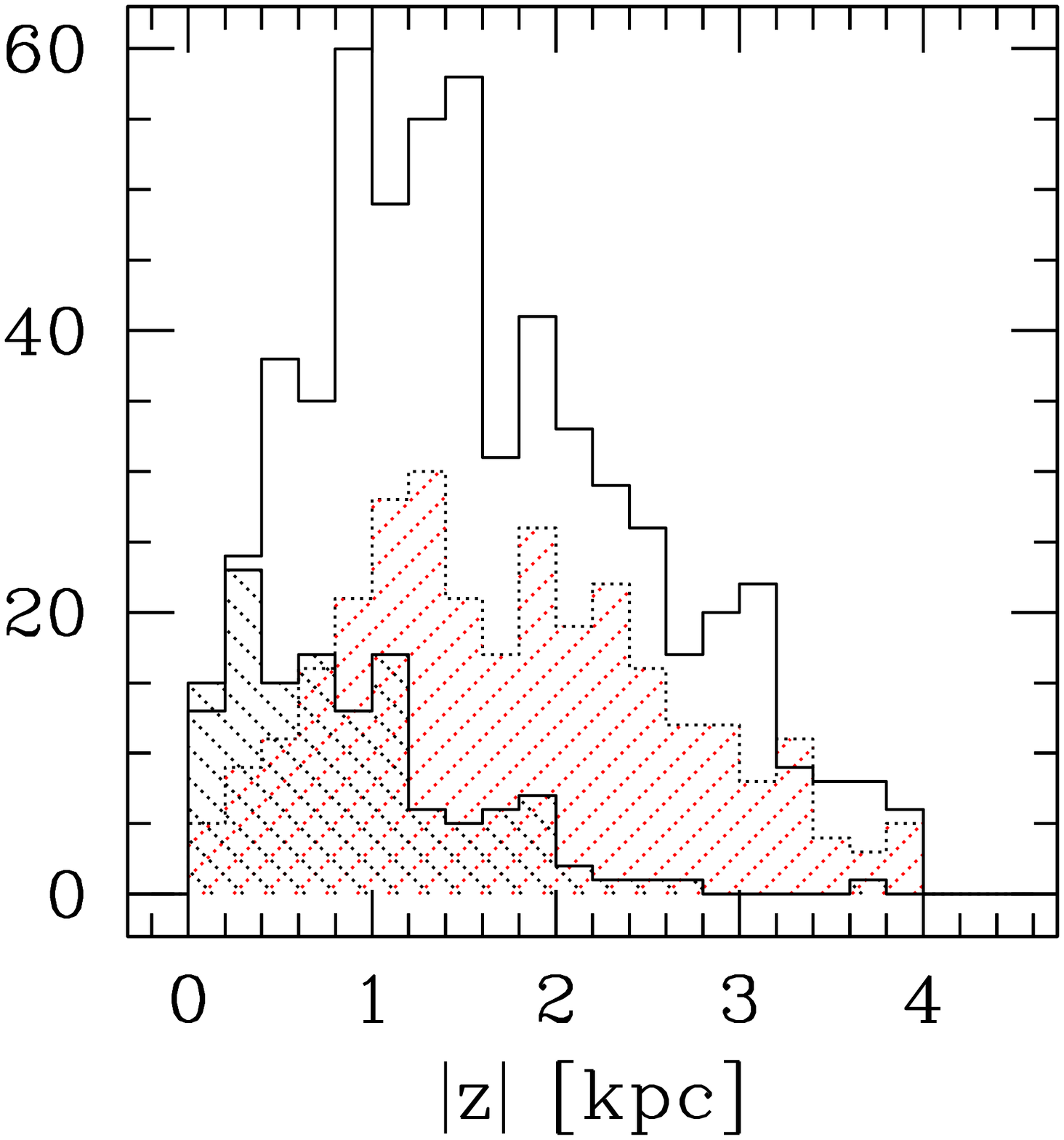} &
\includegraphics[width=4cm]{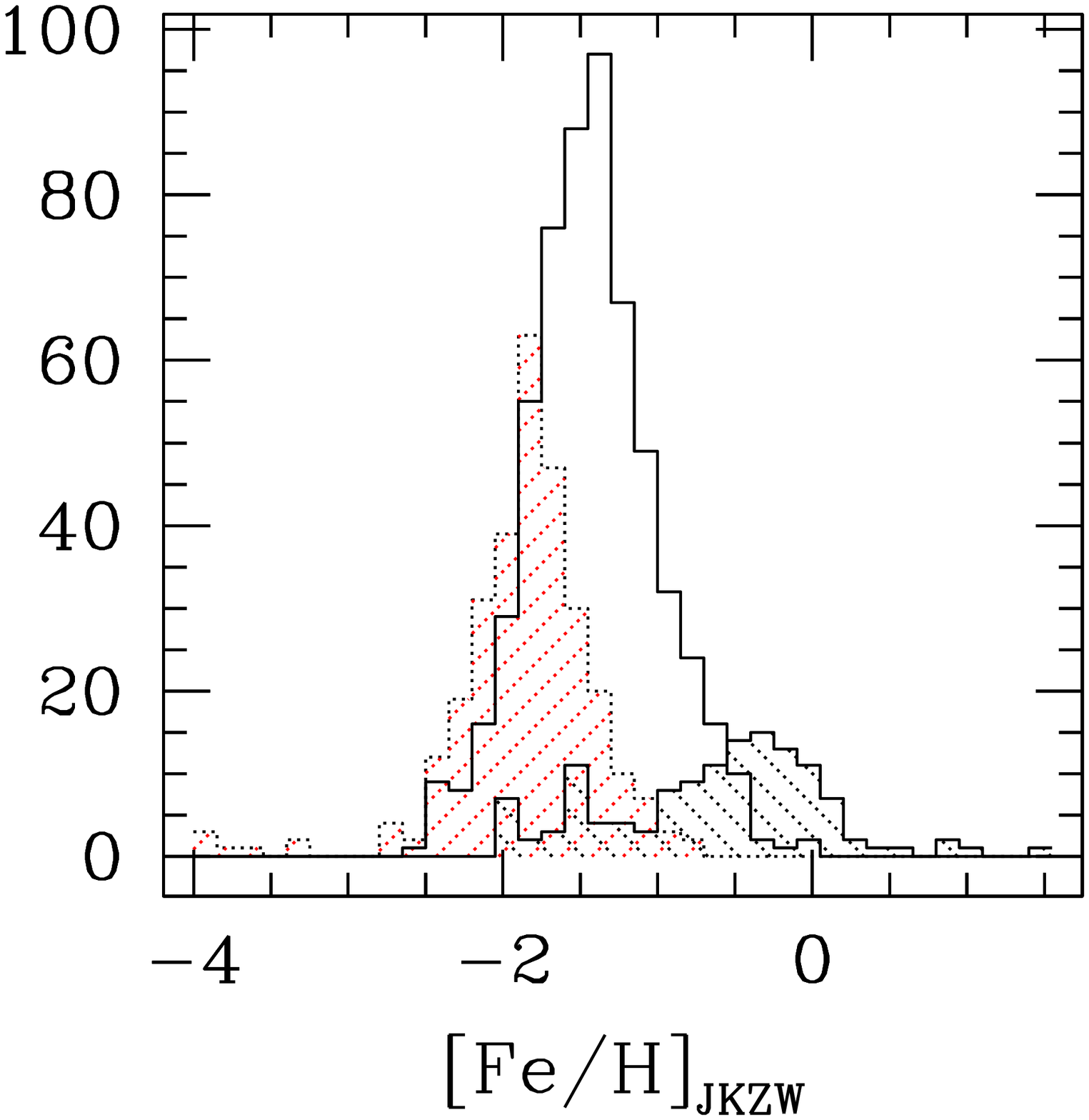} \\
\end{tabular}
\end{center}
\vspace*{-11pt}
\FigCap{The distributions of distance $d$ from Sun ({\it left panel}) and
distance $|z|$ from the Galactic plane ({\it middle panel}) of ASAS RRab
stars. {\it Right panel} presents metallicity distributions. On all three
histograms plain area corresponds to Oo~Ia group, the red shaded area to
Oo~II group, and the black shaded area to Oo~Ib group.}
\end{figure}

The distribution of distances $d$ from the Sun is presented in the left
panel of Fig.~14. If we assume that ASAS is complete to $V=13.5$~mag, then
the distance to which we should observe all RRab stars is about 4~kpc. This
is in agreement with distance distributions for Oo~II and Oo~Ia groups --
the maximum of Oo~Ia and Oo~II histograms is at 4~kpc and detection falls
rapidly for larger distances. However, in the case of Oo~Ib stars (most
metal rich RRab stars, as seen in the right panel of Fig.~14) the maximum
of the histogram is at 2.5~kpc. At the same time this group of metal rich
stars is concentrated close to the Galactic plane (middle panel of
Fig.~14), with the majority of stars having $|z|<1$~kpc. These two
properties coincide with the location of the spiral arms of the Milky Way
at a distance approximately 2~kpc (Xu \etal 2006) and it would be tempting
to locate Oo~Ib RRab stars there. However, we must remember that distances
were calculated without correction for extinction which is not negligible
at low galactic latitudes, thus their errors may be significant. The middle
panel of Fig.~14 also shows that both Oosterhoff~I and II groups are
present at all values of $|z|$, that is, the local neighborhood is a
mixture of older (metal poor) and younger (metal rich) stars, whereas the
youngest stars are found only at low values of $|z|$, that is in the thick
disk region.

\begin{figure}[h]
\begin{center}
\begin{tabular}{@{}cc@{}}
\includegraphics[width=6.3cm]{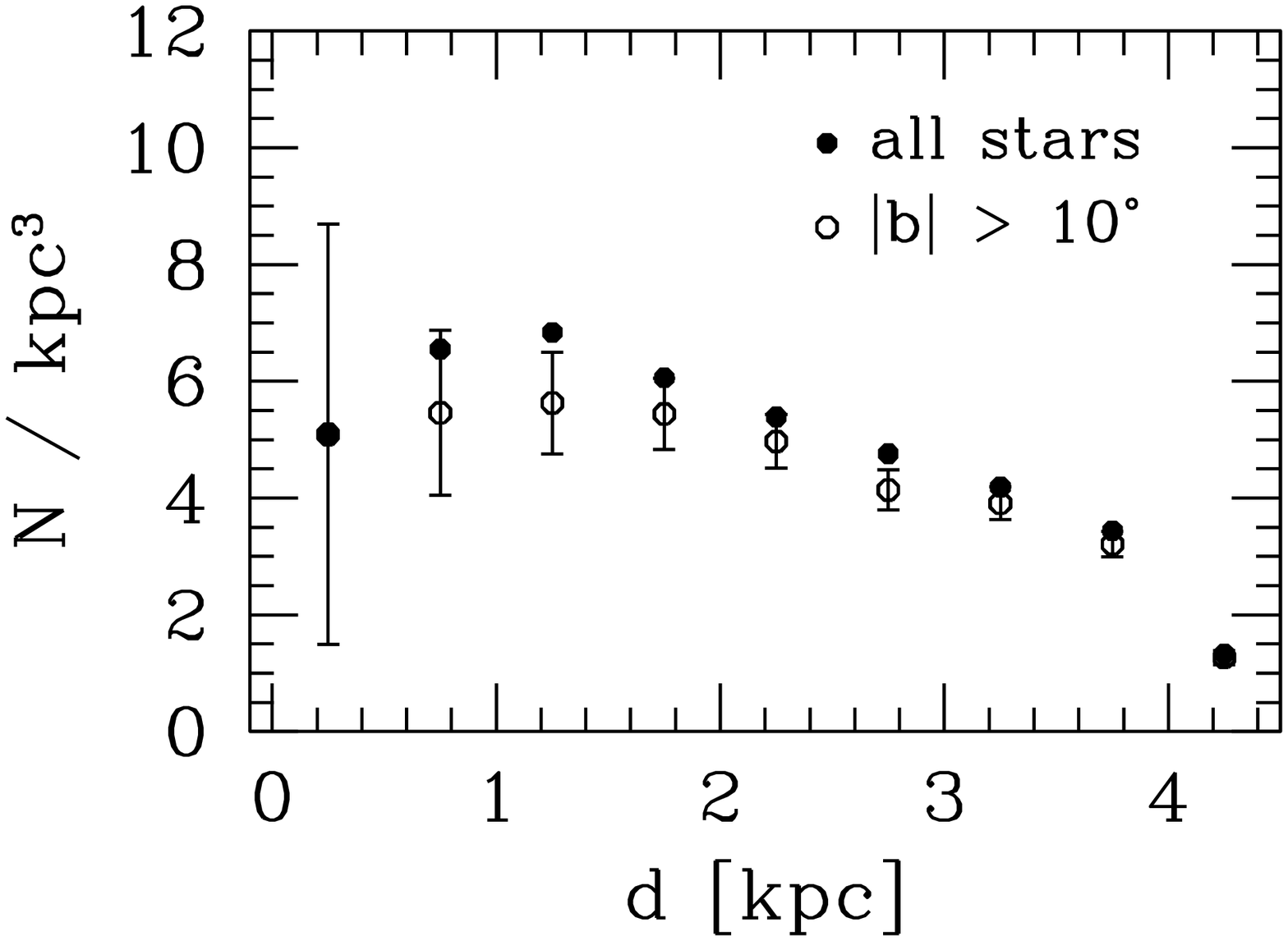} &
\includegraphics[width=6.3cm]{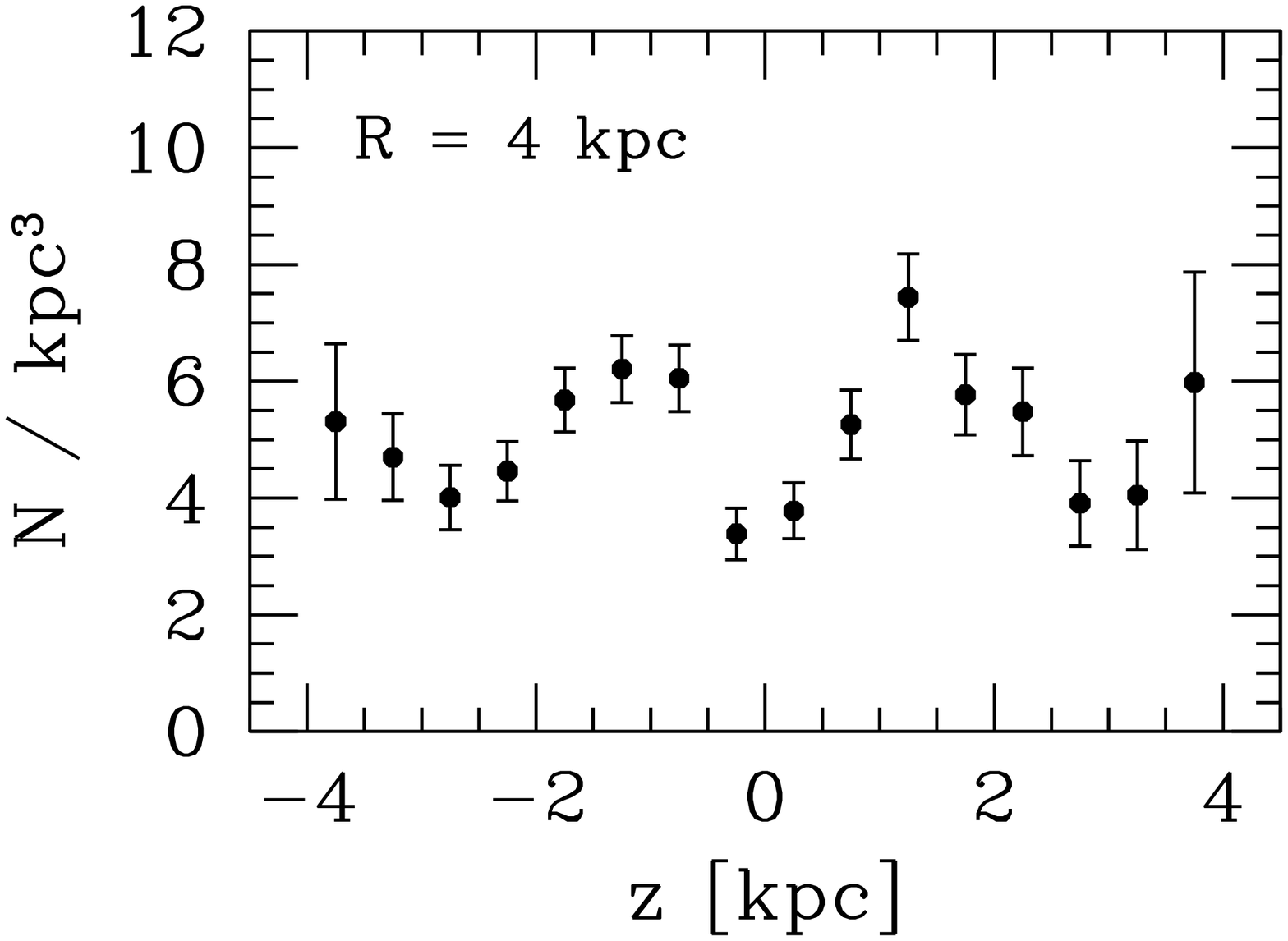} \\
\end{tabular}
\end{center}
\FigCap{Densities of ASAS RRab stars presented as a number of stars per 
kiloparsec cube. The {\it left panel} presents the density change with
distance $d$ from the Sun and the filled circles represent a whole sample
of 1008 RRab stars, while the open circles a subsample of 922 RRab stars
with absolute galactic latitude greater than 10\arcd. The {\it right panel}
shows the density as a function of distance $z$ from the Galactic
plane. The error bars were calculated assuming the Poissonian distribution
of star numbers in each bin. For details on the density calculation see
text.}
\end{figure}
In addition, we calculated densities of ASAS RRab stars depending on the
distance $d$ from the Sun and from the Galactic plane $z$ and the results
are presented in Fig.~15. In case of the distance from the Sun (left panel
of Fig.~15) the distance bin used to calculate the number of stars per
kiloparsec was set to 500~pc, so the densities were calculated for objects
within concentric spheres with the difference in radius equal to
500~pc. The calculations were performed for two data sets: one contained
all 1008 ASAS RRab stars for which the distances were calculated (filled
symbols in the left panel of Fig.~15) and the other contained objects for
which the absolute galactic latitude was greater than 10\arcd (open
symbols). As expected, the density is fairly constant within the Sun's
proximity and the slope of the decrease is small. In case of the distance
$z$ from the galactic plane (right panel of Fig.~15) the densities were
calculated within the slices of the 4~kpc radius sphere, and the thickness of
a slice was set to 500~pc. The density fluctuations are higher due to
smaller number of stars per bin, but the density stays roughly constant in
higher $z$ regions within the error bars. With decreasing $z$ an increase
in density is observed, as expected, but it drops in the low region of $z$
in which the ASAS resolution fails.  When calculating the densities
described above the correction for a limited ASAS vision was included. The
error bars in Fig.~15 were calculated assuming the Poissonian distribution
of star numbers in each bin.

\subsection{Metallicity Distribution in the Galaxy}
Kinemuchi \etal (2006) noticed that metal rich stars (with ${\rm
[Fe/H]>-1}$) are associated with the Galactic disk and confirmed this
conclusion by calculating their scale heights. As shown in the previous
Section (Fig.~14), our Oo~Ib sample is evidently concentrated toward the
Galactic plane, but it does not contain all metal rich RRab stars. In
Fig.~16 we plot metallicity values of all 1008 ASAS RRab stars against
absolute distance from the Galactic plane $|z|$. Here it is clear that
almost all metal rich RRab stars (${\rm [Fe/H]>-1}$~dex) are located at low
$|z|$ ($<2$~kpc) and they belong both to Oo~Ia and Oo~Ib groups. There is
no clear separation line between the Oosterhoff types like the one visible
in Fig.~14 of Kinemuchi \etal (2006) at ${\rm [Fe/H]=-1}$ because we did not
separate Oo~Ia and Oo~Ib groups with a constant [Fe/H] line, mostly a
result of using JK96 metallicity estimation method which does not have
constant [Fe/H] lines in $\log P{-}{\rm Amp}$ plane (for details refer to
Section~4).

\begin{figure}[htb]
\begin{center}
\includegraphics[width=9.7cm]{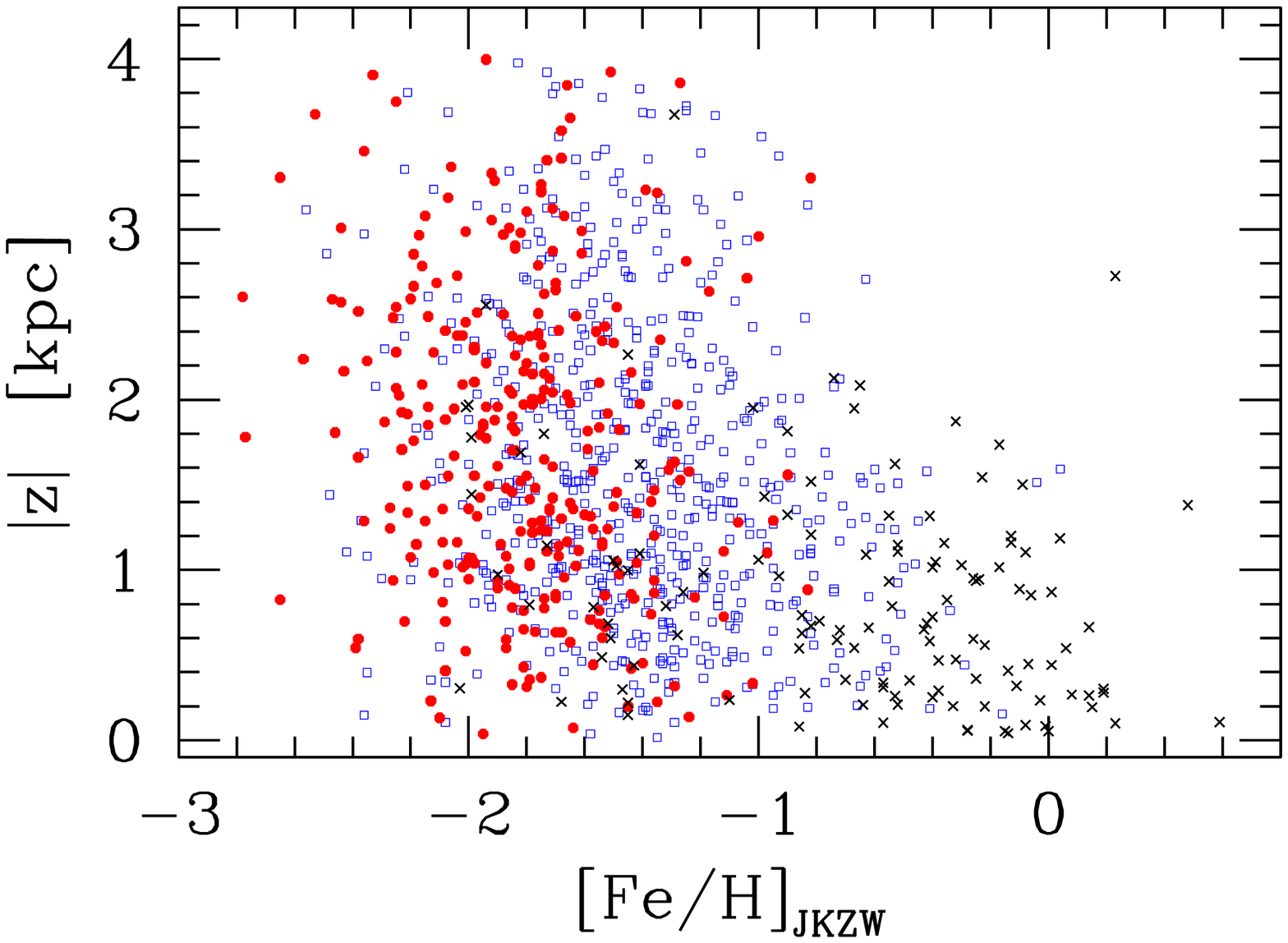}
\end{center}
\vspace*{-11pt}
\FigCap{The distribution of ASAS RRab stars metallicities against distance
from the Galactic plane. Red filled circles represent Oosterhoff~II group,
blue open squares stand for the main part of the Oosterhoff~I group (Oo~Ia)
while black crosses represent Oo~Ib stars.}
\end{figure}

Within the main part of Oosterhoff~I group (Oo~I, blue squares) we notice a
decrease in $|z|$ with increasing metallicity, which does not seem to be
present among Oo~II type stars. In the intermediate and low metallicity
region, for ${\rm [Fe/H]}\in(-2.5,-1)$~dex the cloud is a mixture of both
Oosterhoff~I and II types, uniformly distributed at all values of $|z|$.

In Fig.~17 we plot metallicity values averaged in several distance bins in
order to investigate global changes in metallicity across the Galaxy.  The
first row of images presents average metallicity values of all ASAS RRab
stars against distance $d$ from the Sun (left panel), absolute distance
$|z|$ from the Galactic plane (middle panel) and distance from the Galactic
center $R$ (right panel). The distance of the Sun from the Galactic center
was assumed to be 8~kpc. The second row presents analogous images, but
metallicities are averaged within Oosterhoff groups -- red filled circles
stand for Oo~II RRab stars, blue open squares for the main part of
Oosterhoff~I group, namely Oo~Ia, and black crosses represent an Oo~Ib
subgroup.

\begin{figure}[htb]
\begin{center}
\begin{tabular}{@{}c@{}c@{}c@{}}
\includegraphics[width=4.3cm]{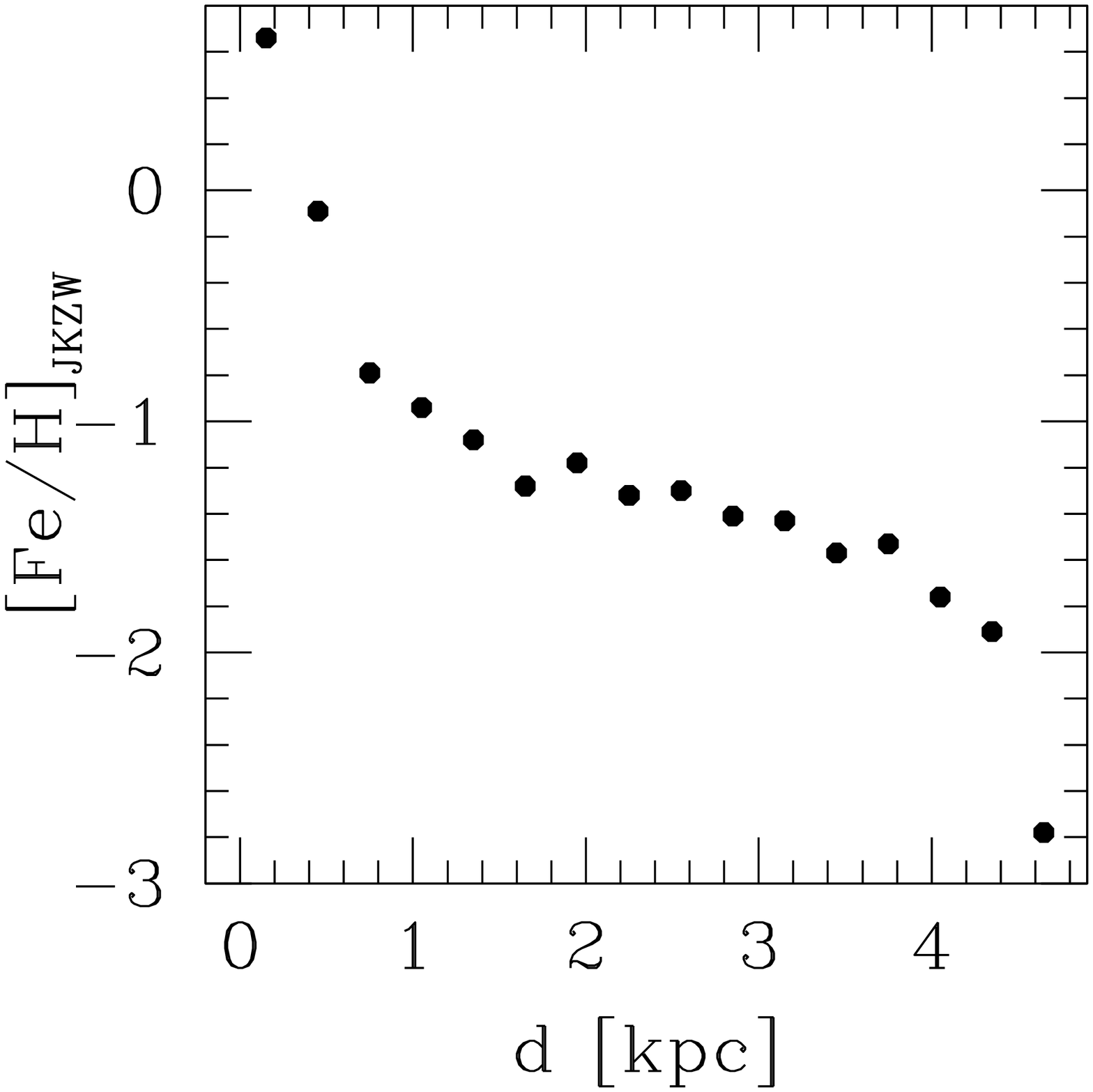} &
\includegraphics[width=4.3cm]{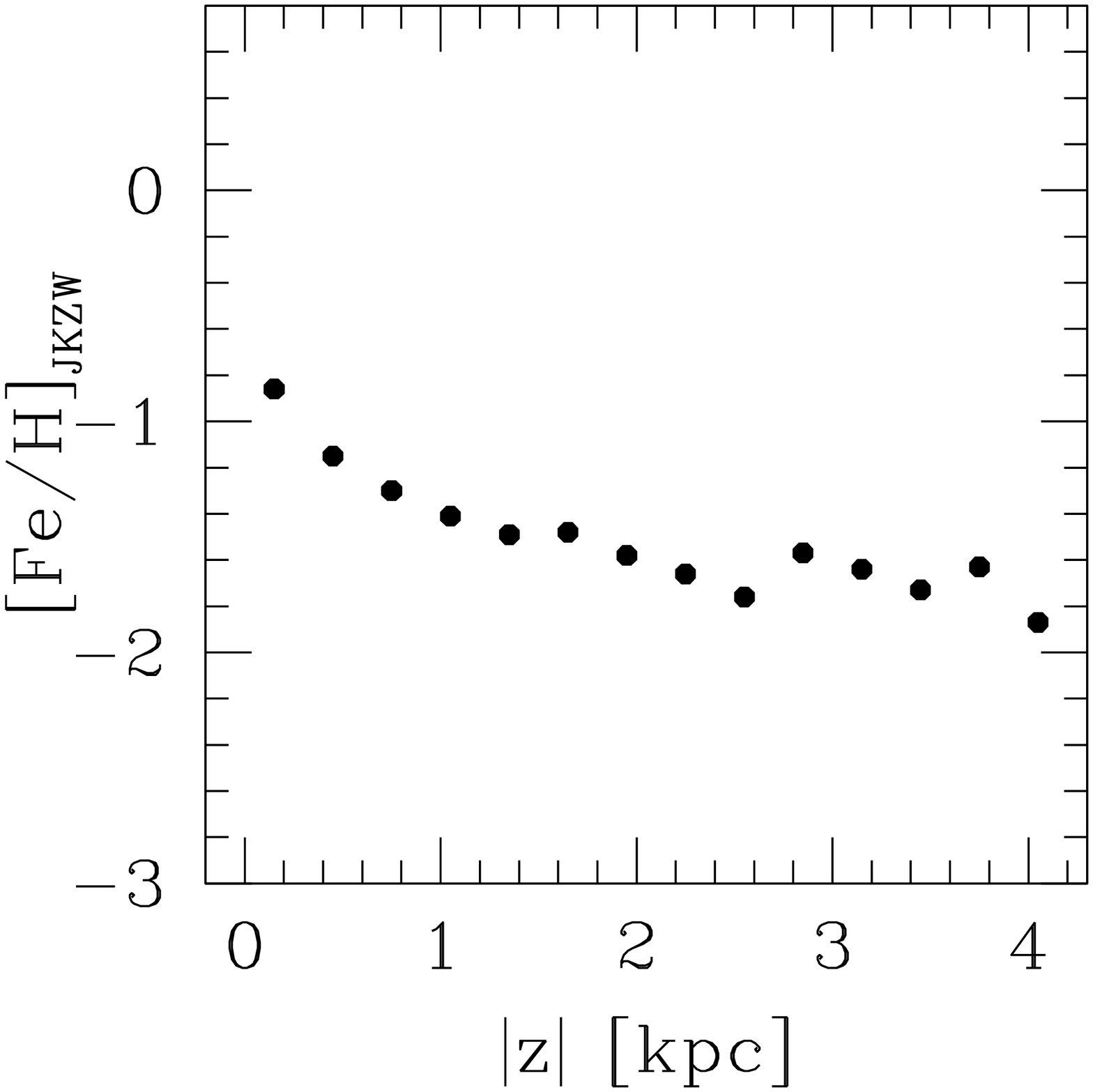} &
\includegraphics[width=4.3cm]{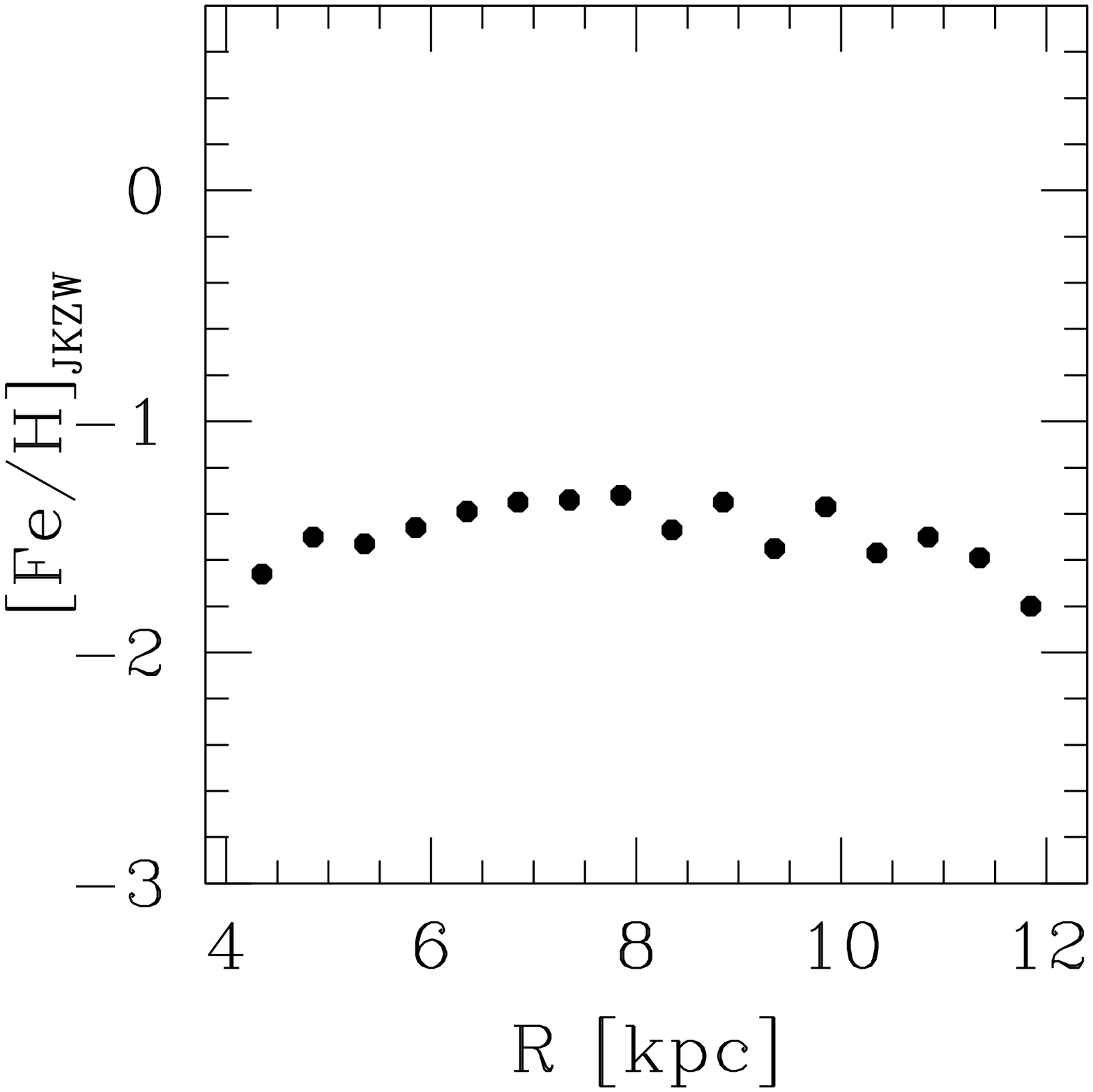} \\
\includegraphics[width=4.3cm]{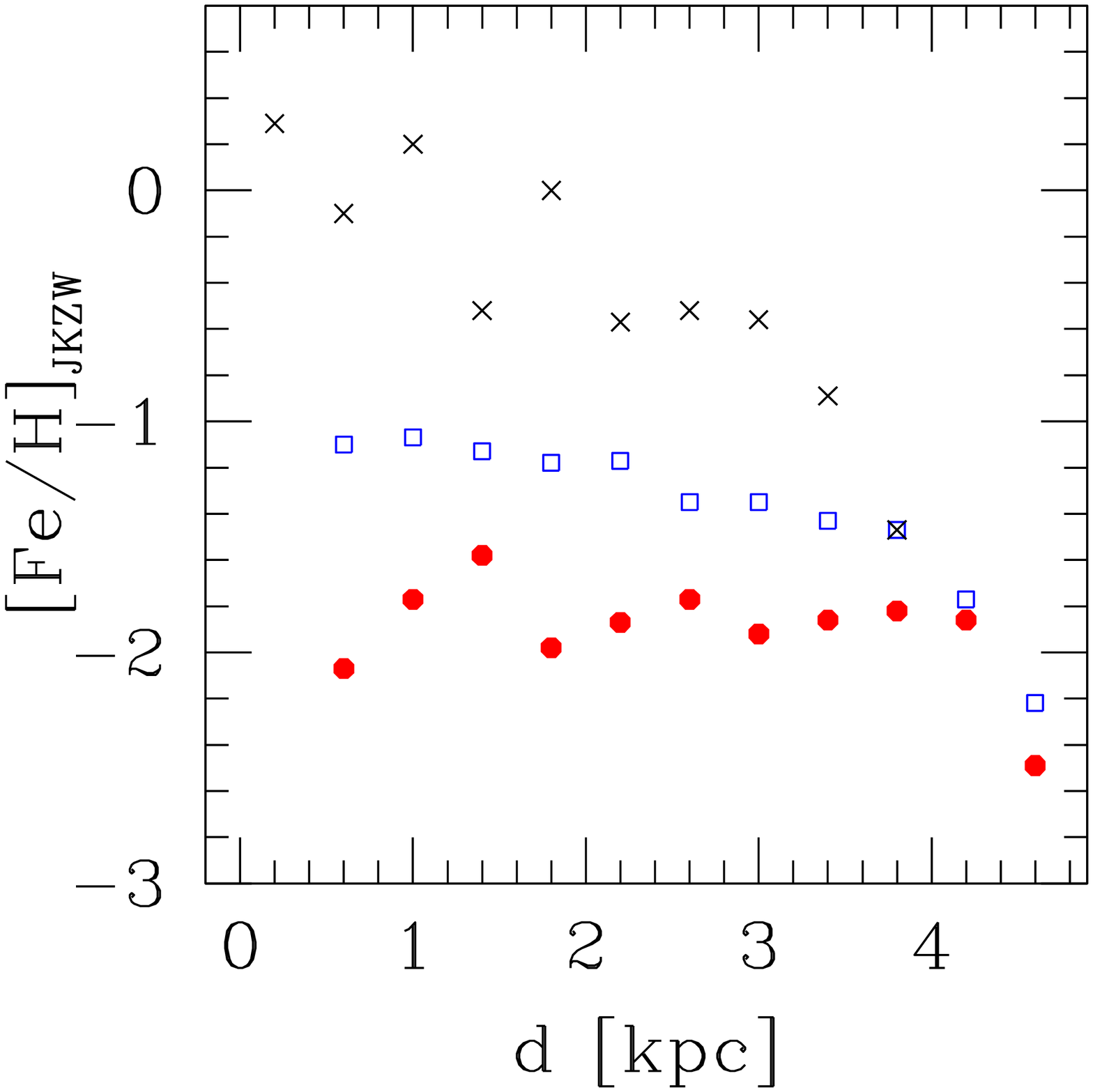}   &
\includegraphics[width=4.3cm]{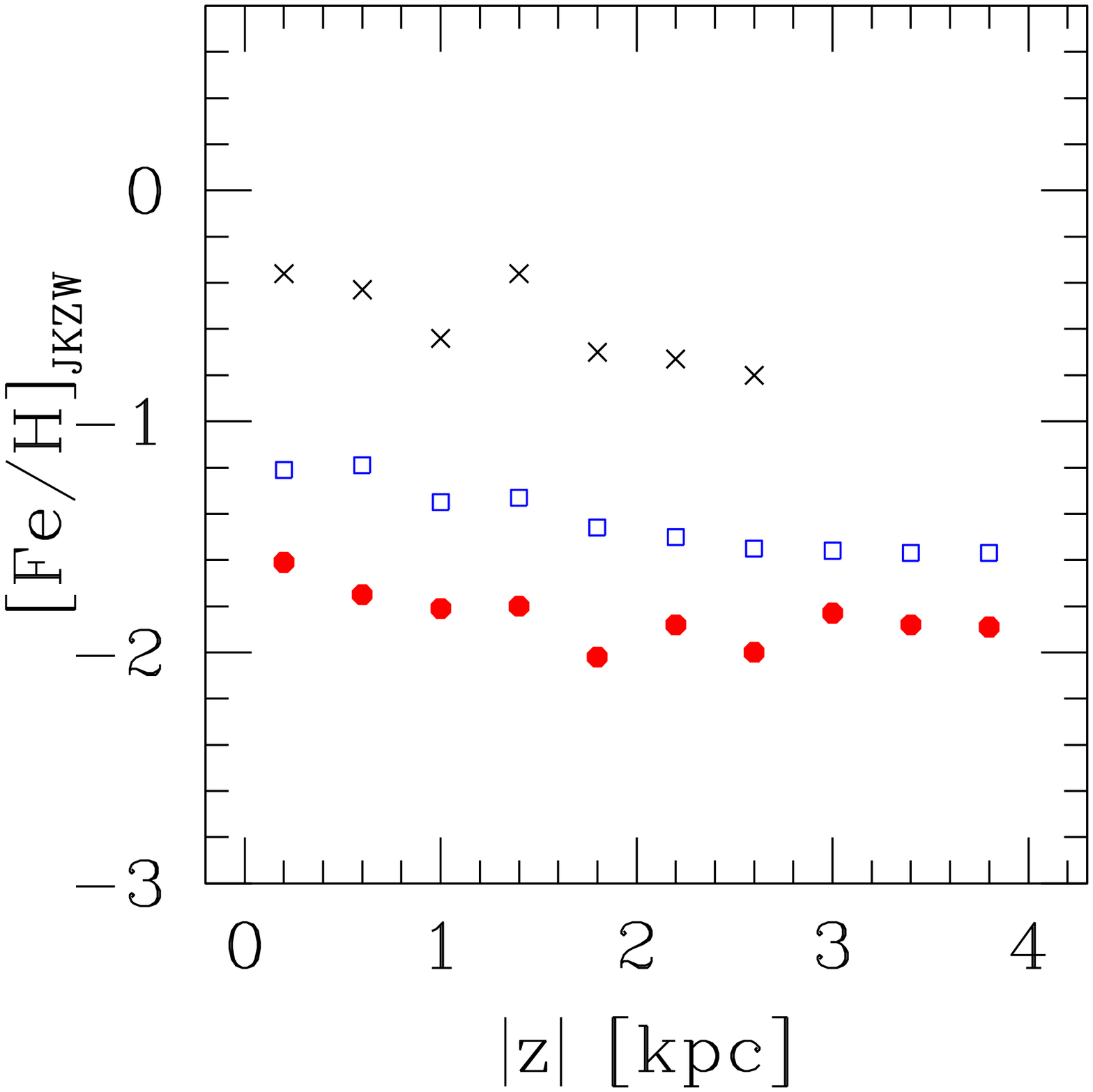} &
\includegraphics[width=4.3cm]{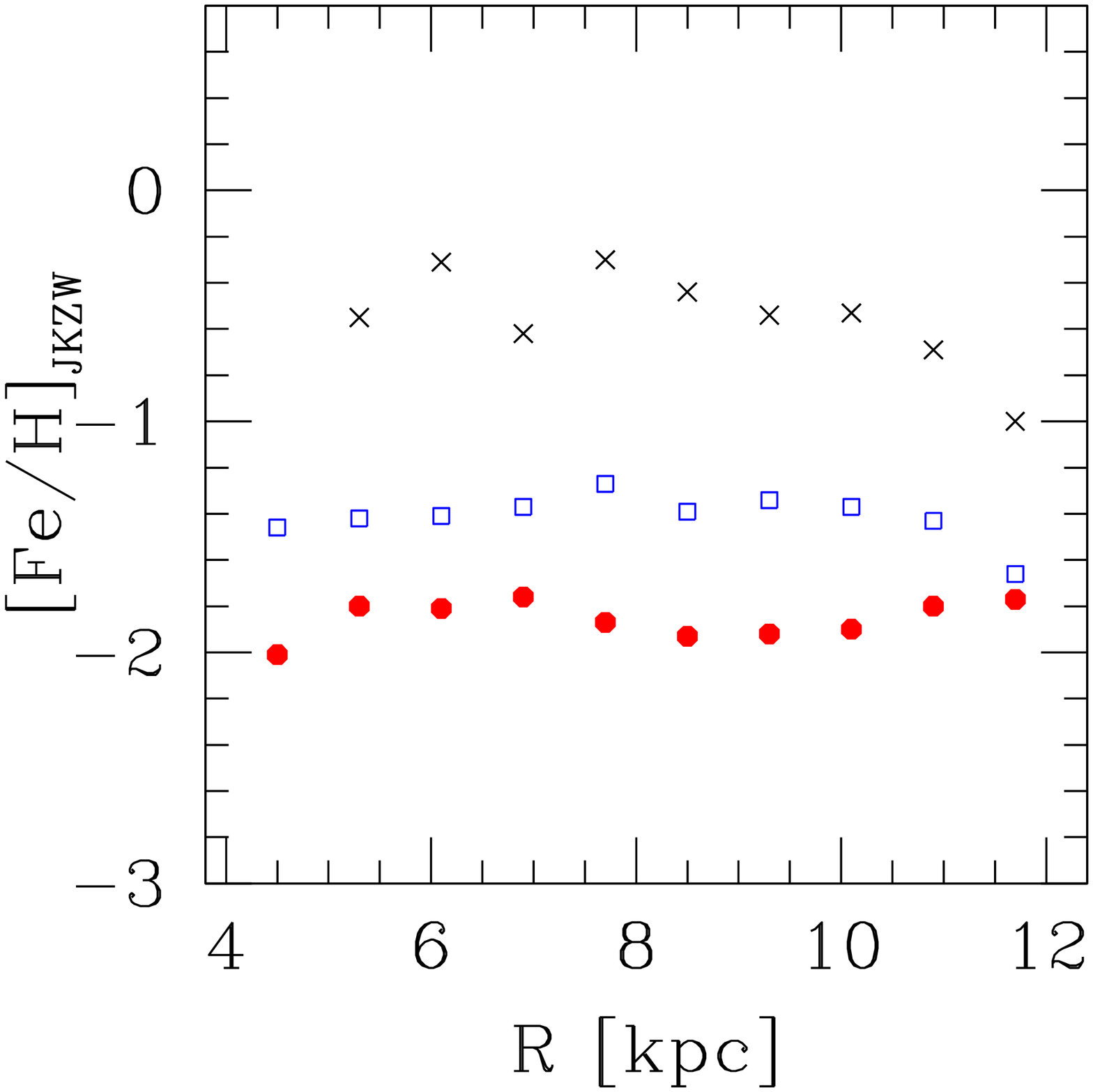}  \\
\end{tabular}
\end{center}
\vspace*{-11pt}
\FigCap{All images present metallicity values averaged in 
distance bins. The {\it first row} plots show metallicity values of all
ASAS RRab stars against distance $d$ from the Sun ({\it left panel}),
absolute distance $|z|$ from the Galactic plane ({\it middle panel}) and
distance from the Galactic center $R$ ({\it right panel}). The {\it second
row} presents analogous images, but metallicities are averaged within
Oosterhoff groups -- filled circles stand for Oo~II RRab stars, 
open squares for the main part of Oosterhoff~I group, namely Oo~Ia, and
crosses represent an Oo~Ib subgroup.}
\end{figure}
The overall image emerging from Fig.~17 is consistent with the general
model of the Galaxy -- the closer to the Sun or to the Galactic
plane we are, the higher are the metallicities of ASAS RRab stars. When we
distinguish between Oosterhoff types it is clear that the largest
contribution to this rise comes from the metal rich Oo~Ib stars which
mostly occur at lower $|z|$. In other words: younger, more metal rich stars
(Oo~I) are concentrated toward the disk, while the older, more metal poor
stars (Oo~II) are on average at larger distances from the Galactic
plane. Both Oo~Ia and Oo~Ib groups display higher metallicities when closer
to the Sun, while the metallicity of Oo~II remains fairly constant
with~$d$. The increase in metallicity with decreasing $|z|$ is present
among all three Oo groups and becomes well defined at
$|z|\approx2.5$~kpc. There is no apparent metallicity trend with distance
from the Galactic center, as shown by Kinemuchi \etal (2006).

\Section{Summary and Conclusions}
The aim of this paper was to analyze 1455 fundamental mode nearby RR~Lyr
stars using the All Sky Automated Survey data. In the process of data
preparation we have inspected all {\it V} and {\it I}-band RRab stars light
curves, cleaned them and refined or recalculated their pulsation
periods. During this stage we rejected stars that had low quality light
curves.

We constructed a period--amplitude diagram for a reduced sample of stars
and unlike in the previous analysis of Galactic field RRab stars
(Kinemuchi \etal 2006) we noticed a clear presence of the Oosterhoff groups
(Oosterhoff 1939) typical for globular cluster RR~Lyr variables. This,
together with the recent study of outer halo RR~Lyr stars (Miceli \etal
2008) implies that the Oosterhoff dichotomy is present in the whole Milky
Way Galaxy including a very close neighborhood of the Sun. It follows that
the accretion of close-by dwarf galaxies as a scenario of early Galaxy
formation can be excluded.

The $\log P{-}{\rm Amp}_V$ relation for Oosterhoff~I RRab stars becomes
strongly flattened at large amplitudes, which was not observed for globular
cluster and outer halo field RR~Lyr before. We also observed a shift of the
two Oosterhoff relations with respect to those of globular clusters. Since
the Oosterhoff dichotomy is believed to be caused by RR~Lyr stars age
differences (about 2~Gyrs for typical globular cluster), and the location
on the Bailey diagram can be connected with the formation time, then this
could imply different creation scenarios for nearby RR~Lyr stars than for
globular cluster RR~Lyr stars.

The density maps of Oosterhoff groups on the Bailey diagram reveal
higher concentrations of stars at certain period and amplitude values,
but we do not know whether they are of any significance. However, we
could speculate that since the location on the period--amplitude diagram
is an indicator of evolutionary state, metallicity or age,  prefered
certain locations would mean that some states are more probable than the
others.

We also put ASAS RRab stars on a period--color diagram. There is no evident
dependence of $V-I$ color on period at minimum light, but when using color
at maximum light two separate groups of stars are visible. These groups are
formed by Oosterhoff~I and II type variables.

Using methods of Jurcsik and Kov\'acs (1996) and Sandage (2004) we
calculated photometric metallicity values for 1008 best quality light
curves. We noticed high discrepancies between [Fe/H] values from different
methods, being also a reflection of the Oosterhoff dichotomy. This suggests
that different metallicity relations should be constructed for two
Oosterhoff groups, which seems reasonable if we assume that RRab stars of
Oosterhoff types are of different age. This observation is supported by
comparison of photometric and spectroscopic [Fe/H] values.

Comparison of both methods with spectroscopic metallicities from Layden
(1994) favors the method of Jurcsik and Kov\'acs (1996) as significantly
better for photometric metallicity estimation. The method of Sandage (2004)
produces an un-physically bimodal distribution of metallicities which is
not seen among RRab stars with spectroscopically obtained [Fe/H] values,
meaning one can expect high errors in metallicities calculated with this
method. Thus the Sandage (2004) method although attractive because of
simplicity of its application, should be used with great caution.

The metallicity distributions of ASAS RRab stars show that Oosterhoff~II
type stars have lower metallicity values, below $-1$~dex, while
Oosterhoff~I type stars can be found in the whole metallicity range. The
more metal rich, younger stars (Oo~I) are observed on average closer to
the Galactic plane than older, more metal poor stars (Oo~II) as
expected, but both Oosterhoff groups are present at all values of $|z|$
implying that the disk is a mixture of young and old RR~Lyr stars.
However, the increase in metallicity with decreasing $|z|$ is present
among both Oosterhoff groups as well as in a short period Oo~I subsample,
and becomes well defined at $|z| \approx2.5$~kpc. We did not observe any
significant metallicity trend with distance from the Galactic center,
which is in agreement with the results of Kinemuchi \etal (2006).

\Acknow{We are thankful to the anonymous referee for useful suggestions.
This research has made use of NASA's Astrophysics Data System. The authors
were supported by the Polish MNiSW grants N203~007~31/1328 and
N~N203~304235.}

\newpage 
\centerline{\bf Appendix}
\vskip9pt
{\it Comparison of ASAS and NSVS Catalogs}
\vskip7pt
{\it I. The NSVS Project}
\vskip3pt
The NSVS project was based on data obtained with the Robotic Optical
Transient Search Experiment (ROTSE-I) during one year observations between
April 1999 and March 2000. ROTSE-I telescopes were located in Los Alamos,
New Mexico and scanned the sky north of declination $-38\arcd$. The catalog
contains stars brighter than unfiltered 15.5~mag (about 15~mag {\it V}-band
equivalent), with about 100--400 good measurements per star. For details
on the ROTSE-I experiment refer to Akerlof \etal (2000) and on the NSVS
catalog to Wo¼niak \etal (2004).  The NSVS catalog does not provide
variability classification itself, but there have been many studies doing
so. In particular, RR~Lyr type variables have been identified by
Kinemuchi \etal (2006) and Wils \etal (2006).

Thanks to the similarity of ASAS and ROTSE telescopes, the ACVS and NSVS
catalogs have almost the same sky resolution, detection efficiency,
magnitude range and incidence of variable stars, that should allow us to
join both databases to create a complete and uniform galactic field RR~Lyr
stars catalog.

\vskip7pt
{\it II. RR~Lyr Stars Sample from NSVS}
\vskip3pt

As already mentioned, there had been at least two independent studies of
RR~Lyr variables from NSVS, one by Wils \etal (2006), and the other one by
Kinemuchi \etal (2006), hereafter K06. The primary difference between them
is the selection criteria used to extract the RR~Lyr stars sample from NSVS
data. Wils \etal (2006) used well known RR~Lyr stars from GCVS that are
present in NSVS database, determined statistical properties of their light
curves and used them to identify all RR~Lyr stars in NSVS. The search
resulted in 785 RR~Lyr stars, 714 of them being RRab Bailey type stars. K06
made use of a cross-correlation of NSVS variable candidates with 2MASS
database and used 2MASS colors as one of the selection parameters. This
resulted in 1563 RR~Lyr stars, among them 1188 RRab stars. In this work we
are using the results of K06 for two reasons: because their selection
procedure was more similar to the one of ASAS (also using 2MASS color
information), and more important, they determined photometric metallicities
for 589 RRab variables, being an important RRab stars property investigated
in this study.

\vskip7pt
{\it III. Combining ASAS and NSVS Samples}
\vskip3pt

While it is true that both ASAS and NSVS RR~Lyr stars data have been
obtained with similar equipment, there are several differences that could
affect the conformity of both catalogs. First, ROTSE-I observed without
filters, thus gathering light in a wide photometric region from $B$ to $I$,
and the peak sensitivity being similar to the {\it R}-band. Given that
RR~Lyr are blue stars, this could result in not detecting the bluest of
them. Second, different algorithms were used to identify and classify
RR~Lyr stars in the database.

In order to verify the conformity of the ASAS and NSVS RRab stars samples
we investigated variables in an overlapping region, that is for
declinations between $-38\arcd$ and $+28\arcd$. The are 967 ASAS objects
and 806 NSVS objects in this area ($806/967=83\%$), of which only 446 are
common ($446/967=46\%$, $446/806=55\%$). The first conclusion is that ASAS
probably has a better detection efficiency, since it detected more RRab
stars than NSVS, while in fact it should detect less -- ASAS magnitude
range is shifted towards brighter objects with respect to NSVS and the
number of stars rises rapidly with larger magnitudes. The reason that NSVS
missed a number of objects could be the fact that ROTSE-I observations are
made without filters (and only later transformed to their $V$ equivalents)
and are less sensitive in the blue regime, occupied by RR~Lyr stars. This
could be a reason for missing the bluest RR~Lyr stars by NSVS.

Another explanation for these high discrepancies would be that RR~Lyr stars
search and classification algorithms of ASAS and K06 differ significantly,
as only about half of objects in each group has their counterparts in the
other group (45\% in case of ASAS and 55\% in case of NSVS). As a further
check, we want to see how these numbers look for a more conservative sample
in the narrower magnitude range.

\vskip7pt
{\it IV. Comparison of ASAS and NSVS Magnitudes}
\vskip3pt

\begin{figure}[htb]
\begin{center}
\begin{tabular}{cc}
\includegraphics[width=6cm]{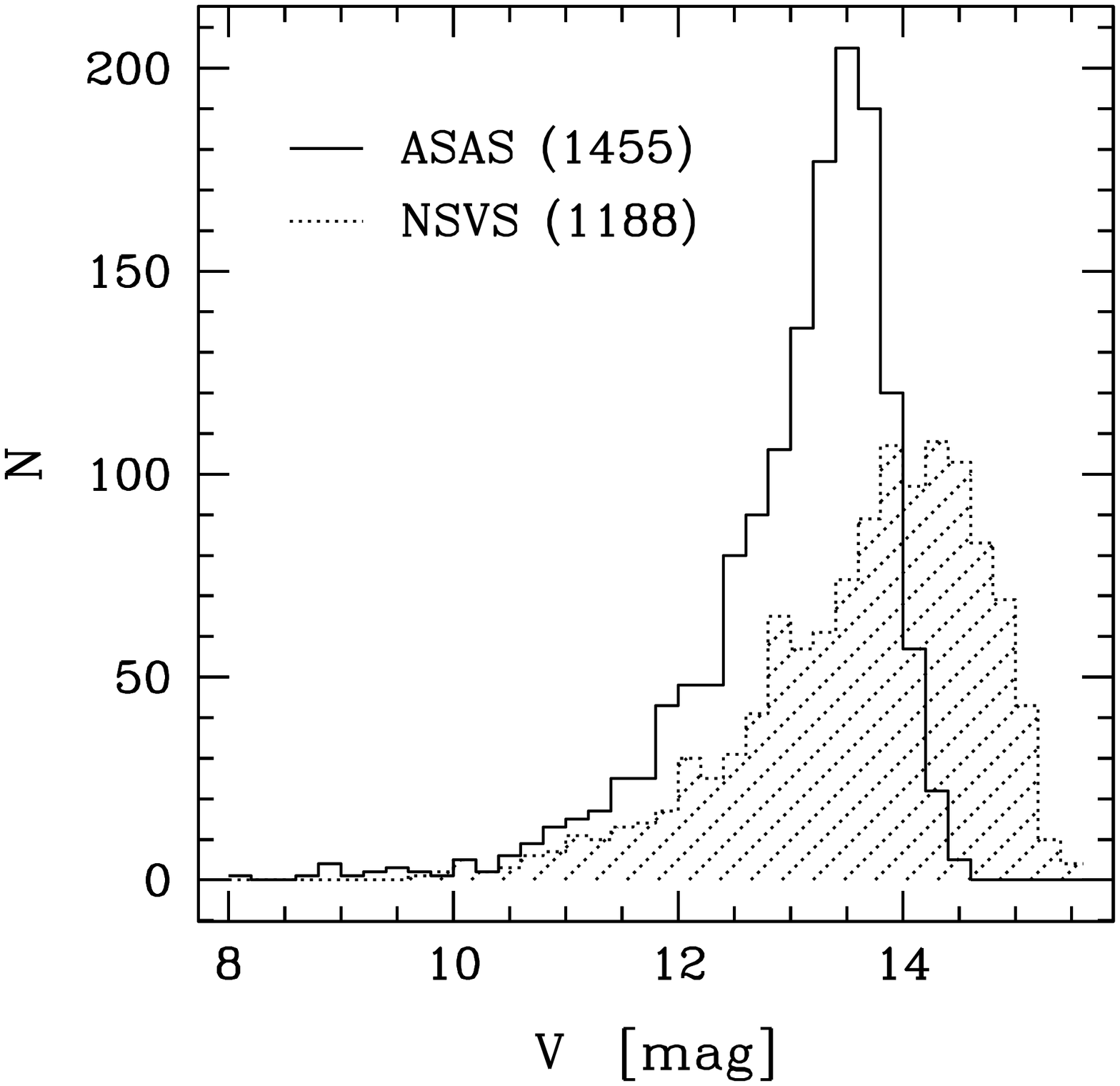} &
\includegraphics[width=6cm]{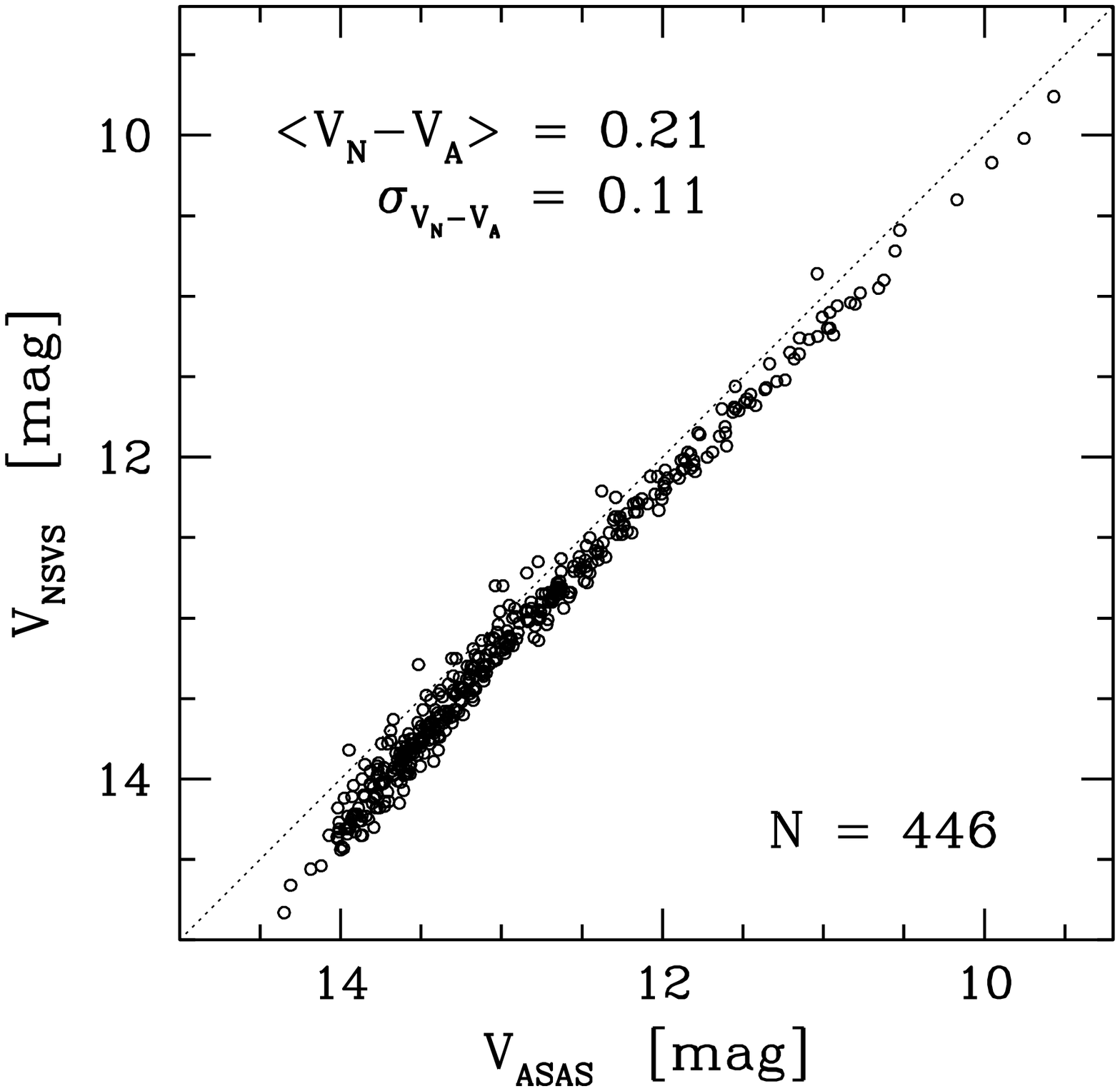} \\
\end{tabular}
\end{center}
\FigCap{{\it Left panel} shows a histogram of mean magnitudes of all ASAS 
(plain) and NSVS (shaded) RRab variables. In the {\it right panel} there is
a comparison of {\it V}-band magnitudes of ASAS and NSVS counterparts from
the overlapping region between declinations $-38\arcd$ and $+28\arcd$. NSVS
magnitudes are on average 0.2~mag larger than their ASAS equivalents.}
\end{figure}
In order to see RRab stars magnitude distributions, we construct a
histogram of mean magnitudes of all RRab stars (the whole sphere) for both
samples in the left panel of Fig.~18. It is clear that for a given
magnitude bin there are less RRab stars in the NSVS database, and at the
same time NSVS observes fainter stars than ASAS. This supports the argument
that NSVS, while having a larger magnitude limit probably misses the bluest
RR~Lyr stars due to lower sensitivity in this photometric region.

The {\it V}-band magnitudes of ASAS and NSVS counterparts from the
overlapping region between declinations $-38\arcd$ and $+28\arcd$ are
plotted in the right panel of Fig.~18. The average magnitude difference in
star to star comparison is $\langle V_{\rm NSVS}-V_{\rm ASAS}\rangle=0.21$~mag and
the dispersion is $\sigma_{\langle V_{\rm NSVS}-V_{\rm ASAS}\rangle}=0.11$~mag. In
other words, NSVS magnitudes are on average 0.2~mag higher that their ASAS
equivalents. Given this, we shift magnitudes of NSVS sample by 0.21 when
combining with the ASAS sample.

Now, we narrow our sample to the magnitude range 10--12~mag. There are 72
NSVS objects and 114 ASAS ($72/114=63\%$), with 63 common objects (which
constitutes 87.5\% for NSVS and 55\% for ASAS). That means, most stars
observed by ROTSE-I telescopes are visible for ASAS, while about half of
ASAS variables are in the NSVS data. This suggests a better completeness of
ASAS catalogs, either due to better sensitivity or more efficient detection
procedures.

\vskip7pt
{\it V. Comparison of ASAS and NSVS Amplitudes}
\vskip3pt

\begin{figure}[htb]
\begin{center}
\begin{tabular}{cc}
\includegraphics[width=6cm]{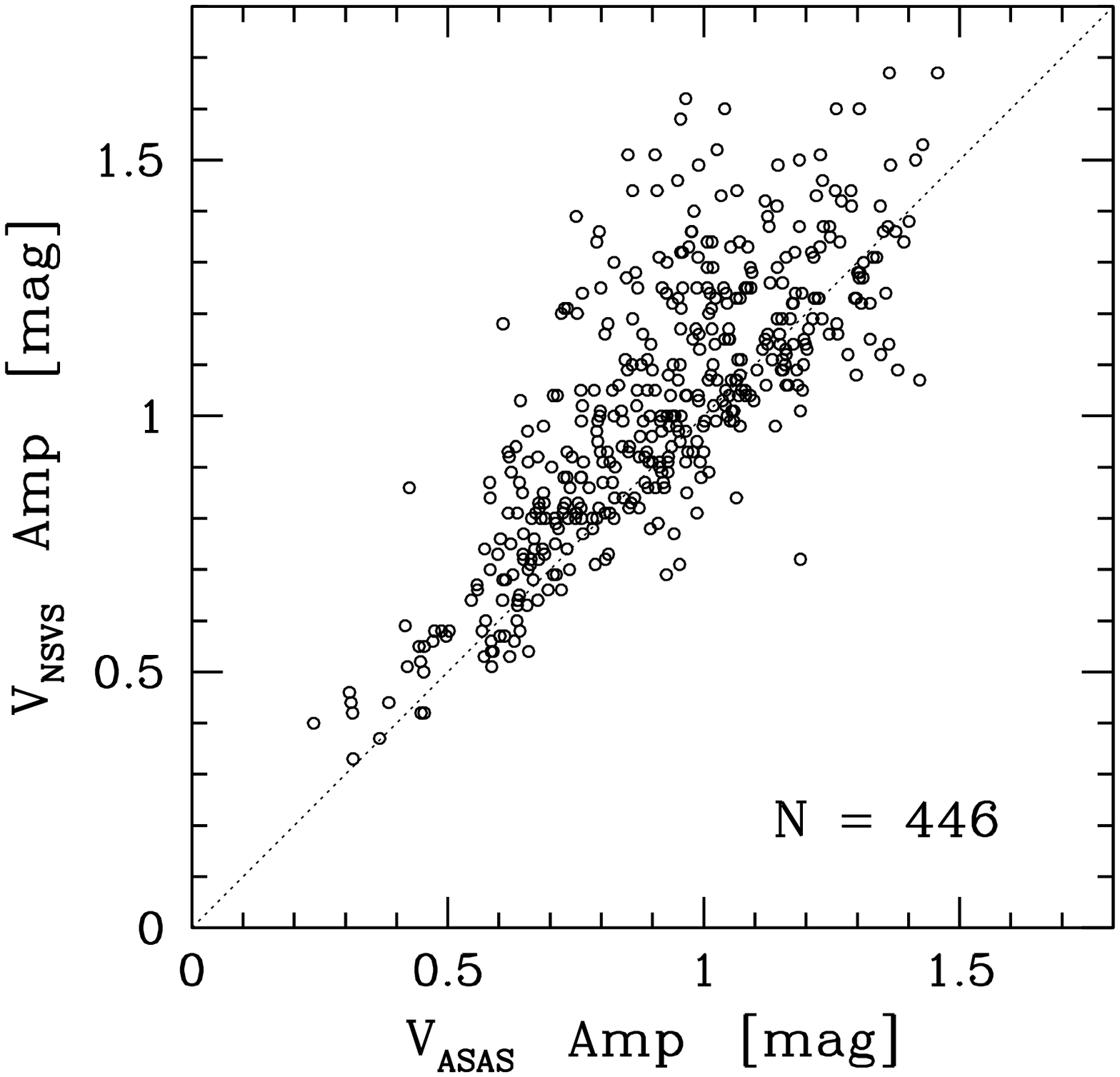} &
\includegraphics[width=6cm]{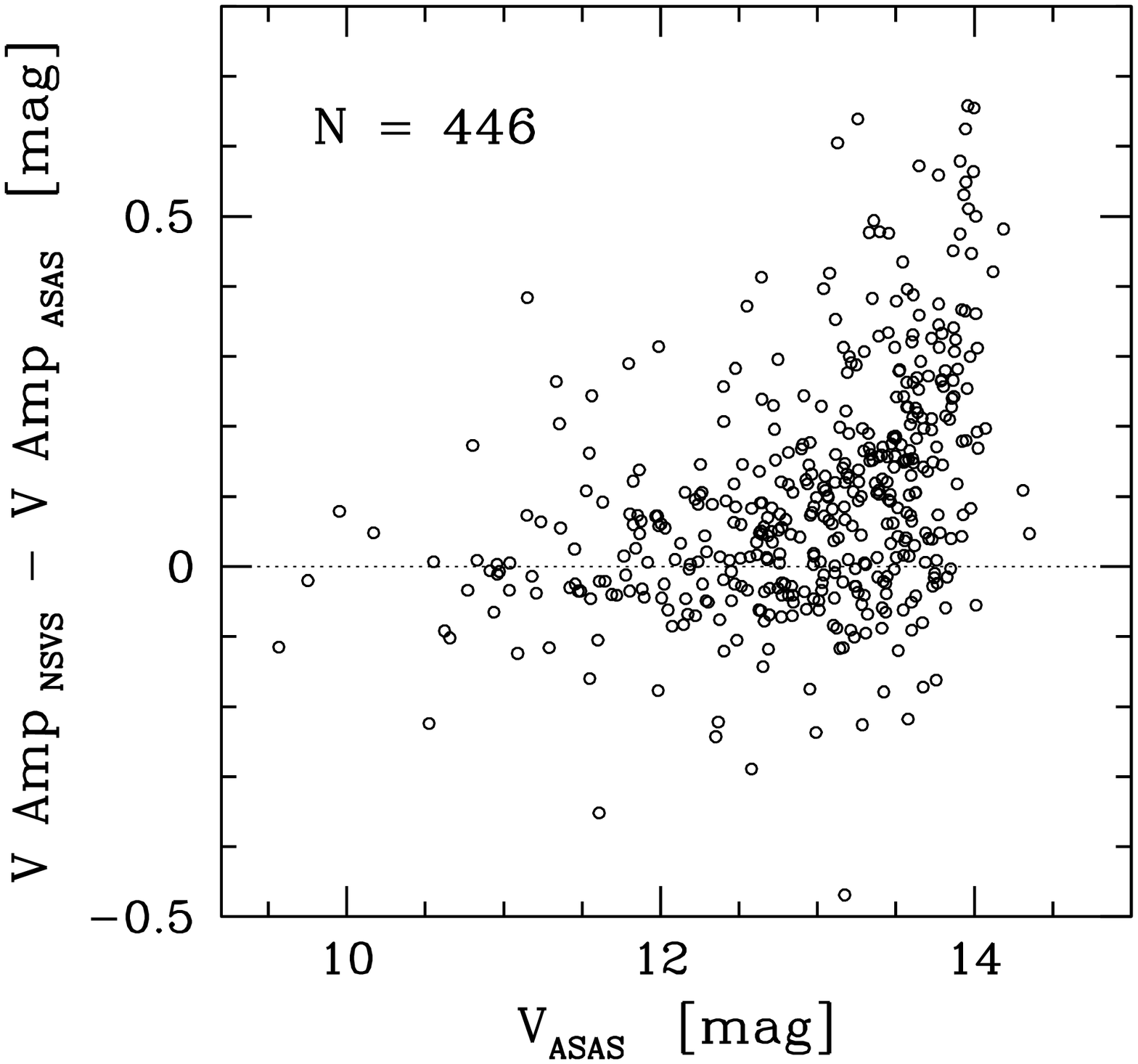} \\
\end{tabular}
\end{center}
\vspace*{-11pt}
\FigCap{In the {\it left panel} there is a star to star {\it V}-band amplitude
comparison of ASAS and NSVS RRab stars in the overlapping region. NSVS
objects have significantly larger amplitudes than their ASAS counterparts
and this difference is largest for largest amplitudes. The {\it right
panel} shows how the amplitude difference depends on the observed magnitude
value -- the fainter the star, the larger the difference is.}
\end{figure}
We have also checked the conformity of {\it V}-band amplitudes for stars
from the overlapping region. The left panel of Fig.~19 presents a star to
star amplitude comparison. It is clear that the scatter is fairly large,
with NSVS amplitudes often much larger than that of ASAS, and the highest
scatter is at largest amplitudes. The right panel of the same figure shows
how the difference between NSVS and ASAS amplitudes ($A_{\rm NSVS}-A_{\rm
ASAS}$) depends on the magnitude -- the fainter the star, the larger the
difference. It follows, that if we want to combine both samples in the
means of amplitudes or any amplitude dependent parameters, we should put
the magnitude limit at about 12~mag on both samples (catalogs).

Some amplitude differences at the faint ASAS limit, below $V=13.5$~mag are
most probably due to amplitude underestimation for ASAS RRab stars -- the
faintest points are missed because of the observing limit. But this does
not explain all other amplitude differences. We have visually inspected a
number of light curves with large amplitude differences and noticed that
almost all of them display a high scatter and look like multiperiodic
RR~Lyr stars. ASAS amplitudes are defined as the magnitude difference
between the minimum and maximum of the fitted model, which in these cases
will be smaller than the difference between the faintest and the brightest
data points. If NSVS amplitudes are calculated in the latter way, this
would explain the differences.

\vskip7pt
{\it VI. Metallicities of NSVS RRab Stars}
\vskip3pt

K06 calculated metallicities for 589 RRab stars from their original sample of
1188 objects, which had sufficient number of observations and a good phase
coverage. They used both methods described in Section~4.1, but the method
of JK96 was not applied to a subsample of 175 RRab stars because of large
$\varphi_{31}$ errors resulting from a poor Fourier decomposition of the
light curves.

\begin{figure}[htb]
\begin{center}
\includegraphics[width=7cm]{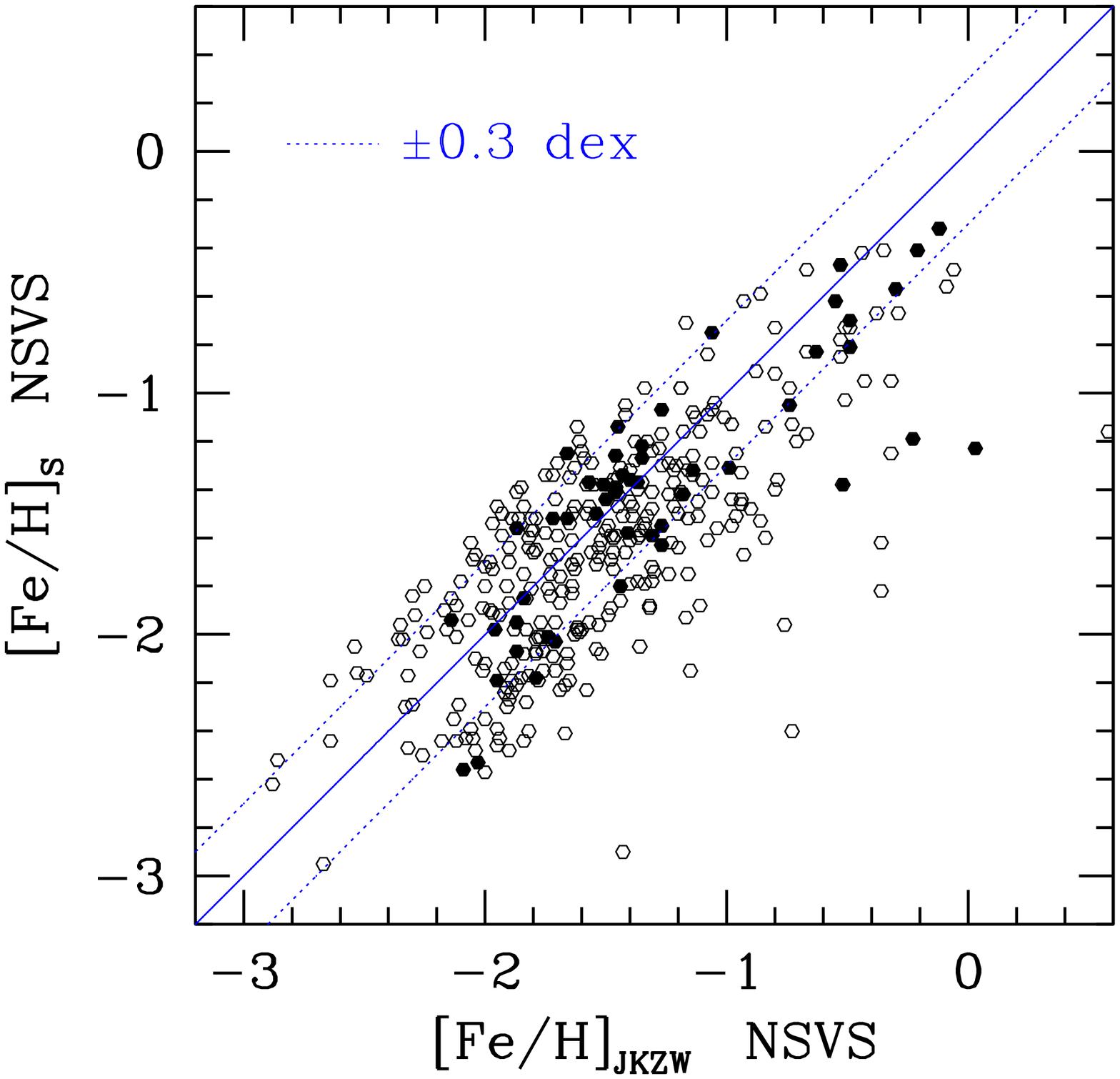}
\end{center}
\vspace*{-11pt}
\FigCap{Comparison of NSVS RRab stars metallicities computed with two
methods: JK96 and S04. Filled symbols stand for variables brighter than
12~mag.}
\end{figure}
The comparison of two methods within NSVS sample, similarly to Fig.~10 is
presented in Fig.~20. There are 363 RRab stars that have photometric
metallicities calculated with both methods. The mean difference is $\rm
[Fe/H]_{diff}=[Fe/H]_{JKZW}-[Fe/H]_S=0.15$ and the scatter is $\sigma_{\rm
[Fe/H]_{\rm diff}}=0.39$.  We see that S04 metallicities are on average
lower than JK96 and the shift towards larger ${\rm[Fe/H]_{\rm JKZW}}$ in
the region of lower metallicities is observed similarly to
Fig.~10. However, there is no clear evidence of the group shifted towards
larger ${\rm[Fe/H]_S}$ that is visible in Fig.~10. We suspect this is the
result of careful data selection for JK96 method by K06.  Nevertheless, we
observe a similar discrepancy between methods as was seen in ASAS data
(Fig.~10), even though K06 chose only best light curves for JK96
metallicity estimation.

As a final metallicity value they adopted a weighted average of both
metallicities for 363 objects and the metallicity of S04 alone for the
remaining 226 objects. Among these were 175 stars with large $\varphi_{31}$
errors and a few dozens of stars that had very big differences of
metallicities computed from JK96 and S04, even as large as 3~dex; for them
the S04 method was assumed to be more reliable.

\vskip3pt
{\it VII. Comparison of ASAS and NSVS Metallicities}
\vskip3pt

Now we compare metallicities of ASAS and NSVS RRab stars in the overlapping
region. Out of 589 NSVS RRab stars for which the metallicities were
obtained 332 stars fall in this area, among them there are 190 objects that
have metallicity values from both methods (${\rm [Fe/H]_S}$ and ${\rm
[Fe/H]_{\rm JKZW}}$) and 142 with ${\rm [Fe/H]_S}$ only. As shown in
Section~V NSVS amplitudes are on average bigger than ASAS amplitudes and
the fainter the star the bigger the difference. Since the method of S04
depends on amplitude, we should be careful calculating ${\rm [Fe/H]_S}$
when $V>12$~mag. Therefore in the following images we will distinguish a
subsample of bright RRab stars with filled symbols.

\vspace*{-15pt}
\MakeTableee{|c|c|c|c|c|c|c|c|}{12cm}{Comparison of ASAS and NSVS metallicities}
{\hline
\douprule Method & ${\rm [Fe/H]_{\rm diff}}$ & $\sigma_{\rm [Fe/H]_{\rm diff}}$ &
${\rm N_{\rm obs}}$
& & ${\rm [Fe/H]_{\rm diff}}$ & $\sigma_{\rm [Fe/H]_{\rm diff}}$ & ${\rm
N_{\rm obs}}$ \\
\hline
JK96             &$-0.02$& 0.50 & 190 &\multirow{4}{*}{${\rm | [Fe/H]_{\rm diff}| \le  0.3 }$} & 0.01 & 0.15 & 124\\
S04              &$ 0.09$& 0.42 & 332 & & 0.06 & 0.13 & 260\\
JK96 ($V \le 12$) &$-0.07$& 0.66 & 47 & & 0.03 & 0.13 & 36 \\
S04 ($V \le 12$)  &$-0.01$& 0.38 & 54 & & 0.01 & 0.10 & 48 \\
\hline
\noalign{\vskip2pt}
\multicolumn{8}{p{12cm}}{Star to star comparison of NSVS and ASAS
photometric metallicities determined with methods of JK96 and S04.
${\rm [Fe/H]_{\rm diff}}$ stands for ${\rm [Fe/H]_{\rm ASAS}{-}[Fe/H]_{\rm NSVS}.}$}}

\vspace*{-15pt}
Left panel of Fig.~21 presents a comparison of JK96 metallicities computed
in this paper with the ones computed by K06 for NSVS RRab stars. The sample
consists of 190 stars. The right panel shows the same comparison for 332
variables with S04 metallicities. The mean difference between ASAS and NSVS
values is $-0.02$~dex for JK96 method and 0.09~dex for S04 and the scatter
is 0.50~dex and 0.42~dex respectively. When we take into account objects
with $V\le12$~mag the average S04 metallicity difference is reduced to
$-0.01$~dex and the dispersion is 0.38~dex.  Comparison of the two methods
(summarized in Table~4) shows that the method of JK96 produces more
consistent results for the whole sample, but it displays quite a big
scatter which is mainly the result of a difficulty of $\varphi_{31}$
parameter estimation for distinct light curves from different projects. S04
metallicities are on average visibly higher for ASAS than for NSVS as
expected, when taking into account the whole sample, but the effect
disappears for bright variables.
\begin{figure}[htb]
\begin{center}
\begin{tabular}{c c}
\includegraphics[width=5cm]{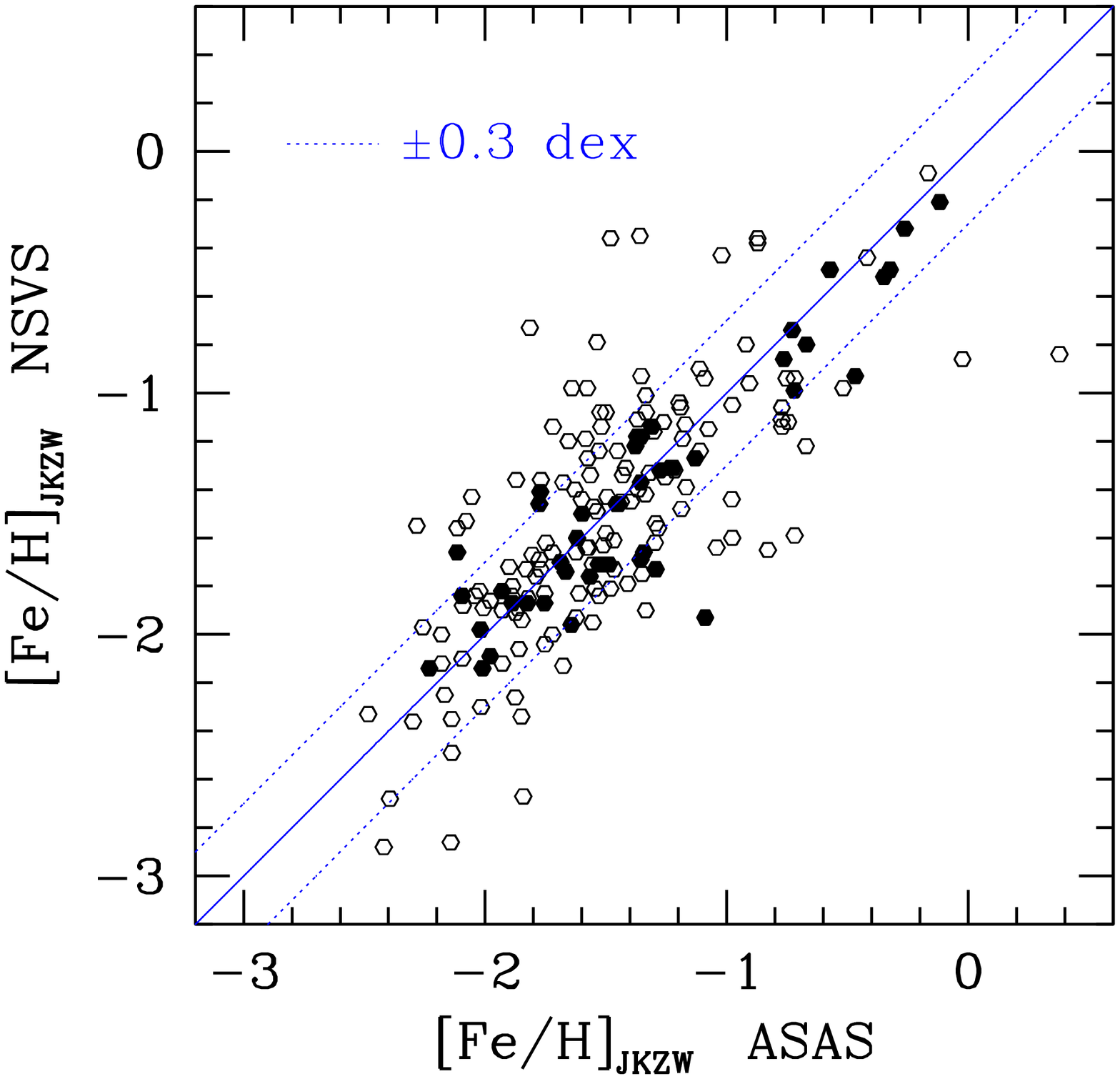} &
\includegraphics[width=5cm]{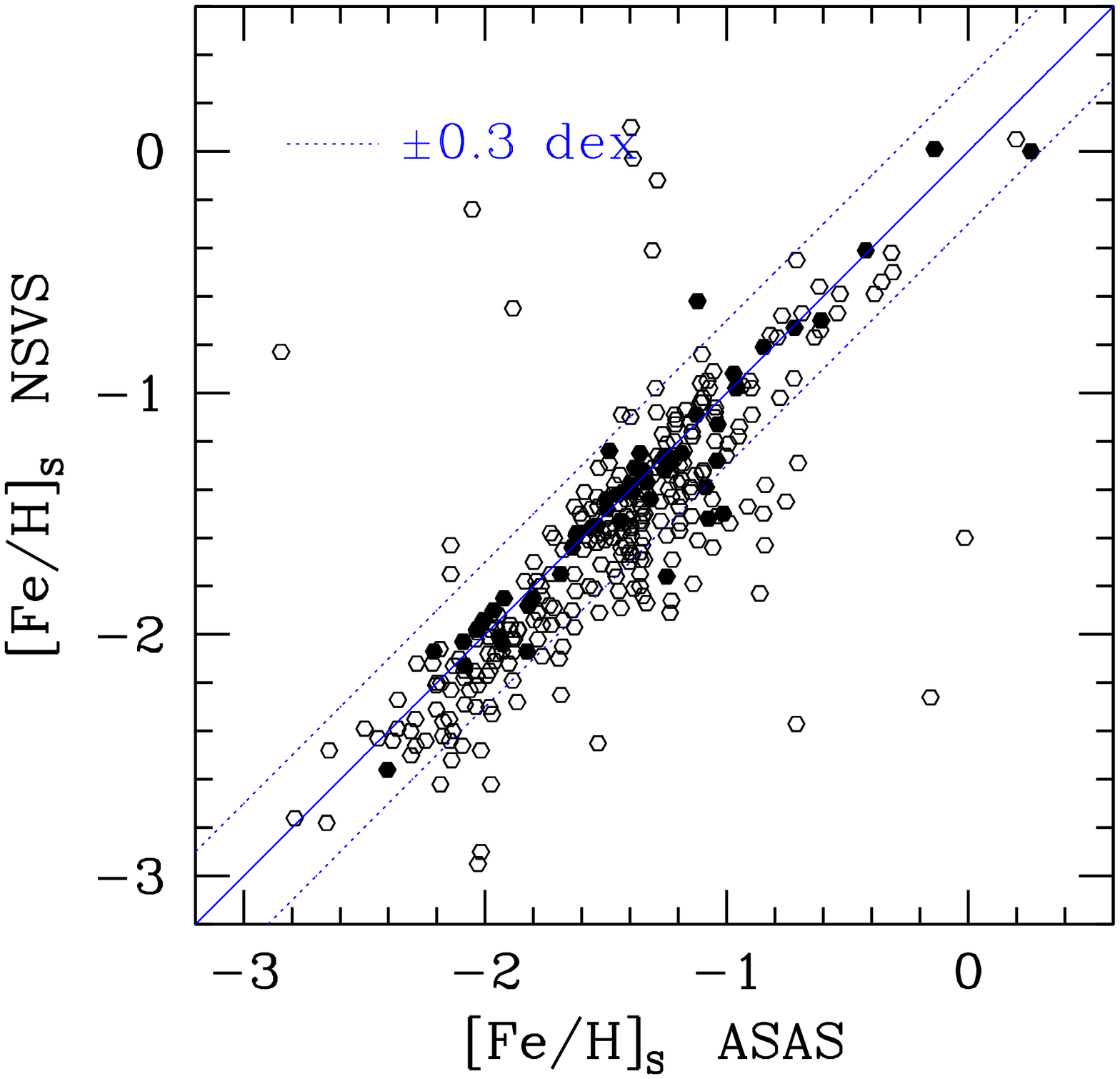} \\
\end{tabular}
\end{center}
\vspace*{-11pt}
\FigCap{Comparison of ASAS and NSVS metallicities computed from the
method of JK96 ({\it left panel}) and S04 ({\it right panel}). Filled
symbols stand for RRab stars with $V\le12$~mag.}
\end{figure}

Finally, in Fig.~22 we compare NSVS average metallicities with ASAS
photometric metallicities. As described in Section~4.5, to determine ASAS
photometric metallicities, we used JK96 method, as we believe they are more
reliable. The mean difference between ASAS and NSVS is $-0.03$~dex and the
dispersion is 0.66 dex for all 322 objects, while for a subsample of 190
stars the average is 0.04 dex and the dispersion is 0.43~dex. Such large
differences in metallicities noticeably affect absolute magnitudes and thus
distances: 0.3~dex difference in metallicity produces 50~pc difference in
distance for a star of $V=10$~mag and 130~pc for a $V=12$~mag star.
\begin{figure}[htb]
\begin{center}
\includegraphics[width=6.3cm]{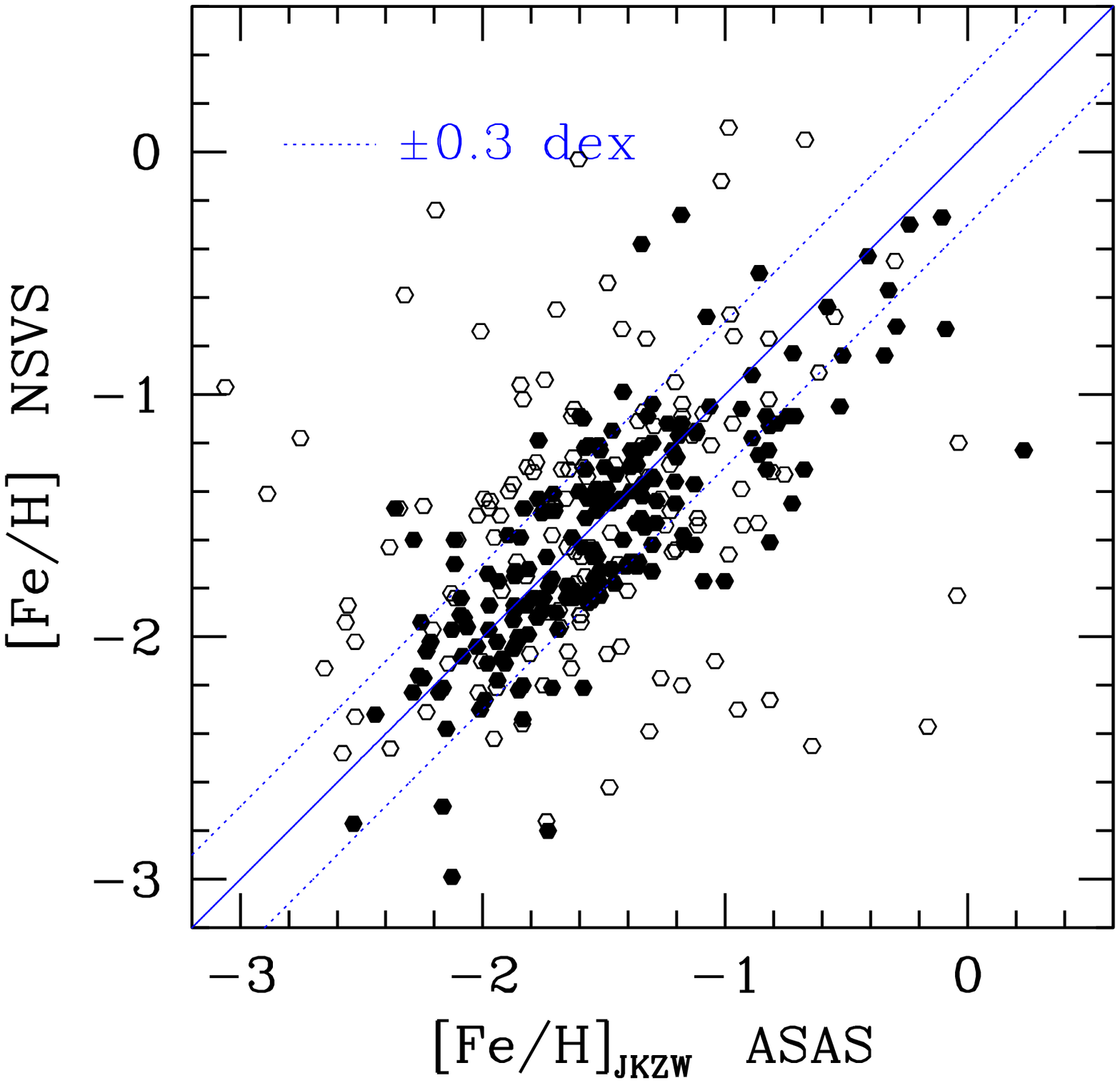} 
\end{center}
\vspace*{-11pt}
\FigCap{Comparison of ASAS JK96 metallicities and NSVS best estimate
metallicities for 332 RRab stars in the overlapping region. Filled circles
stand for 190 RRab stars for which NSVS metallicities from both methods
exist.}
\end{figure}
\end{document}